\documentclass[12pt]{article}

\usepackage{longtable}
\usepackage{amssymb,amsmath,amsfonts,eurosym,geometry,graphicx,caption,color,setspace,sectsty,comment,footmisc,caption,array,hyperref, natbib, subcaption, float, lscape, booktabs, multirow, wrapfig, pdflscape, tabu, threeparttablex, threeparttable, makecell, xcolor, nicefrac}
\usepackage[normalem]{ulem}
\usepackage{setspace}

\definecolor{fairpurple}{RGB}{108, 32, 50}
\definecolor{fairpink}{RGB}{204, 48, 122}
\definecolor{fairorange}{RGB}{231, 86, 66}
\definecolor{fairgray}{RGB}{171, 172, 176}
\hypersetup{colorlinks,citecolor=fairpurple,filecolor=fairorange,linkcolor=fairpurple,urlcolor=fairgray}

\usepackage{tabularx}
\usepackage{times}

\renewcommand{\theequation}{\arabic{equation}}
\newcommand{\subeqn}[1]{\refstepcounter{equation}\tag{\theequation#1}}

\usepackage{scrextend}

\providecommand{\keywords}[1]
{
  \small	
  \textbf{\textit{Keywords---}} #1
}

\providecommand{\jelcodes}[1]
{
  \small	
  \textbf{\textit{JEL---}} #1
}

\newcolumntype{L}[1]{>{\raggedright\let\newline\\arraybackslash\hspace{0pt}}m{#1}}
\newcolumntype{C}[1]{>{\centering\let\newline\\arraybackslash\hspace{0pt}}m{#1}}
\newcolumntype{R}[1]{>{\raggedleft\let\newline\\arraybackslash\hspace{0pt}}m{#1}}

\begin{document}

\title{Intergenerational Mobility Trends and the Changing Role of Female Labor \vspace{-2ex} \thanks{Thanks to Adrian Adermon, Eirik Berger, Aline Bütikofer, Claus Thustrup Kreiner, Søren Leth-Petersen, Mårten Palme, Kjell G. Salvanes, David Seim, Jósef Sigurdsson, Alexander Willén, participants at seminars in Bergen, Copenhagen, Stockholm and Uppsala, and to Marianne Page and three anonymous referees for constructive feedback on this paper. Ahrsjö gratefully acknowledges funding from Handelsbankens Forskningsstiftelser (Jan Wallanders och Tom Hedelius Stiftelse). Karadakic gratefully acknowledges funding by the Research Council of Norway through project No. 275800 and through its Centres of Excellence Scheme, FAIR project No. 262675 (Karadakic). The activities at CEBI are financed by the Danish National Research Foundation, Grant DNRF134 (Rasmussen). All errors are our own.}}

\author{Ulrika Ahrsjö, René Karadakic and Joachim Kahr Rasmussen \vspace{-1cm} \thanks{Ahrsjö: Department of Economics, Stockholm School of Economics, ulrika.ahrsjo@hhs.se; Karadakic: Department of Health Policy \& Management, Harvard T. H. Chan School of Public Health, rkaradakic@hsph.harvard.edu; Rasmussen (formerly affiliated with): CEBI, Department of Economics, University of Copenhagen, jkr@econ.ku.dk.}}
\date{\today}

\maketitle
\thispagestyle{empty} 
\vspace{-1cm}
\begin{abstract}
\onehalfspacing
Using harmonized administrative data from Scandinavia, we find that intergenerational rank associations in income have increased uniformly across Sweden, Denmark, and Norway for cohorts born between 1951 and 1979. By gender, father-son mobility remains stable, while correlations for mothers and daughters rise. Similar patterns appear in US survey data, albeit with different timing. We show that the decline in income mobility reflects a stronger link between female income and underlying productivity, rather than stronger intergenerational transmission of human capital or changes in assortative mating. Finally, we show that parent-child correlations increase mainly when women gain access to jobs that match their productivity.
\end{abstract}

\begin{center}
 \jelcodes{J62, J21}
\end{center}

\newpage
\pagenumbering{arabic} 
\section{Introduction}
\linespread{2}

A central theme in the social sciences is how the childhood family environment shapes economic fortune in adulthood. Empirical studies on the topic often estimate the relationship between the income of parents and their children and interpret low correlations as a sign of social mobility. Early work by \cite{BeckerTomes} and \cite{Solon1999Handbook} highlights that in such estimations, it is essential to account for the role of idiosyncratic labor market conditions. Accordingly, variation in labor market conditions may shape estimates of intergenerational mobility across space \citep{Solon2002CrossCountry, chetty2014land, bratberg2017comparison} and time \citep{corak2013income}. While substantial time-variation in mobility has been documented \citep[see e.g.][]{Solon_lee, Olivetti2015Names, Chetty_still_land}, little is known about how changes in labor market conditions shape these patterns \citep[see e.g.][for a notable exception]{Song2020PNAS}. 

In this paper, we ask how the ``grand convergence'' in labor market conditions between men and women \citep{goldin2014grand} has influenced estimates of intergenerational income mobility. Historically, income has not been reflective of inheritable productivity among women for at least three reasons: we don't observe income for non-workers, selection into market work has made the average productivity among working women unrepresentative of the population average \citep{blau2021selection}, and discrimination against female workers has hampered their occupational choices and career advancement. Over the past 50 years, women in all Western economies have become more likely to participate in market work \citep{Olivetti2016Gendergaps}, and occupational segregation of men and women has decreased \citep{Blau2013Segregation}. These changes have increased the degree to which labor income reflects productivity, thus potentially strengthening the estimated correlation in income between mothers and their children, daughters and their fathers, and indeed between children of both sexes and their parents' household income.\footnote{Previous research has largely excluded women from estimates of intergenerational income persistence precisely because female income has not been a reliable indicator of social status historically \citep{chadwick2002intergenerational, bjorklund2009family, blanden2004britain}. Others have highlighted that cross-sectional estimates of intergenerational mobility may differ substantially by gender due to different opportunities for men and women in the labor market \citep{corak2013income, Solon_lee}.}

Two points are important to make regarding this. First, the recent literature on intergenerational income mobility often studies parent-child income correlations using the average of the mother's and the father's household or individual income as the parental income measure \citep{chetty2014land, bratberg2017comparison, connolly_haeck_laliberte_NBER}. In times of rapidly changing female labor force attachment, caution is warranted when estimating mobility trends using such income definitions in a context like ours \citep[see e.g.][]{markussen2020economic, Harding2019}.\footnote{Importantly, we show that in our setting, results using the \citet{chetty2014land, Chetty_still_land} definition of parent and child income when estimating mobility, i.e., average household income of the focal person and their spouse, lead to the same conclusions as results using individual income.} Second, it is not immediately obvious that increased female labor force participation \textit{per se} results in increased mother-child and father-daughter income correlations, if women are unable to access jobs and job positions that generate pay to match their productivity.

We explore how changes in female labor supply affect intergenerational mobility by turning to the three Scandinavian countries. The high-quality Scandinavian administrative data include individual income, allowing us to study separately how mother-child and father-child correlations in labor income have evolved. Scandinavia provides an ideal setting, as the development towards gender equality precedes that of other countries \citep{kleven2019children}. We document trends in intergenerational mobility in Sweden, Denmark, and Norway for cohorts of children born between 1951 and 1979, using administrative income data from 1968 until 2017. By applying a unified approach to long panels of full-population administrative data for three countries, we can ensure that any differences in findings are not related to country-specific developments, the choice of data period, or the income definition.\footnote{In particular, we show that the results are largely unchanged when studying intergenerational correlations in log income rather than within-cohort income ranks, when considering intergenerational correlations in gross or net-of-tax income rather than labor income, and when using household instead of individual income. This suggests that the observed trends were not driven by simultaneous rank-distorting tax or social transfer reforms across Scandinavia.} 

Our results reveal a substantial decline in intergenerational income mobility in Scandinavia that remains robust across a large set of common empirical specifications. However, this decline is only present among mother-child and father-daughter pairs. In the earliest cohorts in our analysis, child income --- in particular, the income of sons --- were virtually uncorrelated with the mothers' income, while exhibiting a clear and economically significant correlation with the fathers' income. Over time, these estimates between sons and daughters and their mothers and fathers, respectively, have all converged to similar levels. 

Conducting a similar analysis on Panel Study of Income Dynamics (PSID) data from the US, we find comparable patterns in mother-child and father-child income persistence, albeit with a slightly lagged timing. This is suggestive evidence that the observed patterns are not solely a Scandinavian phenomenon. Despite its small sample size by birth cohorts (averaging 150 parent-child pairs) \citep{davis_mazumder_2022}, the PSID is regularly used to study parent-child income persistence \citep{hertz2008group, hartley2022welfare}. See \citet{mazumder2018psid} for an overview of the PSID in research on intergenerational mobility.

Thus, income persistence seems to be rising because women's \textit{realized income} (i.e. observed income) better reflects the income-generating skills and values that are transmissible between generations (from now on referred to as ``productivity''). We denote the income that a person would have if they applied (and earned the market return to) their productivity in the labor market as \textit{potential income}. Our key hypothesis is that the intergenerational correlation in potential income has remained unchanged, while the intergenerational correlation in realized income has strengthened. 

An alternative, more worrying, possibility is that persistence has risen because the skill transmission between generations has increased over time. We do two things to rule this out. First, we contrast mother-child and father-daughter correlations in realized income (that is, our standard income rank associations) with correlations in estimated potential income, which we get by combining the informational value of income, years of education, and occupation using the proxy variable index suggested by \citet{LW2006}. Parent-child rank correlations in this measure are constant or only weakly increasing over time.\footnote{This method has previously been applied by \citet{VostersNybom2017} and \citet{Vosters2018} to understand family correlations beyond income, i.e., how intergenerational persistence measures increase when supplementing the informational value from income with education and occupation, for men as well as women. \citet{Adermon2021Dynastic} use the LW method to estimate multigenerational correlations in economic status.} The increase in intergenerational persistence is thus driven by correlations in income, rather than other skill measures. Mobility also remains constant when correlating the income of sons with those of their maternal uncles, as another proxy for maternal skills.

Second, we set up a model of gender-specific mobility and latent productivity, inspired by the model in \citet{BeckerTomes} and similar to \citet{collado2022estimating}. We decompose the trend in intergenerational mobility into parts that reflect assortative mating, gender-neutral skills transmission, gender-specific skills transmission, and gender-specific return on skill. Calibrating the model for birth cohorts 1962-1979, we show that the downward mobility trend is largely compatible with increasing returns to (inheritable) productivity among women, relative to men. We find no evidence of significant changes in the degree of assortative mating among parents. The decomposition thus suggests that the observed trends in income mobility are due to changes in how women participate in the labor market. 

We finally show that rising female labor force participation on the extensive margin is not the main driver of the observed trend. Most women who entered the labor force when female labor force participation was rapidly expanding worked in low-skill and low-paying occupations, whereby their labor income did not reflect their innate productivity.\footnote{We also show that women in market work were initially a highly positively selected group and that the degree of positive selection subsequently declined as more women entered the labor force.} Instead, most of the trend can be explained by women gaining access to more skilled occupations, which increased the gradient between productivity and realized income.
  
In sum, we show that intergenerational income mobility declined consistently in Scandinavia across cohorts born between 1951 and 1979, and that this decline was driven by female income becoming more reflective of productivity. The return on latent productivity of women has converged towards that of men. At the same time, we do not find evidence supporting increasing skill transmission between parents and their children, or increasing assortative mating among parents. The observed development in intergenerational mobility can thus be seen as an implication of gender equality in the labor market, rather than a sign of declining social mobility.

The literature on trends in intergenerational mobility includes results from a wide range of settings and empirical specifications, all of which affect conclusions on the existence and direction of trends \citep{engzell2023}. Trend estimates focusing on birth cohorts similar to those included in our study generally find increasing levels of income persistence, including evidence from Canada \citep{connolly2019trends}, the US \citep{jacome2021, davis_mazumder_2022}, the UK \citep{blanden2004britain}, Sweden \citep{brandennybom2019, engzell2023}, Norway \citep{markussen2020economic} and Denmark \citep{Harding2019}.\footnote{A set of studies also document lower intergenerational income persistence over time for cohorts born in the 1930s and 1940 in the US and the Nordic countries \cite{pekkala2007differences, bjorklund2009family, pekkarinen2016evolution, Song2020PNAS, jacome2021}. Moreover, \cite{Solon_lee} find no change in income persistence between birth cohorts 1952 to 1975 in US survey data.} Explanations for declining mobility put forward include returns to education \citep{davis_mazumder_2022}, maternal education \citep{connolly_haeck_laliberte_NBER}, and family formation, including assortative mating \citep{Harding2019}.

We make two contributions to our understanding of time variation in intergenerational mobility. First, we provide clear evidence of a uniform decline in intergenerational income mobility across Scandinavia for cohorts born between 1951 and 1979. We demonstrate that this trend persists regardless of methodological choices and does not stem from country-specific policies, thereby corroborating the results in \citet{Harding2019, pekkarinen2016evolution, brandennybom2019} and \cite{markussen2020economic}. We further show that these trends are robust even when applying the widely used definition of household income from \citet{chetty2014land}. In addition, we extend this cross-country comparison to gender-specific trends of mothers, fathers, sons, and daughters, and provide evidence of a similar pattern in the United States from panels of linked survey data. Our findings suggest that not only do mobility \textit{levels} vary by gender \citep[as previously discussed by][]{chadwick2002intergenerational, Solon_lee, corak2013income}, but that secular changes in gender-specific income determinants have also caused \textit{trends} to differ substantially, in turn causing levels to converge.\footnote{Studies on gender and trends in intergenerational mobility mainly differentiate between sons and daughters \citep{blanden2004britain, pekkala2007differences, connolly2019trends}. \citet{Harding2019} find sharply increasing parent-daughter and parent-son correlations for Danish cohorts born between 1957 and 1977. For Norway, \citet{markussen2020economic} and \citet{pekkarinen2016evolution} find declining levels of mobility for daughters, but stable trend estimates for sons.}

Our second and main contribution is to show that the decline in mobility can be fully (Sweden, Norway) or mostly (Denmark) accounted for by changing female labor supply. Others have studied mobility trends by gender of both parents and children \citep{jacome2021,engzell2023}, but we show explicitly how the gradual integration of women into the labor force has affected estimates of intergenerational income mobility. In particular, we show that female labor force participation \textit{per se} does not induce higher parent-child correlations, but that the crucial element is women gaining access to jobs that match their skills. Our results align with those in parallel work by \citet{branden_nybom_vosters_jole}, with the important difference that we show how both mothers and daughters are crucial for the observed trend. In contrast to previous research from Scandinavia \citep{Harding2019, markussen2020economic}, we provide a reassuring explanation for the declining levels of income mobility: that female labor market conditions have caused parental income to be substantially better reflected in child income --- hence inducing a \textit{real} downward shift in intergenerational mobility --- despite the between-generation correlation in latent potential income remaining constant.

The remainder of the paper is structured as follows. Section \ref{SEC:institution} provides a brief overview of the key features of the Scandinavian welfare states, and Section \ref{SEC:data} describes our data sources. In Section \ref{SEC:main_results}, we describe the common methodology used to estimate intergenerational income mobility and present our main results. Section \ref{SEC:realized_potential} discusses whether the results can be attributed to increased female income alone, and Section \ref{SEC:mechanisms} discusses how sorting between skills and occupations affects our estimated trends. Section \ref{SEC:concl} concludes.

\section{Institutional Context} \label{SEC:institution}

The Scandinavian countries share similar traits in terms of economic development, political culture, and institutions. The welfare states are of universal character, with access to social security benefits, health care, subsidized childcare, and tuition-free higher education \citep{6155250}. In order to finance the provision of these public goods, marginal tax rates at the top of the income distribution and the average tax burden are substantially higher in Scandinavia than in other developed countries \citep{kleven2014can}. Employees are to a large degree organized in unions, and wages are often collectively bargained \citep{Pareliussen2018}. Historically, all three countries have been characterized by low levels of inequality and high levels of income mobility, in comparison to other Western countries \citep{egholt2018, bratberg2017comparison}.

During the second half of the 20th century, the role of women in society, and particularly in the labor market, converged towards that of men \citep{goldin2014grand}. Individualization of the tax system \citep{selin2014rise}, the introduction and expansion of paid paternity leave \citep{ruhm1998economic}, and the expansion of compulsory and higher education \citep{meghir2005educational, black2005apple} all contributed to this development. As a result, female labor force participation increased from the early 1950s and is currently higher in Scandinavia than in most other Western economies (see Appendix Figure \ref{fig:lfp_comparison} for a comparison between Scandinavia and the United States).

In Figure \ref{fig:labor_development}, we provide some descriptive evidence on the development of female labor for the parent and child generations separately. Panels A and B show how labor force participation among women converged to the male level. Participation rates of mothers with children born in the 1950s were less than half the rate of fathers, but this gap had closed almost entirely for mothers of children born in the 1970s. It is even less pronounced when we compare sons and daughters of a given birth year. 

Panels C and D show the development of occupational segregation, i.e., the extent to which men and women work in the same occupations. The segregation index is calculated as the difference in the share of all women and men in the labor force who work in a given occupation, summed over all observed occupations \citep{Duncan1955}. To make comparisons of trends easier, we normalize the index with 1962 as the base year, allowing for an interpretation of occupational segregation relative to the 1962 level.\footnote{The occupational segregation index is defined by three-digit occupation codes for Norway and Sweden and one-digit codes for Denmark due to data limitations. Therefore, the cross-country difference in trends should not be interpreted as hard evidence of deviating patterns of occupational segregation.} Evidently, occupational segregation has declined persistently over time, similar to the development in the United States, as documented by \citet{Blau2013Segregation} and \citet{blau2017gender}. In contrast to the development of female labor force participation, the decline in occupational segregation is to a larger extent present in the child generation, rather than the parent generation. 

In addition, the intensive margin labor supply of women increased during the time period under study. Panels E and F provide female-to-male ratios of the share of individuals working full-time, by birth year of the child. For Denmark and Sweden, full-time work is defined as at least 27 working hours per week, while full-time in the Norwegian data is defined as at least 31 weekly hours. In Panel E, we present this for mothers relative to fathers. Similar to the development in labor force participation, mothers have continuously caught up to the rate at which fathers work full time, although a sizable difference remains toward the end of our sample period. The level differences between Sweden/Denmark and Norway stem from different full-time definitions in the data. Moreover, the convergence in intensive margin labor supply in all countries is almost entirely driven by increases in female full-time shares; male full-time shares are almost constant over the entire time period under study. Especially in Norway, our data reveals a significant remaining gender gap in hours worked. Notably, the Swedish female-to-male full-time ratio has increased at a higher rate since around 1960. The full-time ratio of daughters compared to sons, in Panel F, shows two conflicting observations. On the one hand, looking at Sweden and Denmark, the female full-time rate in the 1960s cohort was already quite high, with only small changes after that. On the other hand, the Norwegian series, using a stricter definition, suggests that a significant gender gap in working hours persists also among the child generation of our sample. 

Overall, the three labor market measures presented in Figure \ref{fig:labor_development} show a substantial gender convergence in labor market participation. Convergence in extensive margin labor force participation of mothers happened faster before the 1960s birth cohort, and was almost entirely equal to the fathers' level for the 1979 cohort. Changes in occupational choice and intensive margin labor supply of mothers were, however, more predominant in the second part of our sample, after the 1960 cohort, compared to those born before 1960. We will later argue that both the extensive and intensive margin developments in labor market participation have key implications for our measures of trends in intergenerational mobility.

\section{Data} \label{SEC:data}

For our main analysis, we use register data from Denmark, Norway, and Sweden that cover the whole population of each country. This is available from 1968 to 2017 for Norway and Sweden, and from 1980 to 2017 for Denmark. The data consist of linked administrative records that provide information on birth year, educational attainment, income measures, family status, and various demographic variables. Individuals can be linked to their parents, which allows us to create data sets containing all child-parent pairs in a given time frame with relevant individual income measures. For more details about the registers we use, see Appendix \ref{APP:data_registers}.

Our Scandinavian estimation sample consists of all children born between 1951 (1962 for Denmark) and 1979, who (i) have a valid personal identifier, and (ii) have at least one parent with a valid identifier. As this means that we remove a significant share of immigrants from our samples --- in particular in early years --- we remove all foreign-born individuals and all children with foreign-born parents. Sample sizes per birth year are approximately 70,000 child-parent pairs in Denmark, 60,000 pairs in Norway, and 100,000 pairs in Sweden, with variation over time. See Appendix Tables \ref{tab:summary_stats_children} and \ref{tab:summary_stats_parents} for sample sizes and shares of mothers and fathers with missing IDs by birth year of their child.

Moreover, we look at parent-child income correlations in the US using data from the Panel Study of Income Dynamics (PSID). The PSID is a nationally representative survey that covers information on employment, income, occupation, education, and family links, starting from 1968. The PSID follows families and individuals across time and has a relatively low attrition rate. With this data, we create a sample of child-parent pairs for the US in a comparable, yet more limited, fashion than our analysis on the main Scandinavian samples. In total, the US sample contains about 5,000 child-parent pairs; see Appendix Table \ref{tab:psidira_count} for sample sizes by birth years. 

The PSID offers a nationally representative sample from the Survey Research Center (SRC) as well as a non-representative sample that predominantly comprises low-income families, known as the Survey of Economic Opportunities (SEO). Our primary dataset merges information from both of these sources. However, we present supplementary evidence in the Appendix that supports the trends observed in the SRC sample alone, though with slightly broader confidence intervals. Despite its utility, the PSID has certain constraints for estimating intergenerational mobility trends \citep{mazumder2018psid}. Consequently, we put greater emphasis on our results from the higher-quality Scandinavian administrative data.  

The main income specifications are chosen for easy comparisons with much of the recent literature (see e.g. \citet{chetty2014land} and \citet{Solon_lee}). Income for the child generation is defined as three-year averages of annual labor income at ages 35-37, which balances the need for a measure of permanent income rank with the need for measuring child income relatively early to maximize the number of cohorts that can be included in the analysis \citep{nybom2016heterogeneous, bhuller2017life}.\footnote{Averages are calculated including zeroes. Individuals registered as residents in the population files of the respective administrative datasets, but without reported income, are assigned zero income. We provide summary statistics on this in the Appendix (see Table \ref{tab:summary_stats_children} for children and Table \ref{tab:summary_stats_parents} for parents).} Like previous studies from the Scandinavian context, such as \citet{Harding2019} and \citet{markussen2020economic}, we base our income measure on individual income, which is well measured in our administrative (tax record) data due to individual taxation. However, we also calculate household income according to the definition in \citet{chetty2014land} (average household income of the mother and father), with the caveat that this measure is only as good as the registers on spousal and co-habitation relationships. See Appendix Table \ref{tab:inc_def} for an overview of the income components and how these compare across countries.

Parental income is defined as the average of maternal and paternal individual income, measured as three-year averages of annual labor income around age 18 of the child. In general, this means measuring the parents' income at age 40 or later, which is considered a meaningful proxy for lifetime income in the literature \citep{nybom2016heterogeneous}. In our Appendix, we provide robustness checks to different income definitions for child and parent income variables, such as estimating trends in total factor (gross) income or net-of-tax income, and evaluating the sensitivity to the exact age at which we measure child and parent income. Ranking parent income within the birth year of both the child \textit{and} the parent jointly, we verify that the observed mobility trends are not driven by changes in the timing of childbirth, which affects the age at which we measure income among the parents.

\section{Trends in Intergenerational Mobility} \label{SEC:main_results}

In this section, we first describe the empirical method we use for measuring child-parent rank associations. We then present aggregate trends for Scandinavia, followed by trends split by the gender of the parent and child for Scandinavia and the US.

\subsection{Empirical Method} \label{SEC:method}
In order to measure intergenerational income persistence, we transform observed income into cohort-specific percentile ranks, as in \citet{dahl2008association} and \citet{chetty2014land}, among others. Using ranks, rather than levels or logs, offers certain advantages in this context: rank correlations are less prone to life cycle bias than other measures \citep{Nybom2017BiasesIS}, and the use of ranks allows the inclusion of null income. However, to ensure that our results are not driven by the rank transformation, we also present mobility trends in intergenerational income elasticities (IGE) in the Appendix. 

Rank correlations are estimated with the following regression, separately by birth cohort and country:
\begin{equation} \label{rank_equation}
\text{Rank}_{it}^{\text{C}}=\alpha_{t}+\beta_{t}\text{Rank}_{it}^{\text{P}}+\varepsilon_{it}, \qquad \forall \quad t \in \left[1951, 1979\right],
\end{equation}
where $\text{Rank}_{it}^{\text{C}}$ is the percentile rank of child $i$'s average income at age 35-37 within the distribution of all children born in year $t$. When we analyze sons and daughters separately, we calculate their income rank separately by gender. $\text{Rank}_{it}^{\text{P}}$ is the percentile rank of the same child's parents' income within the distribution of all parents with children born in year $t$, averaged over the age 17-19 of the child.\footnote{When delineating income based on the gender of either the parents or offspring, we prioritize re-ranking, to gain full support in our ranks for all specifications, given the notably lower income levels for mothers in the earliest cohorts. The re-ranking of children by gender does not alter the estimated trends significantly, but results in small upward shifts in the level of the IRA.} The coefficient $\beta_{t}$ captures the average cohort-specific parent-child correlation in income ranks, sometimes referred to as the intergenerational rank association (IRA). Lower values of $\beta_{t}$ are interpreted as lower rank associations in income, and thus higher levels of intergenerational mobility. 

Intuitively, one can think of the IRA as an estimate of the correlation in inheritable skills and values that are transmitted across generations, influenced by the extent to which income reflects these. Changes in the IRA over time are thus not necessarily driven by transmissible factors, but rather by the importance of income determinants that cannot be passed on to children. When analyzing how changing female labor market participation may have affected the intergenerational association in income, this is a relevant consideration.

\subsection{Estimated Trends}  \label{SEC:results}

\begin{figure}[H]
\linespread{1}
\begin{center}
\includegraphics[width=1\linewidth]{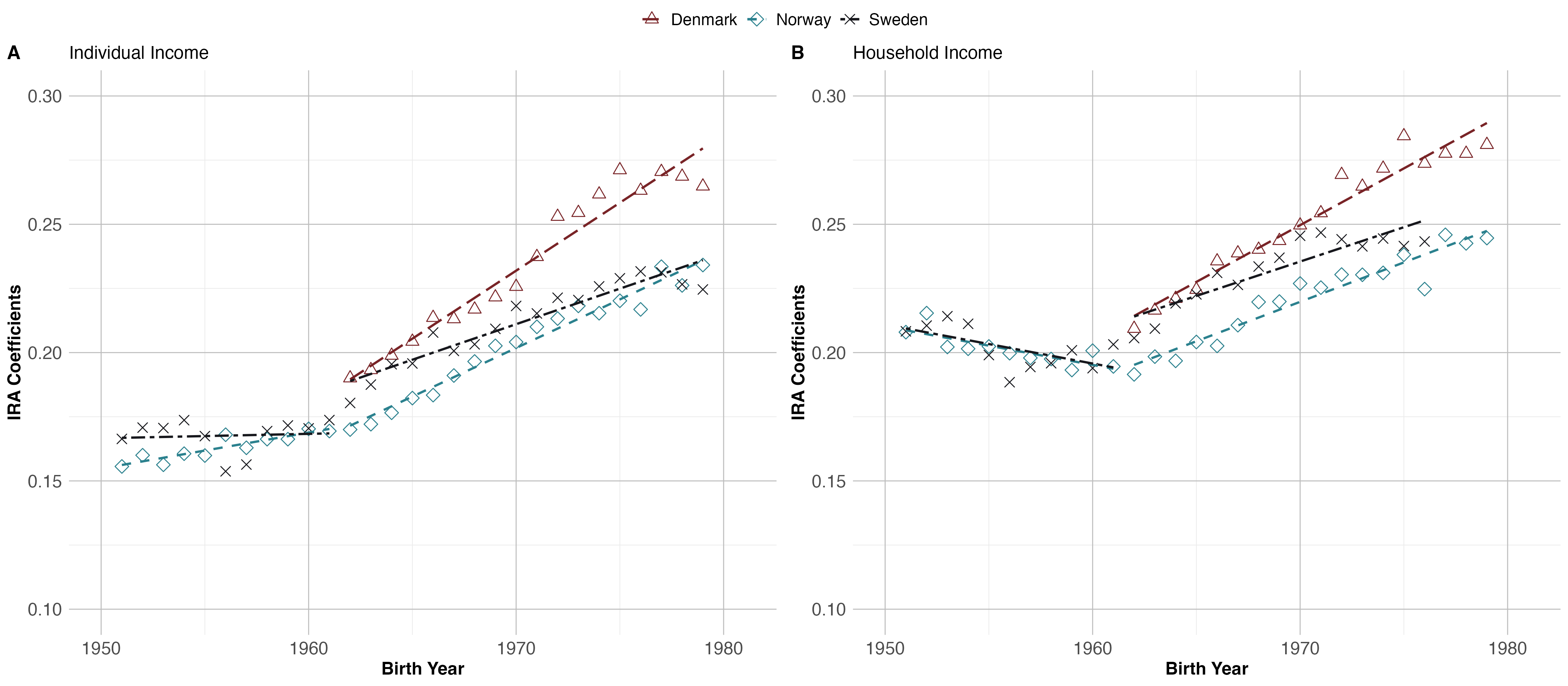}
\end{center}
\caption{Trends in Intergenerational Mobility.}
\label{fig:trends_overall}
\vspace{0.3cm} 
\scriptsize{\textit{Notes:} The figure plots the coefficients for the intergenerational rank association (eq. \ref{rank_equation}) for Sweden, Denmark, and Norway for birth cohorts 1951 (1962) to 1979. ``Birth Year'' refers to the birth year of the \textbf{child} in each parent-child pair. Each panel shows fitted trend lines separately by country and for the periods 1951 to 1962 and 1962 to 1979. Panel A plots the intergenerational rank association in individual labor income, and Panel B in household income between parents and children, where household income in the offspring generation is the sum of the child's and the child's spouse's income. Household income in the parent generation is defined as the sum of the mother's and father's income (incl. their respective spouses' income) divided by two.}
\end{figure}

In Figure \ref{fig:trends_overall}, we present estimates for trends in intergenerational rank associations in individual labor income (Panel A) and household income (Panel B), for Denmark, Norway, and Sweden. Each point in the graph represents a slope parameter for a cohort-specific regression of Equation \eqref{rank_equation}. The figure also includes linear trends, which are estimated separately for 1951-1961 and 1962-1979 to facilitate comparisons between Denmark, Norway, and Sweden for the cohorts where all countries have available data. Appendix Table \ref{tab:ira_coefs} provides an overview of the IRA coefficients for different specifications and tests whether trends are statistically different between countries.

Panel A of Figure \ref{fig:trends_overall} shows that intergenerational mobility, measured using the IRA, has declined in all three countries, with the fastest rate of decline in Denmark. There, the rank association in income increased by 7.5 rank points (39\%) from 1962 to 1979 --- equivalent to an average annual increase of 0.44 rank points. The trends in Norway and Sweden are less pronounced. From 1962 to 1979, the rank association in income increased by 6.4 and 4.4 rank points (38\% vs. 25\%) in Norway and Sweden, respectively, yielding annual increases of 0.38 and 0.25. From 1951 to 1979, the total change in IRA for Norway is 7.8 rank points (50\%) and 5.8 rank points for Sweden (34\%).

Abstracting from nonlinearities in the relationship between parent and child income ranks, a straightforward interpretation of these estimates is the following: for two children born by parents in the bottom versus the top percentile, the difference in the conditional expectation of their income ranks as adults increased by up to 0.44 each year, amounting to as much as 4.4 rank points over a decade. Taking the Norwegian results as an example, another interpretation of the observed trends is that in the earliest observed birth cohort, a ten rank points difference in parental income corresponded to an average difference in income ranks of 1.6 between their children. The same difference was 2.3 rank points for children born in the last cohort. While still indicating relatively high levels of mobility by international standards, such changes over a relatively short time are economically substantial. 

Importantly, Figure \ref{fig:trends_overall} Panel B indicates that trends in household income are similar, albeit slightly smaller, for the three countries. The level of the IRA in household income is higher, but trends in the intergenerational rank association are robust to the choice of income measures. 

We document similar trends for a large set of different specifications in Appendix \ref{APP:figures}. Most importantly, we show that the trends remain similar when measured in net-of-tax income and gross income (Figure \ref{fig:income_types}), and when measuring child income at various ages (Figure \ref{fig:ages_income}). In Figure \ref{fig:ira_excluding}, we restrict the sample to parent-child pairs with labor income surpassing 10,000 USD (2017). In other words, we calculate rank associations for the subset of the population that is fully active in the labor market. Figure \ref{fig:ira_ige} compares mobility trends in rank associations (IRA) to trends in the Intergenerational Income Elasticity (IGE). The mobility trends persist and are similar in magnitude across all specifications.\footnote{However, some cross-country differences are also revealed. Rank associations in Denmark and Norway are lower when excluding non-participating workers from our samples, indicating that \emph{intergenerational correlations in labor market participation} contribute greatly to intergenerational persistence in income --- or at least that children of non-participating parents do disproportionately bad in the labor market themselves. In Sweden, on the other hand, the level of mobility largely remains the same after excluding non-participating parents from the estimation sample (Panel B), and even increases slightly when excluding both non-participating parents and children (Panel C).}

\subsection{Trends by Gender of the Child and Parent} \label{SEC:trends_gender}

\begin{figure}[H]
\linespread{1}
\begin{center}
\includegraphics[width=1\linewidth]{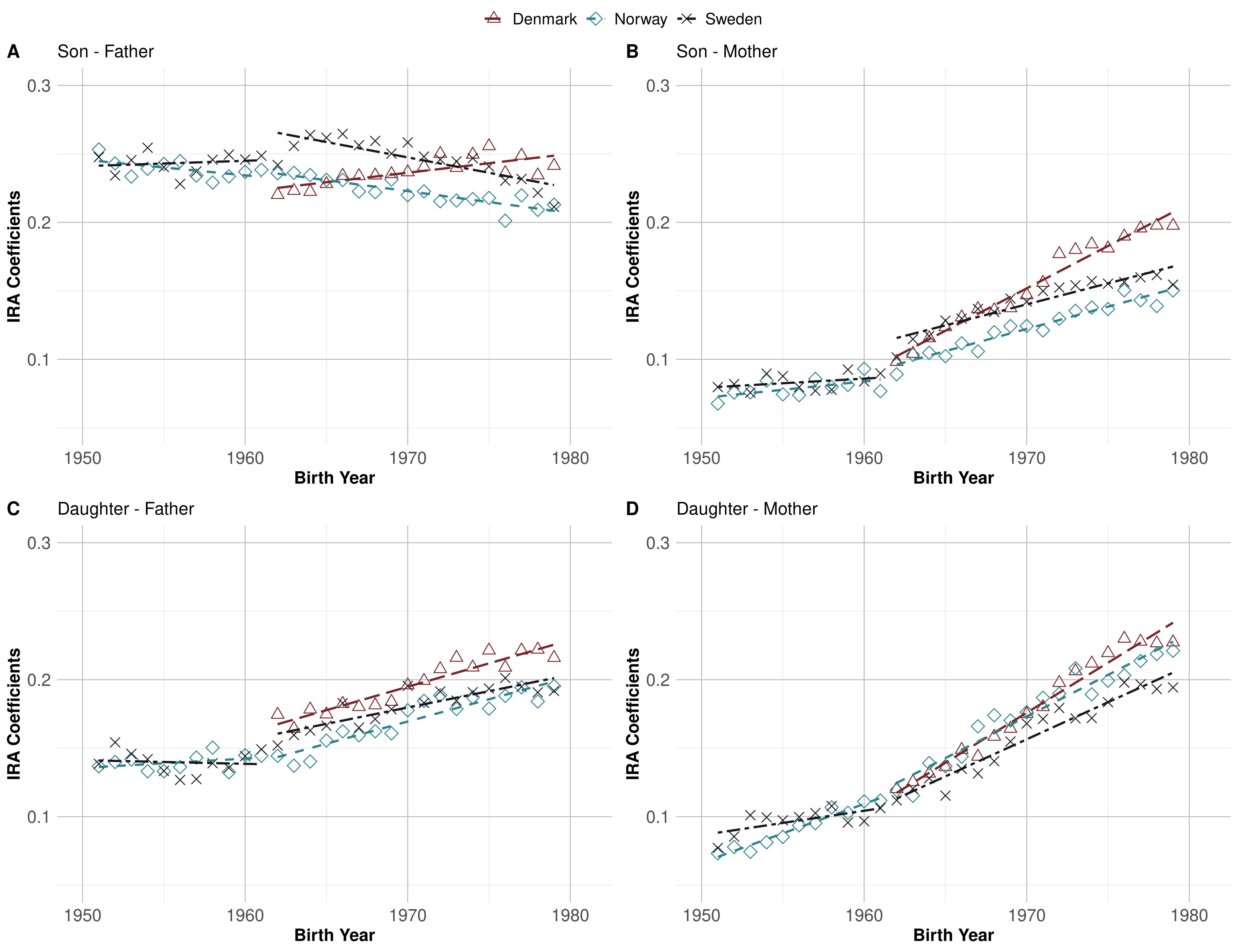}
\end{center}
\caption{Trends in Intergenerational Mobility by Gender of the Parent and Child.}
\label{ira_sep}
\vspace{0.3cm}
\scriptsize{\textit{Notes:} The four panels plot coefficients for the intergenerational rank association (eq. \ref{rank_equation}) in individual labor income for Denmark, Sweden, and Norway, by child year of birth. Each panel provides estimates separately by gender of the parent and child. Each marker indicates the coefficient of a separate regression, and each line indicates fitted trend lines separately for birth cohorts 1951 to 1962 and 1962 to 1979. ``Birth Year'' refers to the birth year of the \textbf{child} in each parent-child pair.}
\end{figure}

Figure \ref{ira_sep} presents estimates of country-specific IRA coefficients for pairs consisting of, in turn, sons and fathers (Panel A), sons and mothers (Panel B), daughters and fathers (Panel C), and daughters and mothers (Panel D). Each coefficient is again obtained by estimating Equation \eqref{rank_equation} for each birth cohort, for the respective combination of child and parent, and with ranks of individual mother or father income instead of the parental average. In Appendix Table \ref{tab:ira_coefs}, we test several hypotheses regarding the trends and also report slope coefficients for different specifications. This table also provides IRA estimates and trends separately by gender of the child, using the parental average income ranks.

The four sets of graphs make clear that while rank associations are increasing for all parent-child pairs involving women, the rank association between fathers and sons is generally \textit{decreasing} (Sweden, Norway) or displays a relatively flat trend over time (Denmark). The strongest trends in IRAs are found for mother-daughter correlations (DK:0.7, NO:0.6, SE:0.55 rank points per year), followed by mother-son correlations (DK:0.6, NO:0.3, SE:0.3 rank points per year). Father-daughter correlations display slightly weaker trends in Sweden and Denmark, but the trend in Norway matches that of mother-son. 

Starting with the 1962 cohort, the trends in IRA for all combinations of child and parent are similar in Sweden, Denmark, and Norway. Estimates for birth cohorts 1951-1979 are strikingly similar in Norway and Sweden: the trends are statistically indistinguishable for all combinations and years except for the trend in the mother-daughter IRAs after 1961. Notably, father-son correlations in Denmark display a weakly increasing pattern in 1962-1975; a deviant pattern compared to Sweden and Norway that is also found for trends in \textit{absolute} mobility in \citet{manduca2020trends}. 

To understand whether the observed mobility patterns are found outside Scandinavia, we compute comparable mobility estimates for the US for birth cohorts from 1947 to 1983. Results from this exercise are presented in Table \ref{tab:psidira}.\footnote{In Appendix Table \ref{tab:psidirafull}, we provide similar estimates with alternative sample specifications and weighting procedures. In Table \ref{tab:psidira_count}, we document the cohort-specific number of parent-child pairs used to compute these trends. Due to the small sample sizes, trends have been estimated directly on the underlying micro data by regressing cohort-specific child ranks on cohort-specific parent ranks interacted with a linear time trend.} US mobility trends are steepest for pairs involving women, and in particular daughters, while father-son rank associations appear to be relatively constant in the US, suggesting a comparable development to that observed in Scandinavia \citep[similar results are shown in][]{Song2020PNAS}. However, the US trends in mother-son correlations are not statistically significant.

Another feature of Figure \ref{ira_sep} and Table \ref{tab:psidira} is that the income of mothers is more strongly related to daughters' income than to sons' income in later cohorts. In fact, while the association in income ranks is generally higher among sons and fathers than among any other combination of genders, the daughter-mother correlation reaches the same level towards the end of the considered period in Scandinavia. For the US, we only provide a pooled IRA coefficient due to the small sample. Nevertheless, this pattern is also found in US data.\footnote{This finding could have several reasons, such as intergenerational occupational mobility being lower within- than across gender, and the general tendency of men and women to sort into different occupations (see e.g., \citet{blau2017gender} for a review on this latter point). \citet{AltonjiDunn2000} also find within-gender correlations in work hour preferences between parents and children, and a recent working paper by \citet{Galassi2021} highlights how employment correlates between mothers and their children, especially for daughters.}

Since all combinations of parent-child correlations that yield upward trends in IRAs involve women, a close-at-hand explanation lies in that women's increasing integration into the labor force has changed the way that income is correlated across generations. The difference in maternal trends between the US and Scandinavia would also be in line with such an explanation, as developments in female labor force participation started later in the United States and therefore likely impacted mothers only for later-born cohorts, while having a potentially larger impact through changing labor market equality for daughters.\footnote{The validity of this explanation is confirmed in Table \ref{tab:psidira30}. Here, we estimate child income around age 30 rather than 36, allowing us to compute US gender-specific IRAs for cohorts of children born in 1953-1989 rather than 1947-1983. Looking at this set of later-born children, we find that IRAs including mothers exhibit a clear and significant upward trend.}

\section{Explaining the Rise in Intergenerational Income Persistence} \label{SEC:realized_potential}

In the previous section, we documented that the intergenerational rank association has increased rapidly in Scandinavia for parent-child pairs involving mothers or daughters. A close-at-hand explanation for the downward trend in mobility is thus that rank correlations have increased because women's integration into the labor force means that income ranks have become more informative of underlying skills. However, and much more worrying, the same pattern of results could arise because of changes in the family influence (in particular, the influence of mothers) on individual income.\footnote{Changed assortative mating among parents is an additional hypothesis. However, studying the case of Swedes born in 1945-1965, \cite{Holmlund2020} finds that the influence of changes in assortative mating on intergenerational income associations is small. For Norway, \citet{eika2019educational} demonstrate increasing educational assortative mating, while mating patterns using income-based social class are stable over the same period, suggesting changes in the composition of educational groups rather than changes in mating patterns \citep{bratsberg2018trends}.}

Here, we will differentiate between \textit{realized income}, which is observed without error, and \textit{potential income}, which is a latent variable reflecting inheritable human capital. To fix ideas, let $Y_i^I$ denote realized (observed) income for person $i$. This can be thought of as a linear combination of a person's productivity-based income, $Y_i^E$ (i.e., potential income), and an attenuation factor $\varepsilon$ reflecting ``everything else'', such as norms regarding female work and labor market discrimination:
$$ Y_i^I = Y_i^E + \varepsilon$$
Under a paradigm of low female labor market participation, potential income will exceed realized income for many women, meaning that the attenuation factor will be negative on average. As female labor force attachment increases, the attenuation factor declines, and the two concepts will equalize. 

Our key hypothesis can be phrased as saying that the intergenerational correlation in potential income has remained unchanged, while the intergenerational correlation in realized income has strengthened. Next, we test this hypothesis in two ways: first, we construct an estimate of potential income by combining the informational value of realized income, education, and occupation, and second, we build and calibrate a model that we use to deconstruct the mobility trend into components reflecting intergenerational transmission of human capital, assortative mating, and the importance of underlying productivity for realized income (i.e. $Y_i^E$).

\subsection{Intergenerational Correlations in Potential Income} \label{SEC:LW}

First, we estimate intergenerational associations in \textit{potential income}, constructed from a weighted average of several proxy variables. As explained above, one can think of the historical low levels of parent-child income correlations for women as a form of ``attenuation bias'' stemming from the difference between realized and potential income. We follow recent work by \citet{VostersNybom2017}, \citet{Vosters2018}, and \citet{Adermon2021Dynastic} and apply the method laid out in \citet{LW2006} (from now on ``LW''), where ``optimal weights'' for summing over proxy variables are calculated as the covariance between a given outcome variable (in our case, child income) and the proxy variable, which gives an estimator that minimizes attenuation bias among its class of estimators (\citet{LW2006}, p.552).\footnote{This class of estimators includes a straightforward imputation of economic status (or ``income score'') from observable proxies, such as that used in e.g. \citet{abramitzky2021immigrants} and \citet{collins2022african} to study historical mobility trends. Essentially, the LW strategy as we implement it can be thought of as a special case of imputing income from gender, education, and occupation of labor-market active individuals of the same birth cohort and generation, with weights tailored to the intended left-hand-side variable of the estimating equation. As an additional validation exercise, we have imputed female income based on observed average income among men with the same level of education and occupation, within a given birth cohort and generation. The results show stable trends between mother-son and daughter-father pairs in imputed income ranks.} The procedure requires the theoretical assumption that each proxy variable affects the left-hand side variable --- child potential income --- only through inheritable income potential, but it does not assume independence between the proxy variables.

We use income, years of education, and occupation as proxy variables for the potential income of mothers (and, as a validation exercise, fathers). Income is included for two reasons. First, it allows us to assess how much informational value labor income provides about latent income potential, and how this relationship evolves over time. Second, since education and occupation are expressed on fixed, categorical scales, income provides a continuous denominator to which these proxies can be scaled. Importantly, our use of the LW approach offers a transparent interpretive benchmark: if education and occupation offer no additional predictive value beyond income, then the LW estimate will align exactly with the observed rank-rank correlation in income. This feature allows us to gauge the extent to which observed income—especially for women—reflects their potential income, and how this changes across cohorts.

For occupations, we follow \citet{VostersNybom2017} and include dummy variables for ten occupational groups. These are roughly equivalent to 1-digit ISCO codes: professional, managerial, clerical, commerce, agriculture, mining, transportation, manufacturing, military, and services. An additional dummy variable is included for missing occupational information. See Appendix \ref{APP:data_registers} for a full description of how we assign individuals to occupations and divide occupations into groups.\footnote{In Appendix Figure \ref{fig:LW_contributions}, we show that both education and occupations contribute independently of one another to the rank correlations between mothers' potential income and sons' income rank. We calculate the \textit{contribution} of proxy variable $j$ for birth cohort $t$ as $\frac{\rho_j b_j}{\beta^{LW}_t}$, i.e., as the share of the aggregate coming from proxy variable $j$. The contribution of occupations is calculated by summing over the ten different occupational group dummies.}

The proxy variables are denoted $x_j, \quad j \in {1,..,k}$. The LW estimator for the parents with children born in year $t$ is constructed as follows:
\begin{equation}\label{eq:LW}
    \beta^{LW}_t = \sum_{j=1}^{k} \rho_j b_j,
\end{equation}
where $\rho_j = \frac{cov({Income}_{it}^{\text{C}}, x_{jit})}{cov({Income}_{it}^{\text{C}}, {Income}_{it}^{\text{P}})}$ is the weight on proxy variable $j$. This is essentially a Wald estimator: the covariance between child income and proxy variable $j$, scaled by the covariance between child and parent income.\footnote{As in previous applications, we use the logarithm of child and parent labor income to calculate the LW estimates. The reason is intuitive: we first calculate a measure directly comparable with income and then rank it, just like we do with income. In order to use the same full-population sample as in our main analysis, we assign individuals with zero labor income a token low level of log income. Sensitivity checks show that the exact level of income assigned does not alter the conclusions from this analysis.} As such, it can be calculated with a 2SLS estimator. Since parent income serves as the denominator, the weight on the parent income, $\rho_1$, is equal to one. The $\mathbf{b_j}$'s are OLS coefficients from a multiple regression of child income on the set of parent proxy variables. Summing $\rho_j b_j$ over all proxy variables gives the estimate $\beta^{LW}_t$.

In order to compare these estimates to IRAs, we want to correlate parent \textit{ranks} in potential income to child \textit{income ranks}. We thus transform the parental LW estimates into percentile ranks, using the explicit index construction mentioned in \citet{LW2006} (p.554):
\begin{equation}
    x_{it}^{\rho,\text{P}} = \frac{1}{\beta^{LW}_t}\sum_{j=1}^{k} x_{jit}b_{jt},
\end{equation} 
Finally, we regress the child income ranks on these parental index ranks, for mothers and fathers separately, and for each birth cohort. Note that our main goal is not to provide point estimates for mother-child correlations in potential income, but rather to assess the development over time. We estimate:
\begin{equation} 
\text{Rank}_{it}^{\text{C}}=\alpha_{t}+\beta_{t}x_{it}^{\rho,\text{P}}+\varepsilon_{it}.
\end{equation}

The method described so far addresses the problem of unrepresentative maternal realized income. To understand whether daughter-father correlations are subject to the same issue, we repeat the above procedure for daughters and approximate their potential income with realized income, education, and occupations.\footnote{We refrain from estimating mother-daughter correlations in latent income potential, since this would require a method to adjust for measurement error in both the dependent and independent variables.} Since the LW method addresses measurement error in the right-hand-side (independent) variable, this requires ``flipping'' the intergenerational model (eq. \ref{rank_equation}), and estimating rank associations between fathers and their daughters:
\begin{equation} 
\text{Rank}_{it}^{\text{P}}=\alpha_{t}+\beta_{t}\text{Rank}_{it}^{\text{C}}+\varepsilon_{it}.
\end{equation}
This has only minor impacts on the year-specific IRA estimates and does not alter the trend. Apart from this first step, we proceed in an identical manner to the son-mother estimation.

\subsubsection{Results}
\begin{figure}[H]
\linespread{1}
\begin{center}
\includegraphics[width=1\linewidth]{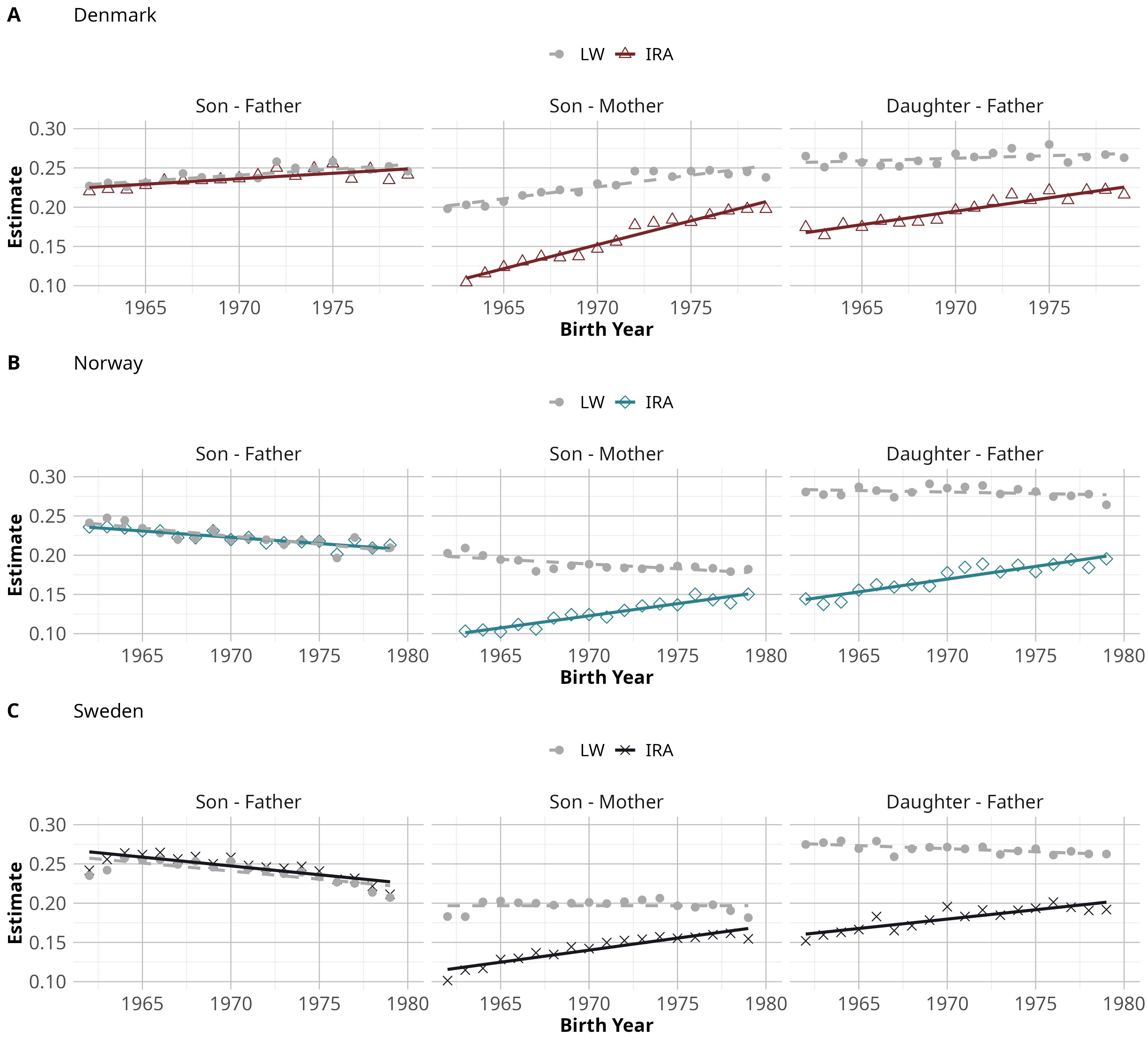}
\end{center}
\caption{Trends in Intergenerational Mobility in Latent Income Potential.}
\label{lw_ira_fig}
\scriptsize{\textit{Notes:} The three panels plot comparisons of intergenerational rank associations and rank associations in latent potential income (calculated using the proxy variable method from \citet{LW2006}) for Denmark, Sweden, and Norway. Each panel shows, in turn, son-father, son-mother, and daughter-father correlations. Each marker indicates the coefficient of a separate regression (eq. \ref{eq:LW} for LW; eq. \ref{rank_equation} for IRA), and each line indicates fitted trend lines for the period 1962 to 1979. ``Birth Year'' refers to the birth year of the child in each parent-child pair.}
\end{figure}

Figure \ref{lw_ira_fig} plots the trend in IRA and LW estimates for birth cohorts 1962 to 1979, separately by country. We focus here on the birth years 1962-79 because we observe the largest increase in rank correlations for this part of the sample. In Appendix Table \ref{tab:trendcomppart}, we also report the difference between the trend estimates and test whether trends in intergenerational rank associations are statistically distinguishable between the IRA and LW approaches. 

The first column of Figure \ref{lw_ira_fig} shows that the son-father trends obtained from the LW method are almost identical to the son-father IRA trends, which validates that the proxy variable method captures the rank-rank association in potential income. When realized income is fully informative of potential income, the LW weighting procedure yields the same result as income rank correlations. For Norway and Sweden, IRA and LW trends are negative, indicating a development towards \textit{increased} mobility, while Denmark's decline in mobility is supported by both the IRA and LW methods. 

The middle column shows son-mother estimates. Compared to the IRA trend, our LW trends are noticeably weaker, and in the case of Norway even negative. This suggests a development similar to that of father-son estimates. The difference between the trends in the IRA and LW coefficients is statistically meaningful and different from zero, and also similar in magnitude across all three countries. Evidently, when using mothers' years of education and occupations - rather than just observed labor income - to proxy for their latent income potential, income persistence between male children and their mothers, as well as their fathers, has remained relatively constant over time. 

In the last column of Figure \ref{lw_ira_fig}, we present comparisons between trends in LW and IRA coefficients for daughters and fathers. Again, the LW trends are significantly less steep than the IRA trends, and the differences between them are almost identical across countries. For Denmark, the adjusted trend still indicates that over time, mobility in potential income decreases, albeit at a lower rate. In Norway, the relationship is stable, while daughters in Sweden experience a small increase in mobility over time. 

Interestingly, the levels of the son-mother and daughter-father LW coefficients are constant, and similar to the IRA coefficients of son-father pairs, a result also supported by findings in \citet{VostersNybom2017}. The transmission of economic potential between parents and their female and male children has thus seen little change across birth cohorts from 1962 to 1979.\footnote{This finding is further supported by our results in Appendix Figure \ref{fig:ige_education}, showing mostly flat trends in educational mobility as captured by educational IGEs for parent-child combinations between 1962 and 1979 in both Norway and Sweden.} This might reflect the Scandinavian setting, with relatively equal schooling opportunities among boys and girls already for individuals born in the 1950s. The fact that father-daughter correlations are as high as the father-son correlations suggests that whatever skills relevant to economic success are transmitted between parents and their children, these are gender-neutral.

Nonetheless, one could argue that the occupational and educational choices of women historically suffer from the same low correlation with underlying skills as income. To corroborate the LW results, we also estimate the intergenerational rank association in labor income between sons and their maternal uncles. Given a constant level of brother-sister correlation in productivity \citep{bjorklund2009family}, this estimated trend captures changes in the importance of parental productivity for child outcomes. Using observed skills of maternal uncles to proxy for unobserved female skills is a strategy previously used by, e.g., \cite{gronqvist2017intergenerational}. Because the data needed for generating parental sibling links is partly unavailable, the sample size used to estimate these correlations is smaller, particularly for the earliest birth cohorts, and Denmark is left out of the analysis. 

\begin{figure}[H]
\linespread{1}
\begin{center}
\includegraphics[width=1\linewidth]{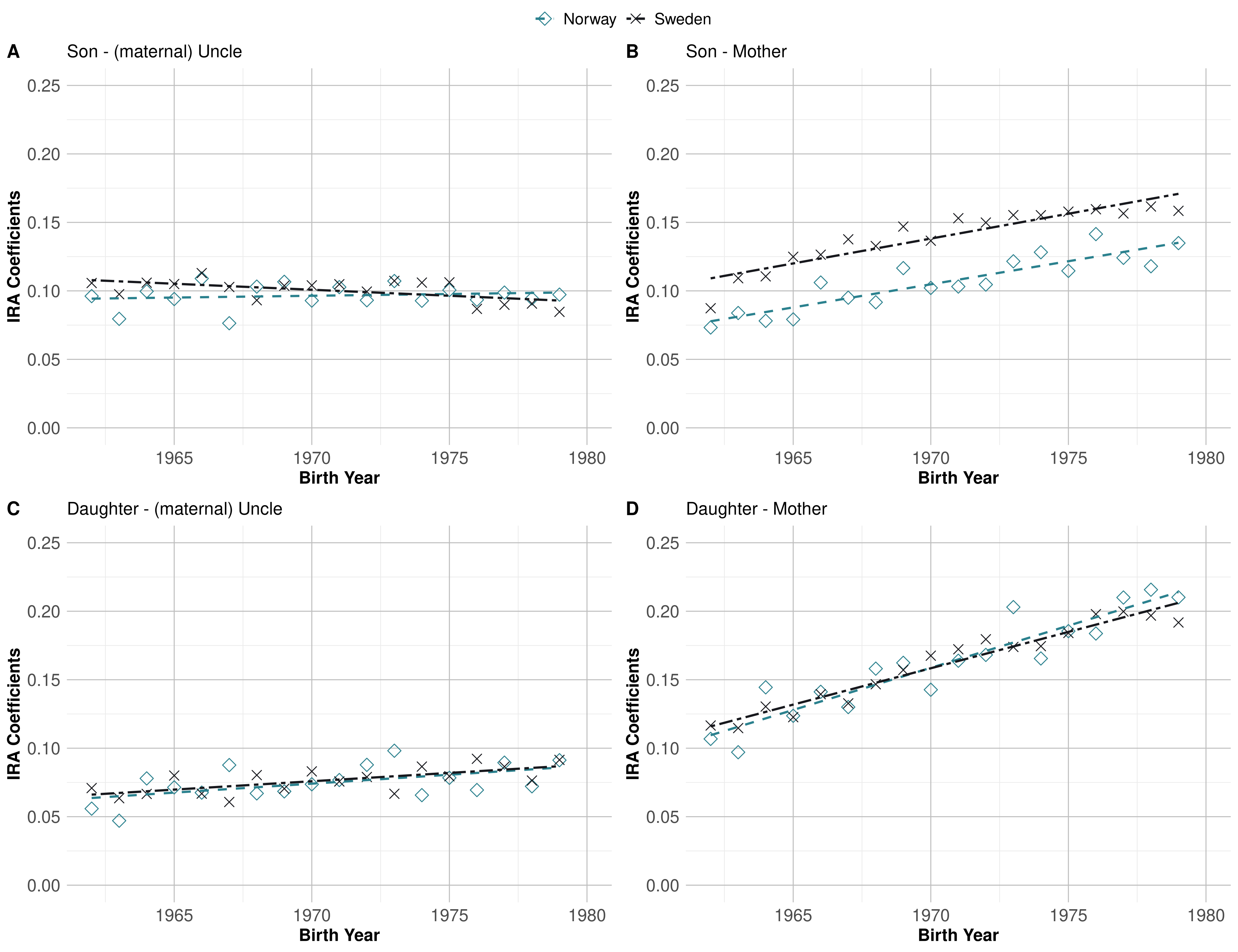}
\end{center}
\caption{IRA Estimates Between Children and Their Maternal Uncles.}
\label{IRA_uncles}
\scriptsize{\textit{Notes:} The four panels depict IRA coefficients (eq. \ref{rank_equation}) for the correlation between children and their mother's brothers, i.e. maternal uncles; Sons in Panel A and daughters in Panel C. Panels B and D show the estimated IRA between sons and daughters and their mothers for the sample where information on maternal uncles is available. ``Birth Year'' refers to the birth year of the child in each parent-child or uncle-child pair. Results for Sweden are depicted in dark X's, Norway in hollow diamonds (Denmark is left out of this analysis due to data availability).}
\end{figure}

Figure \ref{IRA_uncles} presents the results. Panel A shows a constant level of rank associations over time between sons and their maternal uncles. Panel B shows the original mother-son associations for comparison, and in Panels C-D, the same results are shown for daughters and maternal uncles. Daughter-uncle trends are substantially flatter than daughter-mother trends, indicating that a certain part of the mother-daughter trends is driven by mothers. However, the remaining IRA trend indicates that increased labor force attachment among daughters also contributes to the increasing rank correlations between daughters and mothers.

\subsection{Decomposition by Income Determinants} \label{SEC:theory}

Next, we present evidence from an alternative strategy for exploring the drivers of the increase in income persistence. We build a model of intergenerational income persistence and use the gender-specific variation in mobility trends, along with correlations in parental income, to quantify the importance of different potential channels through a decomposition exercise.

Inspired by the canonical model in \cite{BeckerTomes}, income for individual $i$ belonging to birth cohort $t$ (either by own birth year or birth year of their child), denoted $y^k_{it}$, is determined by two factors; inheritable skills (productivity), $x^k_{it}$, and a non-inheritable determinant $\varepsilon^k_{it}$, for fathers, mothers, sons, and daughters, i.e. all $k\in\{F,M,S,D\}$. Like \citet{collado2022estimating}, we separate income into inheritable and non-inheritable components and consider gender-specific processes and assortative mating, with the difference that we allow our set of parameters to change with the birth year of the child.\footnote{\citet{collado2022estimating} develop a model of latent advantage in extended families, applied to data on extended kinship relations. Their model is naturally complex as it aims to capture a large set of kinship parameters. Since our decomposition exercise aims to quantify how changes in these components contribute to observed trends in intergenerational mobility, we allow our more limited set of parameters to change with birth year of the child.}

The model's technical details are presented in Appendix Section \ref{APP:MODEL}. In essence, the model assumes that productivity (in this section used interchangeably with the word ``skills'') is transmitted passively from the parental generation to the child generation through a process governed by both general ($\kappa_t$) and gender-specific ($\alpha_t$) inheritance. Parental income is correlated by a factor $\psi_{t}$ that captures time-varying assortative mating. The final elements of the model are parameters governing the importance of inheritable skills for individual income: $\phi_{t}^{K} \quad \forall \quad k\in\{F,M,S,D\}$. These are measured for mothers relative to the (same-cohort) level of fathers, and for daughters relative to sons. In other words, we assume a generation-specific gender bias in the importance of skills for income. For tractability, we assume that realized income fully reflects inheritable skills for fathers and sons -- an assumption also supported by our results in the previous section showing almost identical father-son rank associations in realized and ``potential'' income. Table \ref{tab:model_parameter_interpret} summarizes the model parameters.

Calibrating the model allows us to understand how country-specific changes in intergenerational mobility can be decomposed into changes in the rate at which inheritable skills manifest themselves in labor income among mothers and daughters relative to fathers and sons, and changes in assortative mating on skills among parents. The calibration exercise is described in Appendix Section \ref{APP:cali}, where we also document model fit. Several points are worth noting about the calibrated values, which are presented in Table \ref{tab:parms}. First, the parameters associated with how skills are reflected in income for mothers and daughters, $\phi_t^M$ and $\phi_t^D$, have increased at a somewhat similar pace across all three countries. This suggests that female income has become more reflective of inheritable skills in both the parent and child generations.

Second, the parameter associated with assortative mating, $\psi_t$, is rather stable over time in all three countries, despite strongly increasing associations in maternal and paternal income over time (which is displayed in Appendix Figure \ref{fig:matching}).\footnote{We note the calibrated levels of assortative mating in latent skills are somewhat lower than those found in \cite{collado2022estimating}, who set up a similar conceptual framework. The reason for this discrepancy, however, is straightforward: it can be attributed to the normalization of $\phi^F=1$. In our model, the income ranks of fathers are solely determined by inheritable skills, whereby the income ranks of mothers will also be attributed to inheritable skills to a significant extent. Now, due to high levels of $\phi^F$ and $\phi^M$, any lack of correlation between the incomes of fathers and mothers need to pass through a low level of $\phi$. Given that the above-mentioned normalization is not used by \cite{collado2022estimating}, their measure of assortative mating will mechanically be higher.} This may result from the increased reflection of maternal inheritable skills in income, which mechanically increases the observational correlation in father and mother income for a given correlation in skills (assortative mating). 

Third, within-gender intergenerational correlations in inheritable skills appear stronger than cross-gender correlations in skills --- $\alpha_t$ is approximately 0.6 in all countries. Finally, the parameter associated with non-gendered skill transmission, $\kappa_t$, is stable over time. 

\subsubsection{Results}

In order to decompose changes in rank associations into effects associated with changes in the modeling parameters, we compute ``counterfactual'' income associations through the following process. First, we re-simulate the model using our calibrated time-specific parameters while holding \textit{one} parameter fixed at the calibrated value from a baseline period (cohort 1962). Second, we re-compute the intergenerational rank associations with this ``counterfactual'' set of model parameters. Differences in estimated intergenerational rank associations are attributed to the parameter that was held fixed. More details are found in Appendix Section \ref{app:decomp}.

The results from this exercise are shown in Table \ref{tab:parm_trends}. The rank associations in income do not exhibit a clear trend for cohorts born between 1952 and 1961 in Sweden and Norway, whereby there is not much to be explained by the decomposition parameters. However, the parameter associated with gender neutral skills transmission, $\kappa_t$, contributes negatively to the change in IRA, while the opposite is true for the parameters associated with the importance of skills for female income, $\phi_t^M$ and $\phi_t^D$. This suggests that the family influence on income may have declined, thereby improving income mobility, but that this effect was mitigated by the increasing extent to which women's income reflects their productivity.

For birth cohorts 1962 to 1979, the model captures the fact that IRAs are increasing uniformly across Scandinavia remarkably well. Parental assortative mating ($\psi_t$) and gender-specific skill transmission ($\alpha_t$) contribute little to mobility trends in this period. Our results suggest a bigger role for gender-neutral skill transmission ($\kappa_t$) --- at least in Denmark, where this component explains almost half of the observed trend in mobility. In Sweden and Norway, however, the contribution of $\kappa_t$ is negative, and the importance is negligible. 

Changes in the extent to which female income (and especially maternal income) reflects inheritable skills ($\phi_t^M$ and $\phi_t^D$) are found to be important drivers of downward trends in mobility in Denmark, Norway, and Sweden. These effects jointly contribute to a yearly increase in the income IRA of between 0.28 and 0.30 rank points, amounting to a total increase in the IRA of between 5 and 6 rank points over the period. Taking this result at face value would thus suggest that increased labor market valuation of female skills alone can explain most of, or even the entire, observed decline in intergenerational income mobility in Scandinavia.

Taken together, the results from this decomposition exercise, along with the latent variable method above, show that the observed downward trend in social mobility is present in intergenerational correlations in realized income but not in potential income. We find no support for an interpretation of the trend as a development toward greater intergenerational transmission of human capital (or greater valuation of transmittable skills across genders), and thus, a society of declining social mobility. 

\section{The Importance of Female Labor Market Developments} \label{SEC:mechanisms}

Next, we look at how female labor force attachment on different margins affect estimates of income mobility. Our results so far show that parent-child rank correlations were relatively constant across birth cohorts 1951-1962, despite a great increase in maternal participation rates. Mechanically, whenever income is informative about productivity, we would expect parent-child income ranks to correlate more strongly at higher female participation rates. The fact that this is not what we find speaks to a development where expansions on the extensive margin of employment happen in occupations where women's productivity is not well reflected in their income.

\begin{figure}[H]
\linespread{1}
\begin{center}
\includegraphics[width=1\linewidth]{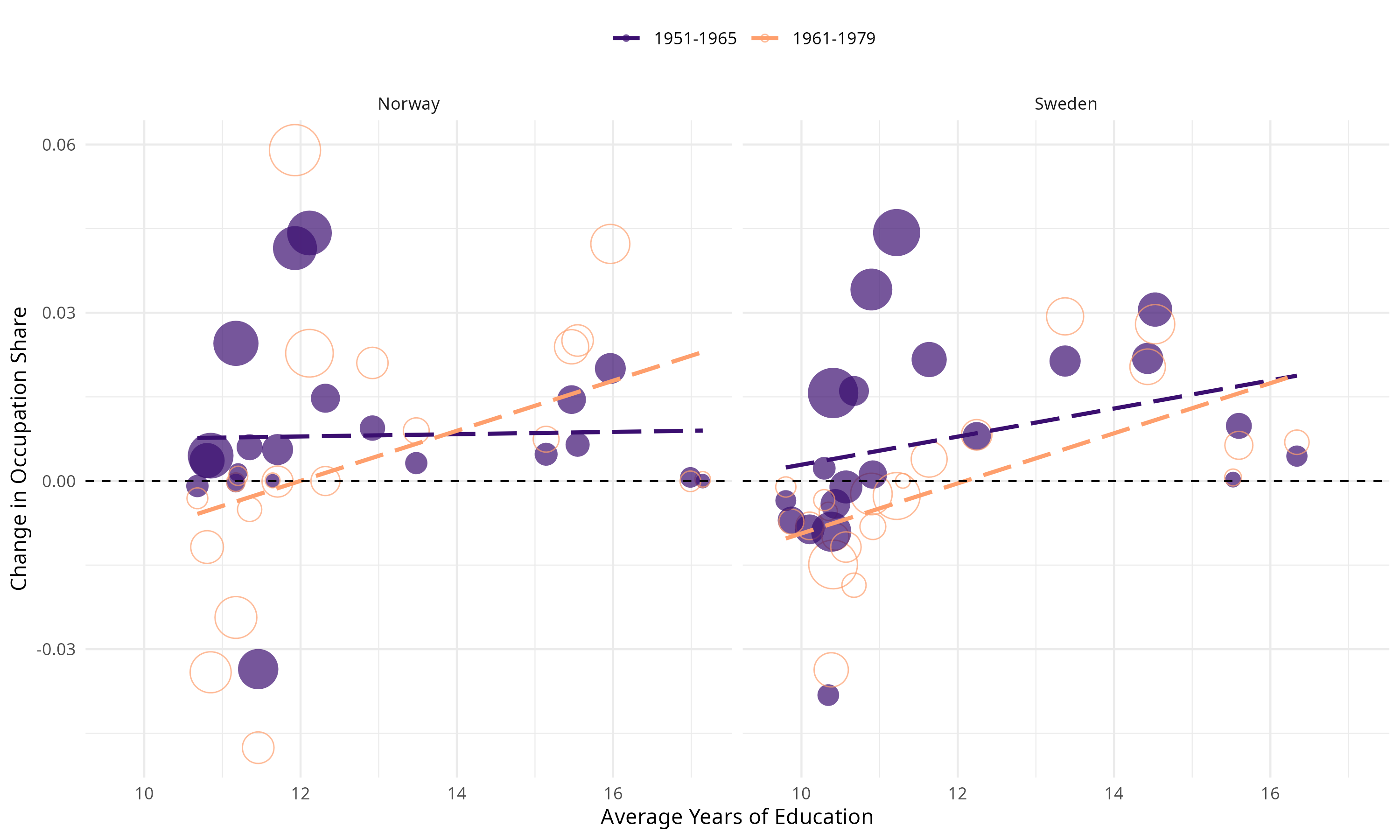}
\end{center}
\caption{Changes in the Occupational Composition of Mothers.}
\label{fig:occ_rank_share}
\vspace{0.3cm} 
\scriptsize{\textit{Notes:} The figure plots the average years of education within an occupational group (a proxy for the skill level of the occupation) against the change in the share of mothers with a job in that occupational group, from one group of birth cohorts to another. Each dot represents an occupational group, and the size of the dot represents the group size. Purple (filled) circles denote the change from mothers of the 1951-1955 birth cohorts to mothers of the 1961-1965 birth cohorts; orange (hollow) circles denote 1961-1965 to 1975-1979. Average years of education calculated in 1975-1979. Dashed lines represent the OLS fitted line between change in occupational share and average years of education, weighted by occupational size; purple for 1951-1965 and orange for 1961-1979. The left panel shows Norway and the right panel shows Sweden. Occupations are grouped as described in Appendix Table \ref{tab:occ_groups_new}.}
\end{figure}

We thus explore how the composition of the female labor force changed during the time of expanding female labor force participation in the 1950s and the early 1960s.\footnote{This section focuses on occupations and education. Ideally, we would also analyze how intensive margin labor force participation affects intergenerational rank associations. However, administrative data on hours worked does not extend sufficiently far back in time for us to include it in a meaningful way in this analysis. Moreover, Denmark is excluded from this analysis for comparability, as the occupation data differs from that in Sweden and Norway.} Figure \ref{fig:occ_rank_share} shows the relation between the change in the share of mothers in a given group of occupations, and the average years of education in that same group, for Norway and Sweden respectively, for birth cohorts 1951-55 to 1961-65, and 1961-65 to 1975-79. Average education is measured in 1975-1979, to prevent changes in incentives for and returns to education from complicating the interpretation. 

The figure shows that between the early 1950s and 1960s cohorts, when labor force participation among mothers rose quickly, mothers mainly entered relatively low-skilled occupations. In fact, in both Sweden and Norway, personal services occupations and secretaries together increased by almost 10 percentage points, which corresponds to about two-thirds of the increase in extensive margin employment among mothers. See Appendix Table \ref{tab:occ_change_list} for a list of the three occupational groups that saw the largest female employment growth in 1951-65 and 1962-79, respectively, for mothers and daughters. Under the assumption that pay in low-skilled occupations is not well differentiated across skills,\footnote{For example, \citet{autor_dorn2009inequality} argue that low-skill service occupations are relatively homogeneous in skill requirements, and as a result, they tend to offer more compressed wage distributions.} the income of mothers remained uninformative of underlying productivity, and mother-child income correlations did not increase. 

Conversely, our estimated IRA trends for birth cohorts 1962 to 1979 show that mother-son and parent-daughter income ranks converged rapidly in this period. From Appendix Figure \ref{fig:ira_excluding} (which shows IRA trends exclusively for individuals active in the labor market), it is evident that extensive margin entry can not explain this, since correlations limited to only participants still display a trend toward lower mobility. Additionally, our descriptive statistics in Figure \ref{fig:labor_development} (Section \ref{SEC:institution}) show rapidly declining occupational segregation and increasing rates of full-time employment.

Figure \ref{fig:occ_rank_share} (and Appendix Table \ref{tab:occ_change_list}) shows that mothers with children born in this period increasingly entered the high-skilled occupations within the ``professionals'' category (any job requiring an academic degree). As extensive margin labor supply changed at a slower pace during this time for mothers and remained constant for daughters, this implies a higher degree of sorting of workers into occupations based on skills. Similar results for fathers indicate that there is almost no change in the way fathers sort into high-skilled occupations in either Norway or Sweden (shown in Appendix Figure \ref{fig:occ_rank_father}). The average income rank of women in a particular occupation would thus become more reflective of their skill level, primarily because of declining mean ranks of low-skilled occupations.\footnote{We demonstrate this in Appendix Figure \ref{fig:edu_rank_occupations}, which shows that over time, the average income rank of mothers in low-education occupations declines relative to those in high-education occupations.}

\begin{figure}[H]
\linespread{1}
\begin{center}
\includegraphics[width=1\linewidth]{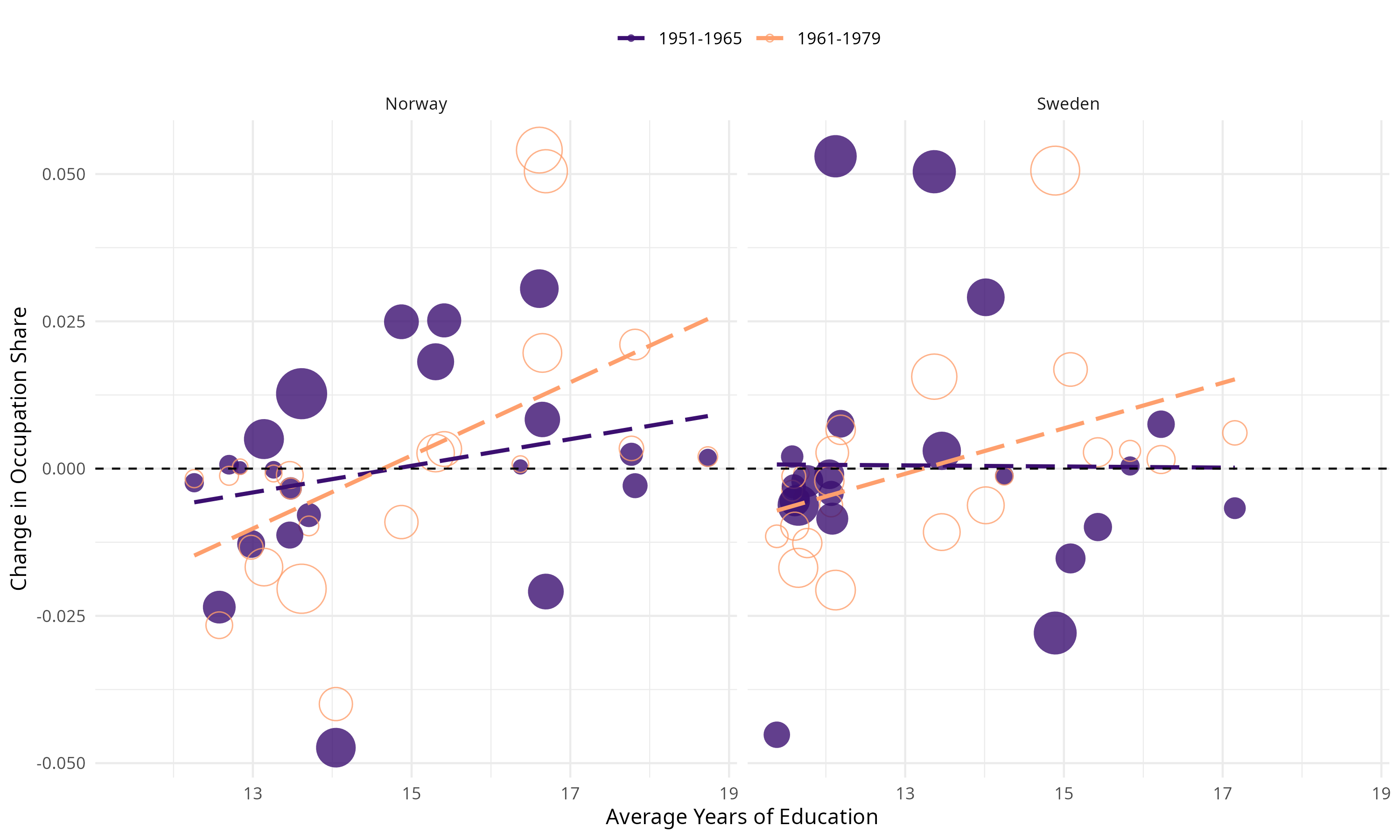}
\end{center}
\caption{Changes in the Occupational Composition of Daughters.}
\label{fig:occ_rank_daughters}
\vspace{0.3cm} 
\scriptsize{\textit{Notes:} The figure plots the average years of education within an occupational group (a proxy for the skill level of the occupation) against the change in the share of daughters with a job in that occupational group, from one group of birth cohorts to another. Each dot represents an occupational group, and the size of the dot represents the group size. Purple (filled) circles denote the change from daughters of the 1951-1955 birth cohorts to daughters of the 1961-1965 birth cohorts; orange (hollow) circles denote 1961-1965 to 1975-1979. Average years of education are calculated for 1975-1979. Dashed lines represent the OLS fitted line between change in occupational share and average years of education, weighted by occupational size; purple for 1951-1965 and orange for 1961-1979. The left panel shows Norway, the right panel shows Sweden. Occupations are grouped as described in Appendix Table \ref{tab:occ_groups_new}.}
\end{figure}

In Figure \ref{fig:occ_rank_daughters}, we show that the same patterns can be found among the daughters in our data set, with some caveats. Women born in the early 1960s are more likely to work than women born in the early 1950s, but they work primarily in low-skilled occupations.\footnote{The results for Norway nuance this picture somewhat, as can be seen in Appendix Table \ref{tab:occ_change_list}. Between the early 50's and early 60's cohorts, employment grew in low-skilled occupations such as caregivers, but also in the ``lower professionals'' group, which includes technicians, sales staff, dental assistants, etc, and in the ``executives'' group, which mainly includes management positions in the public sector.} Conversely, women born in the late 1970s are less likely than those born in the early 1960s to work in low-skilled occupations, while the share in high-skilled occupations (professionals, teachers, and medicine) is higher.

Appendix Figure \ref{fig:avg_rank_by_eduy} shows how the relation between years of education and income ranks becomes stronger from the 1962-65 cohorts to the 1975-79 cohorts for mothers and daughters, indicating that women with higher income potential, as measured by years of education, over time have become more likely to have jobs that reflect their productivity. Meanwhile, the gradient between income and education stays the same for fathers and sons, and for mothers and daughters in the earlier birth cohorts. Taken together, these findings indicate an increased level of sorting along skill levels for women in the latter part of our study period. 

An alternative explanation, where low-skilled women enter the labor market earlier than high-skilled women, does not appear to account for these patterns. Appendix Figure \ref{fig:lfp_mothers_byedu} shows that the share of mothers in the labor force was consistently higher for more educated women. Among mothers of the 1951-55 cohorts with post-secondary education, labor force participation was twice as high as that of mothers with only compulsory schooling. While less marked among the 1975-79 cohort mothers, this gap in labor force participation by educational attainment persisted over time, indicating that selection into market work was positive throughout our sample period. Thus, we may describe the later years as a period in which female workers became better able to ``earn their potential'', i.e., achieve realized income corresponding to their potential income.

\section{Conclusion} \label{SEC:concl}

In this paper, we document trends in intergenerational income mobility in Denmark, Norway, and Sweden for children born between 1951 and 1979. Harmonizing data and definitions, we show that intergenerational rank associations between parents and children in individual labor income and household income have increased significantly in all three countries. These trends are robust to using different types of income measures, as well as to restricting the analysis to labor market active individuals. Splitting trends by gender of parents and children, son-father correlations exhibit weak or flat trends in all three countries, whereas all correlations involving mothers and daughters increase over time. The strongest trend is found between mothers and daughters. A similar, but delayed, pattern of changes in mobility is found for US parent-child pairs from the Panel Study of Income Dynamics.

We further show that this is the result of stronger manifestation of skills in income for women, such that the intergenerational correlation in observed income matches that in ``potential income'', i.e. income reflecting a person's underlying productivity. At the same time, the intergenerational correlation in potential income remained constant, meaning that the transmission of human capital between parents and children did not increase. We also rule out that mobility trends are substantially affected by changes in assortative mating among parents.

These results highlight the importance of accounting for changes in female economic status when estimating trends in intergenerational mobility. The interpretation that higher income rank associations between children and parents reflect a lower degree of social mobility or equality of opportunity is not always applicable in times of structural changes in the labor market. Our findings suggest that over time, female income becomes increasingly determined by their productivity, meaning that the traits and norms that women inherit from their parents are also better reflected in their income. Such a development is a necessary side effect of increased gender equality in the labor market, and it is not clear whether it should be seen as a reduction or advancement in equality of opportunity. 

\newpage
\bibliographystyle{aea}
\bibliography{mobility_literature}


\newpage
\section*{Figures}

\begin{figure}[H]
\linespread{1}
\begin{center}
\includegraphics[width=.8\linewidth]{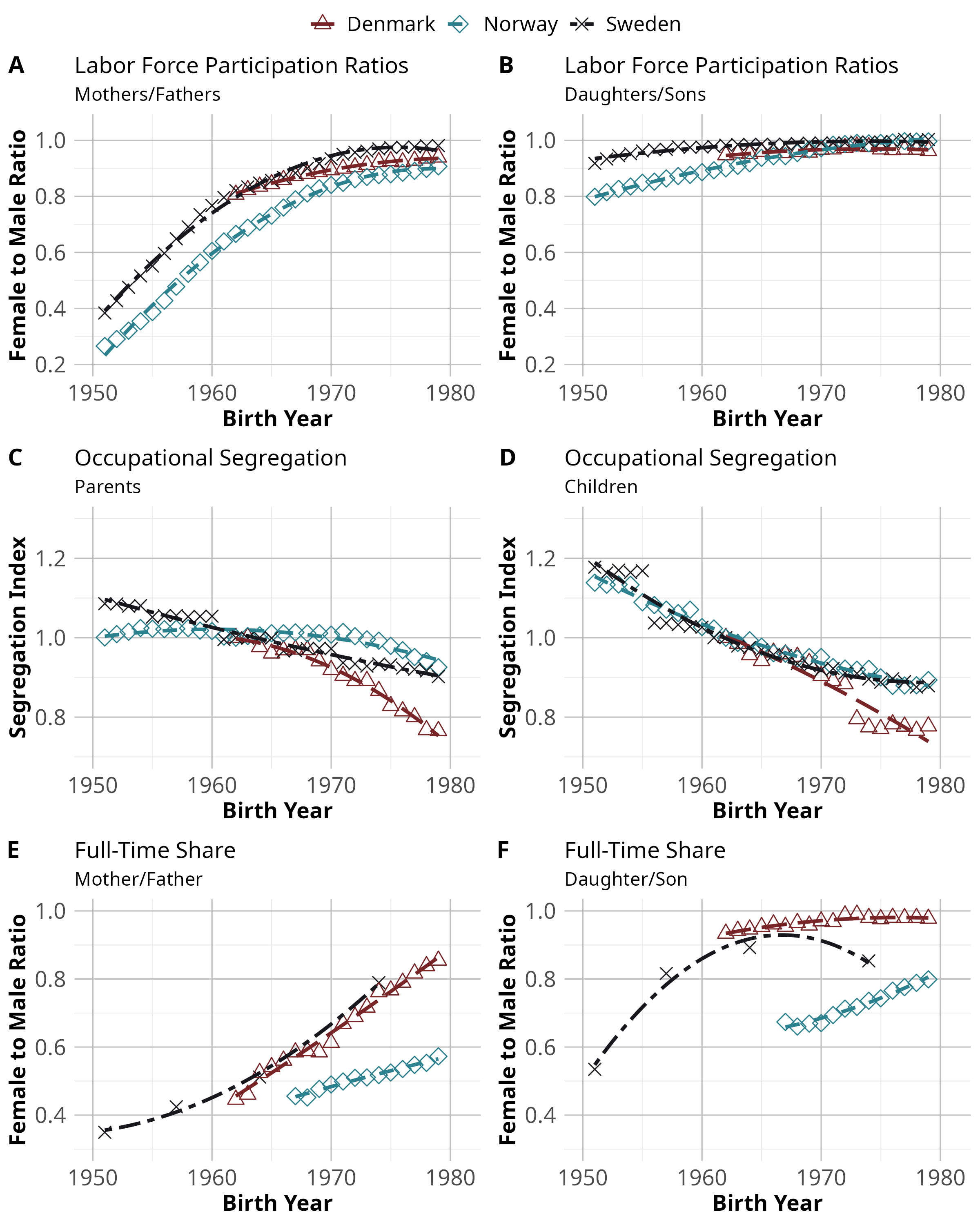}
\end{center}
\caption{Changes in Gender Differences in Labor Force Participation, Occupational Segregation and Working Hours.}
\label{fig:labor_development}
\vspace{0.3cm}
\scriptsize{\textit{Notes:} Panels A and B depict female-to-male ratios of labor force participation in our main samples; for parents in Panel A and children in Panel B. The labor force participation rate is based on annual labor income: a person is considered in the labor force if they have annual income exceeding the equivalent of 10,000 USD (2017). Panels C and D depict an index for labor market segregation, for parents and children, respectively. The index is normalized to the base year 1962. In some years, Danish occupational codes have been imputed from other variables --- therefore, the Danish trend in occupational segregation should be interpreted with caution (see more in Appendix \ref{APP:data_registers}). Panels E and F provide female-to-male ratios of full-time work. Full-time work is defined as working at least 27 hours in Denmark and Sweden, and at least 31 hours in Norway. Data for full-time shares is obtained from linked employer-employee data in Norway and Denmark, and nationally representative surveys in Sweden. In Denmark, there is a significant data break in the full-time definition, which only affects the child cohort --- we attempt to adjust for this appropriately with a simple correction procedure (see more in Appendix \ref{APP:data_registers}). ``Birth Year'' refers to the birth year of the \textbf{child} in each parent-child pair.}
\end{figure}

\newpage
\section*{Tables}

\begin{table}[H]
\centering
\linespread{1}
\begin{threeparttable}
\caption{IRA Coefficients and Trends (United States).}
\renewcommand{\arraystretch}{.5}  
\begin{tabular}{l *{5}{p{2.3cm}}} \toprule
&  \multicolumn{1}{c}{\textbf{Parents}} &\multicolumn{2}{c}{\textbf{Father}} & \multicolumn{2}{c}{\textbf{Mother}}  \\ \cmidrule(l{2pt}r{2pt}){2-2} \cmidrule(l{2pt}r{2pt}){3-4} \cmidrule(l{2pt}r{2pt}){5-6}
   & \multicolumn{1}{l}{Child} & \multicolumn{1}{l}{Son} & \multicolumn{1}{l}{Daughter} & \multicolumn{1}{l}{Son} & \multicolumn{1}{l}{Daughter}  \\
\midrule
Pooled IRA   & 0.317***  & 0.336***  & 0.195***    & 0.097***    & 0.137***  \\
 & (0.017)  & (.022)  & (0.031)   & (0.025)  & (0.029) \\
 \midrule
Trend  $\times$ 100 & 0.603*** & -0.240  & 0.980***  & 0.136     & 1.047***  \\
& (0.149)    & (0.205)   & (0.277)   & (0.253)  & (0.292)         \\
 \midrule
N      & 5,392  & 2,272      & 1,637      & 2,477  & 2,205 \\
\bottomrule
\end{tabular}
\linespread{1}
\scriptsize{\textit{Notes:} The table presents estimates of the IRA and linear trends in the IRA separately for different child-parent combinations. Due to the small sample sizes, trends have been estimated directly on the underlying micro data by regressing cohort-specific child ranks on cohort-specific parent ranks interacted with a linear time trend. The trend coefficients and corresponding standard errors have been multiplied by 100 in order to avoid too many digits after the separator. Estimates are based on the full sample of individuals in the PSID born between 1947 and 1983 using PSID sample weights. Standard errors are in parentheses. P-values indicated by * $<$ 0.1, ** $<$ 0.05, *** $<$ 0.01.}
\label{tab:psidira}
\end{threeparttable}
\end{table}

\begin{table}[H]
\centering
\caption{\label{tab:model_parameter_interpret} Summary of Model Parameters}
\begin{threeparttable}
\renewcommand{\arraystretch}{0.5} 
\begin{tabular}{c>{\raggedright\arraybackslash}p{13cm}}
\toprule
 Parameter & Interpretation \\
\midrule
$\psi_t$ & Correlation in parental inheritable skills (assortative mating) \\
$\kappa_t$ & Intergenerational correlation in inheritable skills \\
$\alpha_t$ & The strength of within-gender skill transmission (relative to between-gender) \\
$\phi_t^F$ & The relative importance$^*$ of inheritable skills for income, fathers \\
$\phi_t^M$ & The relative importance$^*$ of inheritable skills for income, mothers \\
$\phi_t^S$ & The relative importance$^*$ of inheritable skills for income, sons \\
$\phi_t^D$ & The relative importance$^*$ of inheritable skills for income, daughters \\
\bottomrule
\end{tabular}
\vspace{0.2cm}
\linespread{1}
\scriptsize{\textit{Notes:} Details of the model and parameters are available in Appendix Section \ref{APP:MODEL}. $^*$ The $\phi_t^K$} parameters reflect inheritable income-generating potential (``skills'') in relation to the non-inheritable idiosyncratic factors that we referred to above as the attenuation factor $\varepsilon$.
\end{threeparttable}
\end{table}

\begin{table}[H]
\centering
\linespread{1}
\caption{\label{tab:parms} Estimated Values of the Model Parameters for Cohorts 1951, 1962, and 1979.}
\begin{threeparttable}
\renewcommand{\arraystretch}{0.5} 
\begin{tabularx}{\textwidth}{XXXrcrXrrrXlllX} 
\toprule
         & & & \multicolumn{3}{c}{1951} & \multicolumn{1}{c}{} & \multicolumn{3}{c}{1962} &  & \multicolumn{3}{c}{1979} & \\ 
\cmidrule{4-6}\cmidrule{8-10}\cmidrule{12-14}
           & & & SE    & DK & NO          &                      & SE    & DK    & NO       &  & \multicolumn{1}{r}{SE} & \multicolumn{1}{r}{DK} & \multicolumn{1}{r}{NO} &  \\ 
\midrule
& $\psi_t$  & & 0.131 & -  & 0.147       &                      & 0.289 & 0.189 & 0.171    &  & 0.249                  & 0.186                  & 0.174   &                \\
& $\kappa_t$ & & 0.301 & -  & 0.300       &                      & 0.257 & 0.267 & 0.274    &  & 0.261                  & 0.290                  & 0.286 &                  \\
& $\alpha_t$ & & 0.580 & -  & 0.603       &                      & 0.632 & 0.582 & 0.626    &  & 0.561                  & 0.560                  & 0.564 &                  \\
& $\phi^M_t$ & & 0.286 & -  & 0.260       &                      & 0.368 & 0.398 & 0.371    &  & 0.594                  & 0.701                  & 0.622  &                  \\
& $\phi^D_t$ & & 0.511 & -  & 0.501       &                      & 0.591 & 0.721 & 0.619    &  & 0.935                  & 1.011                  & 0.951 &                   \\
\bottomrule
\end{tabularx} \vspace{0.2cm}
\linespread{1}
\scriptsize{\textit{Notes:} The table presents calibrated decomposition parameters for Sweden (SE), Denmark (DK), and Norway (NO) in three selected cohorts. These cohorts represent the start of the Swedish and Norwegian data, the start of the Danish data, and the final cohort in the sample. The coefficients have been obtained by matching a simulated version of the model described in Section \ref{sec:model_setup} to the gender-specific IRA coefficients observed in the data, as well as the relation between father and mother income.}
\end{threeparttable}
\end{table}

\begin{table}[H]
\centering
\linespread{1}
\caption{\label{tab:parm_trends} Results from the Decomposition Exercise for Cohorts 1951-1962 and 1962-1979.}
\begin{threeparttable}
\renewcommand{\arraystretch}{0.5} 
\begin{tabularx}{\textwidth}{XlXrcrXrrrX} 
\toprule
                           & & & \multicolumn{3}{c}{1952-1961} & \multicolumn{1}{c}{} & \multicolumn{3}{c}{1962-1979} &  \\ 
\cmidrule{4-6}\cmidrule{8-10}
                          & & & SE     & DK & NO              &                      & SE     & DK     & NO &          \\ 
\midrule
& Trend in $\beta_t$ &        & 0.013  & -  & 0.140           &                      & 0.277  & 0.530  & 0.379  &       \\ 
\midrule
&Trend in $\tilde{\beta}_t$ & & 0.068  & -  & 0.138           &                      & 0.240  & 0.527  & 0.327  &      \\
& \quad\text{Due to }$\psi_t$ & & 0.189  & -  & 0.000           &                      & -0.056 & -0.001 & 0.001  &      \\
& \quad\text{Due to }$\kappa_t$ & & -0.343 & -  & -0.164          &                      & -0.041 & 0.242  & -0.035 &      \\
& \quad\text{Due to }$\alpha_t$ & & 0.009  & -  & 0.007           &                      & 0.002  & 0.000  & 0.001  &      \\
& \quad\text{Due to }$\phi^M_t$ & & 0.054  & -  & 0.117           &                      & 0.138  & 0.220  & 0.161 &       \\
&\quad\text{Due to }$\phi^D_t$ & & 0.020  & -  & 0.130           &                      & 0.158  & 0.069  & 0.119 &       \\
\bottomrule
\end{tabularx} \vspace{0.2cm}
\linespread{1}
\scriptsize{\textit{Notes:} The table presents trends in observational IRA coefficients, $\beta_t$, in the three countries as well as trends in IRA coefficients obtained from the calibrated models in the three countries, $\tilde{\beta}_t$. The contribution from each parameter is computed as the difference in $\tilde{\beta}_t$ that is obtained from holding one calibrated parameter fixed at a time. The sum of contributions from each parameter need not sum to the trend in $\tilde{\beta}_t$ as part of the trend will be driven by changes in the scale of gender-specific income distributions, which is not modeled.}
\end{threeparttable}
\end{table}

\newpage

\section*{Appendix} \appendix
\renewcommand\thefigure{\thesection\arabic{figure}} 
\renewcommand\thetable{\thesection\arabic{table}}

\section{Data Registers and Variable Definitions} \label{APP:data_registers}
\setcounter{figure}{0} 
\setcounter{table}{0} 

\noindent\textbf{Denmark}

The Danish income registries start in 1980 and contain detailed information on the individual income composition of Danish adults. The registries are based on information from the Danish tax authorities, supplemented with information from other Danish authorities, including unemployment insurance funds and local authorities. 

The measure of labor income that is being used in this paper consists of wage payments (incl. non-wage benefits, non-taxable wage payments, stock options, and more) and any net surplus from own, private company. Gross income is equal to labor income, transfers, property income, and any other non-classifiable income that the individual may have received throughout the year. Net-of-tax income is finally equivalent to gross income net of all taxes that have been paid to either the government, municipalities, or other public authorities. Individuals with no parents in the sample (generally people who moved to Denmark, whose parents have moved abroad, or whose parents are deceased) are naturally dropped from the sample. In order to ensure comparability between income definitions across countries, negative observations of income have been set to income.

When constructing household income measures, individuals are being linked to their spouses. In the Danish sample, a spouse is generally defined by marriage, registered partnership, or registered as a cohabiting couple. Matching individuals to spouses as well as parents is done using the population registries of Denmark.

Occupation data are obtained from the Danish employment classification module, AKM. Occupations are characterized by the Danish ISCO classification system (DISCO)\footnote{In this paper, we focus on the 1-digit classification codes.} and only observable from 1992 and onward.\footnote{There is a series of data breaks occurring from 1992 and onward --- we attempt to handle these appropriately.} In order to impute data from 1980 to 1991, we use a random forest-algorithm in order to map a range of occupation-related variables to occupation codes in 1992.\footnote{Code is available upon request} We then use this mapping in order to characterize individual occupations from 1980 and up until 1990 where the occupation codes are missing. Starting from 1991, there are several data breaks, in the sense that certain types of occupations are re-classified, split up or gathered into one group. In order to make occupations as comparable as possible over time, we re-classify occupations across years so that classifications are approximately constant over time. For certain individuals, the occupational status is either unknown or missing. These are instead assigned to the occupational status that was observed most recently within a window of +\/- 3 years. Due to these imputation and re-classification procedures, any results in this paper that rely on occupation codes from Denmark should be interpreted with some caution.

Information on individuals working part time is obtained from a matched employer-employee data module, IDAP. For a series of individuals (people who are not self-employed and do not have multiple occupations), full time work is assessed in a relatively straightforward way. In cases of doubt (individuals are e.g. self-employed or have multiple part-time occupations), we assign full employment status to a individual whose labor income amounts to at least the \textit{mean} labor income of full-time employed individuals (specifically, full-time employed individuals whose income fall within the range of the 15th and 25th income percentile as observed within the same age-category). Up until 2007, full-time work is classified as at least 27 weekly hours. From 2008 and onward, the equivalent number is 32. This data break does not show up in the parental generation in our sample (as these are observed up until 1998). In the child generation, gender-specific full-time rates are re-scaled in the following way in order to account for the data-break when computing aggregate cohort- and gender-specific statistics: 

\begin{center}
$$ \widetilde{\mathrm{ftr}}_t^g  = \begin{cases}  \mathrm{ftr}_t^g, & \text { if } t<1971 \\ \widehat{\mathrm{ftr}}_t^g, & \text { if } t=1971 \\ 1-\left(1-\mathrm{ftr}_t^g\right) \frac{\left(1-\widehat{\mathrm{ftr}}_{1971}^{g}\right)}{\left(1-\mathrm{ftr}_{1971}^g\right)}, & \text { if } t>1971 \end{cases}$$
\end{center}

where $\widetilde{\mathrm{ftr}}_t^g$ are the full time rates that we report in Figure \ref{fig:labor_development}, ${\mathrm{ftr}}_t^g$ are full time rates as observed in the data (hence, without accounting for the data break) and $\widehat{\mathrm{ftr}}_t^g$ is the fitted gender-specific full time rate in 1971 that is obtained from a linear regression of full time rates, ${\mathrm{ftr}}_t^g$, on $t$, using full time rates from 1966 up until 1970.

An individual's level of education reflects the highest attained level of education. In some years, the duration of the education is not available in the data. In these cases, the duration of the education is either imputed from the duration of that same education in an earlier year, or it is imputed from educations that are characterized as being similar.

\vspace{2cm}
\noindent\textbf{Norway}

Our Norwegian data set combines information from the central population registry with information about income and earnings from the tax registry. Income data in Norway is available from 1967 to 2018. Labor income includes payments related to employment, including overtime pay, taxable sickness benefits, parental leave pay, short-term disability pay, and rehabilitation benefits. This is top-coded for a few years in the 1970s at the maximum amount for contributions to the national social security scheme (folketrygden). Gross income is the sum of labor income and taxable and non-taxable transfers, and income from capital. Disposable income is defined as gross income minus taxes and is also sometimes referred to as net-of-tax income. The definitions change somewhat over time due to reforms of the benefit, insurance, and tax systems. For the net-of-tax and gross income variables, the data series ends in 2014, which is why these income measures are constructed from more detailed income data only available from 1993. Spouses (married couples as well as couples in civil unions) are linked through their personal identifiers

The occupation data used for implementing the method proposed by \citet{LW2006} is pooled from matched employer-employee data (\textit{Registerbasert sysselsettingsstatistikk}) available annually starting with the year 2000. In addition occupation data from the censuses 1960, 1970, and 1980 are added. To achieve a comparable classification of occupations we use the STYRK-98 one-digit code to group individuals into broad occupational groups (see Table \ref{tab:occ_class}). Individuals are assigned the occupation they have at age 36. In cases where this is not possible we use the closest applicable occupation we observe in the data. Due to the long break in occupational data between the 1980 census and the start of the employer-employee data, there might be some differences in the age at which we observe occupations.

Data on intensive margin labor supply is obtained from an earlier version of matched-employer employee data covering the years 1986 to 2010. These data are available annually and include information on individuals' employment characteristics in connection with Norwegian firms. In order to construct the full-time share in Figure \ref{fig:labor_development} we use a variable categorizing employment relationships into three categories depending on the contracted weekly working hours. For Norway all individuals working at least 31 hours a week are counted as full-time workers. In order to account for year-to-year fluctuations in intensive margin labor supply, we take the mode of this variable over three years to capture individual hours worked. In detail, this means that parents hours are constructed as the mode of hours worked during the three years around the time their child turns 18. For children hours are constructed as the mode during the three years around age 30. Due to data limitations the age at which child hours are captured is slightly lower than the age used for our main lifetime income measure.  

The educational data for the LW method is also pooled from different registries. Most individuals we observe are included in the national education database available from 1970. These data include variables for the highest achieved education of all individuals, which we can link via personal identifiers. For individuals who are not included in the national education database, we obtain information about their educational attainment via census data from 1960, 1970, and 1980.

\vspace{1cm}
\noindent\textbf{Sweden}

The Swedish Income and Taxation registry starts in 1968 and holds official records of income for all individuals with any recorded income. In general, it contains all earned income from employment or businesses, capital income, taxable (mostly social insurances), and non-taxable transfers (social welfare, educational grants, child benefits, etc.). Identifiers for biological or adoptive parents are linked to the child identifier through the multi-generational register. Households are constructed by linking individuals (children, mothers, and fathers) to their spouses. This is available only for married couples (and those in registered partnerships) and thus excludes households formed by cohabiting partners. 

Data on occupations are taken from two sources. First, the population censuses, \\ \textit{Folk- och bostadsräkningarna}, contain occupational codes corresponding to the ISCO-58 classification system, called NYK. This information is available in 1960, and then every five years between 1970 and 1990 for the adult population. Individuals who are in the labor force, but whose occupation is unknown are dropped from the sample, while individuals not in the labor force are coded as "no occupation". The census data are used to infer occupations for all parents in our Swedish sample, and we assign each parent an occupational code from the census closest in time to when the child is 18 years old (for example, a mother with a child born in 1951 will primarily be assigned an occupational code from the 1970 census, and occupations for fathers with children born in 1975 will be taken from the 1990 census). If no occupations is observed in this focal year, we search iteratively through the second and third closest waves, and so on. Parents who are missing an occupational code after this procedure, and who are at least 18 years old in 1960, are assigned occupations from that year's census. This mainly serves to capture occupations of women who are out of the labor force continuously after the birth of their first child; about 6.5 percent of the mother sample (3 percent of the fathers). 

Occupational codes for the child generation are primarily taken from the 1985 and 1990 censuses for individuals born in the years 1951-1955, and primarily from population register data for birth cohorts 1956 to 1979. Since these sources use different classification systems, we have created a mapping between the NYK and the SSYK occupational codes at the 3-digit level, available upon request. The population occupation register, contained in the \textit{Integrated database for labour market research (LISA)} uses an adapted version of the ISCO-08 classifications, called SSYK-2012, and is available in our data for the years 2012-2017\footnote{Years 2012 and 2013 are re-coded at the 4-digit level from the earlier SSYK-96 occupation codes, using the official translation key from Statistics Sweden (https://www.scb.se/dokumentation/klassifikationer-och-standarder/standard-for-svensk-yrkesklassificering-ssyk/ 2022-09-08.)}. As a result, the age at which occupations are observed among the child sample varies between 35 and 56, which might induce noise in between-birth cohort comparisons. On the other hand, this age span corresponds to prime working age, and occupational choice is relatively constant, especially given the broad classes we use in our analysis.

Information on parent- and children-specific rates of full time work are calculated using the Swedish Level of Living Survey, which is a nationally representative survey on about 0.1 \% of the adult population from years 1968, 1974, 1981, 1991, 2000, and 2010. Respondents are asked about their contracted number of weekly working hours (1974 and on), alternatively about their number of hours worked last week (1968). Only respondents with non-zero survey weights are kept in the sample. \textit{Parents} are defined as individuals aged 45-55 with at least one child in the survey waves 1968-1991. We set the "birth year of the child" to 1951, 1957, 1964, and 1974, respectively, i.e. 18 years before the observation. \textit{Children} are defined as individuals aged between 30 and 36 in the years 1981 to 2010.

The highest attained level of education is observed in the 1970 census, and in the annual population registers that start in 1990. Each person is assigned the level of education that he or she displays in the year closest in time to when income is observed (age 36 for children; age 18 of the child for the parents). Years of education is then inferred from these categorical data (e.g., completing a three-year secondary education program is coded as twelve years of education, or eleven years if the person completed primary school when it was still only seven years in duration).

\begin{landscape}
\def\arraystretch{0.3}
\begin{threeparttable}
\footnotesize
\centering
\begingroup
\centering
\singlespacing
\caption{Overview Income Definitions by Country.}
\label{tab:inc_def}
\begin{tabular}{p{0.025\linewidth}p{0.075\linewidth}p{0.3\linewidth}p{0.3\linewidth}p{0.3\linewidth}}
    \toprule
    &  & Denmark & Norway & Sweden\\
    \midrule
    1 & Salary & taxable salary incl. fringe benefits, tax-free salary, anniversary and severance pay and value of stock options & all payments related to employment including overtime pay & all payments from employment\\
    \midrule
    2 & Net Profit & net profit from self-employment incl. profit of foreign company and net income as employed spouse & net income from self-employment and income from other businesses & net profit from self-employment, income from other businesses \\
    \midrule
    3 & Transfers & cash benefits, unemployment insurance benefits, sickness benefits, unemployment benefits, pensions, child allowance, and more & taxable sickness benefits, parental leave benefits, unemployment benefits, short-term disability payments, rehabilitation benefits & sickness benefit from employer (sjuklö), value of e.g. car, travel expenses (förmånsvärden)\\
    \addlinespace[0.3em]
    \midrule
    \multicolumn{5}{c}{\textbf{Earnings/Labor Income = Combination of 1+2+3}}\\
    \midrule
    \hspace{1em}
    4 & Transfers & cash benefits, unemployment insurance benefits, sickness benefits, unemployment benefits, pensions, child allowance, and more & taxable transfers: benefits from the national insurance scheme (disability insurance, pensions, etc.) non-taxable: child benefits, housing allowance, scholarships, parental leave benefits, social assistance payments & taxable: social insurances (unemployment, parental leave etc), private pension income, stipends etc. non-taxable: pensions and annuities housing support child support social welfare alimony conscript support grants and loans for students\\
    \midrule
    \hspace{1em}
    5 & Property Income & capital and wealth income excl. calculated rental value of real estate & gross interest income, dividend income, return on life insurance, net realised capital gains (e.g. shares, house, land), other capital income (taxable rental income) & capital income (gross pre-2004, net post-2004) and after-tax rental income\\
    \addlinespace[0.3em]
    \midrule
    \multicolumn{5}{c}{\textbf{Gross Income = Combination of Earnings +4+5}}\\
    \midrule
    \hspace{1em}
    6 & Other Income & other non-classifiable income &  & \\
    \midrule
    \hspace{1em}
    7 & Taxes & taxes on earnings, wealth taxes, property value tax, tax on share dividends/gains and more & taxes, maintenance paid, mandatory insurance premia & all taxes\\
    \addlinespace[0.3em]
    \midrule
    \hspace{1em}
    8 & Negative Transfers &  &  & repayments of study loans, paid alimony\\
    \midrule
    \multicolumn{5}{c}{\textbf{Disposable Income = Combination of Gross Income - 7+10}}\\
    \bottomrule
\end{tabular}
\par\endgroup
\end{threeparttable}
\end{landscape}

\newpage

\begin{table}[H]
\centering
\linespread{1}
\caption{Occupation Classification by Country.}
\label{tab:occ_class}
\begin{threeparttable}
\renewcommand{\arraystretch}{0.5} 
\begin{tabular}{ll} 
\toprule
Code & Definition\\
\midrule
\multicolumn{2}{l}{\textbf{Norway}} \\
\midrule
0 & Armed forces and unspecified\\
1 & Managers\\
2 & Professionals\\
3 & Technicians and associate professionals\\
4 & Clerical support workers\\
5 & Service and sales workers\\
6 & Skilled agricultural, forestry and fishery workers\\
7 & Craft and related trades workers\\
8 & Plant and machine operators and assemblers\\
9 & Elementary occupations\\
\midrule
\multicolumn{2}{l}{\textbf{Sweden}} \\
\midrule
1 & Professional work (arts and sciences)\\
2 & Managerial work\\
3 & Clerical Work\\
4 & Wholesale, retail and commerce\\
5 & Agriculture, forestry, hunting, and fishing\\
6 & Mining and quarrying\\
7 & Transportation and communication\\
8 & Manufacturing\\
9 & Services\\
10 & Military/Armed Forces\\
\midrule
\multicolumn{2}{l}{\textbf{Denmark}} \\
\midrule
0 & Military work\\
1 & Management work\\
2 & Work that requires knowledge at the highest level in the area in question\\
3 & Work that requires knowledge at intermediate level \\
4 & Ordinary office and customer service work\\
5 & Service and service work\\
6 & Work in agriculture, forestry and fisheries\\
7 & Craft and related trades workers\\
8 & Operator and assembly work, transport work\\
9 & Elementary occupations\\
\bottomrule
\end{tabular} 
\linespread{1}
\scriptsize{\textit{Notes:} Occupational categories for Norway are assigned using the STYRK-08 classificaiton provided by SSB. For Sweden the classification follows SSYK-2012 similar to \citet{VostersNybom2017}. For Denmark, we use the first integer from the Danish ISCO classication (\href{https://www-dst-dk.translate.goog/da/TilSalg/Forskningsservice/Dokumentation/hoejkvalitetsvariable/personers-tilknytning-til-arbejdsmarkedet-set-over-hele-aaret--akm-/disco08-alle-indk-13?_x_tr_sl=da&_x_tr_tl=en&_x_tr_hl=en&_x_tr_pto=nui,sc}{link}). In the Danish case, note that this variable is not available for all years in the data. For this reason, we generate it from a set of other available occupation-related variables. Code is available upon request.}
\end{threeparttable}
\end{table} 

\newpage

\begin{table}[H]
\centering
\linespread{1}
\caption{Occupational Groups by Skill Level and Type of Work.}
\label{tab:occ_groups_new}
\begin{threeparttable}
\renewcommand{\arraystretch}{0.5} 
\begin{tabular}{lll} 
\toprule
Code                         & Label                        & Examples                                          \\ \midrule
00                           & missing                             & Undefined or missing occupation                                           \\
01                           & executives                          & Politicians, business executives, managers                                \\
02                           & professionals                       & Professions requiring advanced college degrees                            \\
03                           & nurses                              & Nurses, midwives, physical therapy                                        \\
04                           & teachers                          & Pre-school and elementary school teachers                                 \\
05                           & law                                 & Law professionals                                                         \\
06                           & professionals (lower)                & Professions requiring shorter college education                           \\
07                           & secretary/clerical                  & Secretary, bank clerks, administrators                                    \\
08                           & other administrative               & Sales persons, customer service agents, postal workers, local politicians \\
09                           & services                            & Waiters, beauticians, security personnel, public transport workers        \\
10                           & retail                              & Shop assistants, cashiers, phone marketing                                \\
11                           & care                                & Assistant nurse, personal assistant, nursery staff                        \\
12                           & agriculture                         & Farming, forestry, fishery                                                \\
13                           & construction/craft                  & Carpenters, welders, printers, food processing, tailors                   \\
14                           & machine operators                  & Mining workers, steelworkers, fitters, industry machine operators         \\
15                           & transportation                      & Truck drivers, sailors, bus drivers, train drivers                        \\
16                           & cleaning                            & Cleaning and domestic services                                            \\
17                           & military                            & Military personnel, officers                                              \\
18                           & medicine                            & Medical doctors, veterinarians, dentists, psychologists                    \\
19                           & teaching professionals             & University teachers, high school teachers, vocational teachers            \\
20                           & farm help                         & Planters, croppers                                                        \\
21                           & manual                              & Dock workers, factory workers                                             \\
22                           & other services                     & Garbage collectors, market vendors, fast food workers, janitors        \\    
\bottomrule 
\end{tabular} 
\linespread{1}
\scriptsize{\textit{Notes:} The mapping between these codes and ISCO 3-digit codes is done by the authors, with the overall aim of separating female-dominated occupations from other occupations with the same 2-digit codes in the official ISCO system. These are used in the descriptive exercise in Section \ref{SEC:trends_gender}. The exact mapping between 3-digit codes and these groups can be obtained from the authors upon request.}
\end{threeparttable}
\end{table} 

\newpage
\section{Additional Figures and Tables} \label{APP:figures}
\setcounter{figure}{0} 
\setcounter{table}{0}  


\begin{figure}[H]
\begin{center}
\linespread{1}
\includegraphics[width=1\linewidth]{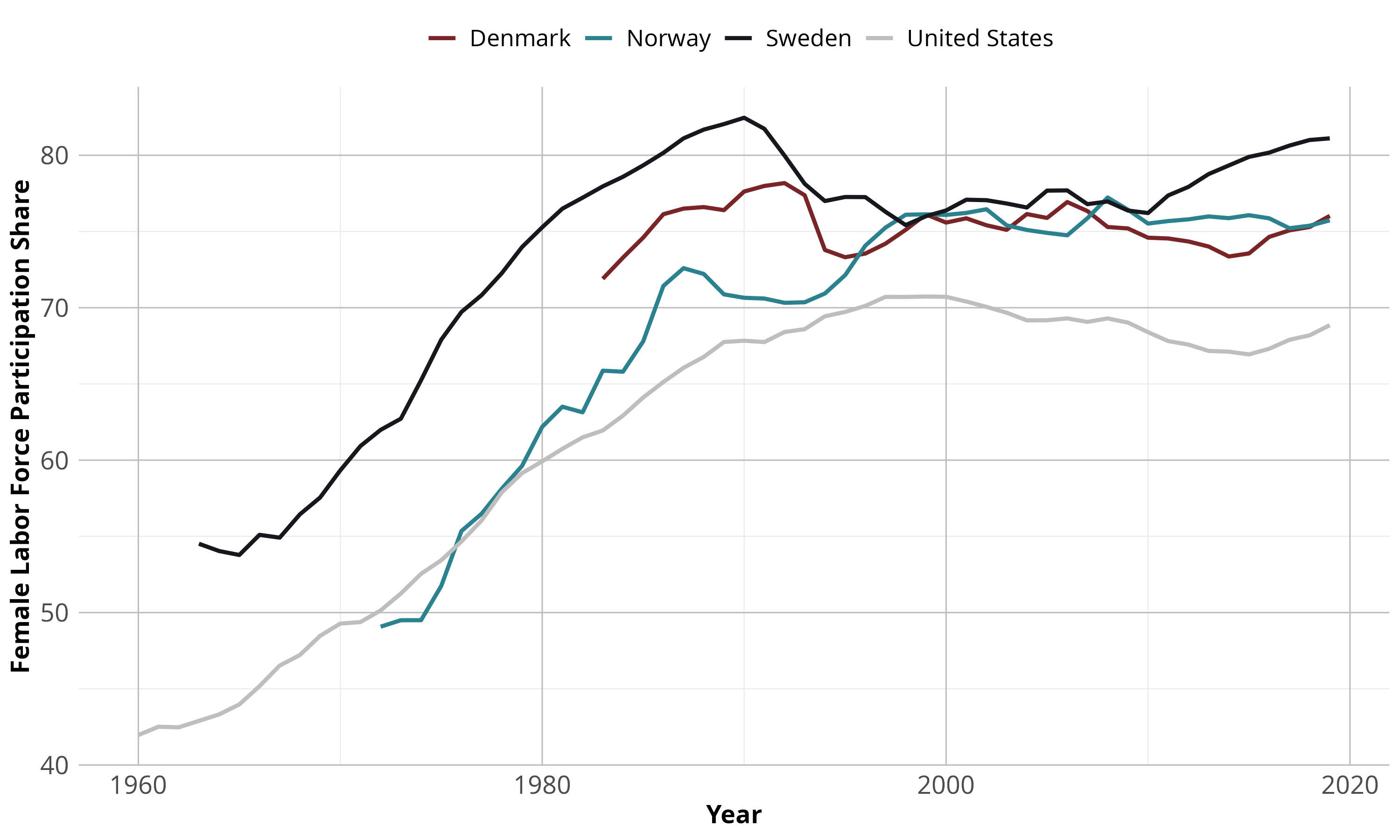}
\vspace{-.3cm}
\caption{Female Labor Force Participation Rate.}
\label{fig:lfp_comparison}
\end{center}
\linespread{1}
\scriptsize{\textit{Notes:} The figure depicts the labor force participation rates of women aged 15 to 64 for Denmark, Norway, Sweden, and the United States. The data was obtained from the \citet{OECD2021LFP} and covers all years available for the respective countries.}
\end{figure}

\begin{figure}[H]
\begin{center}
\linespread{1}
\includegraphics[width=1\linewidth]{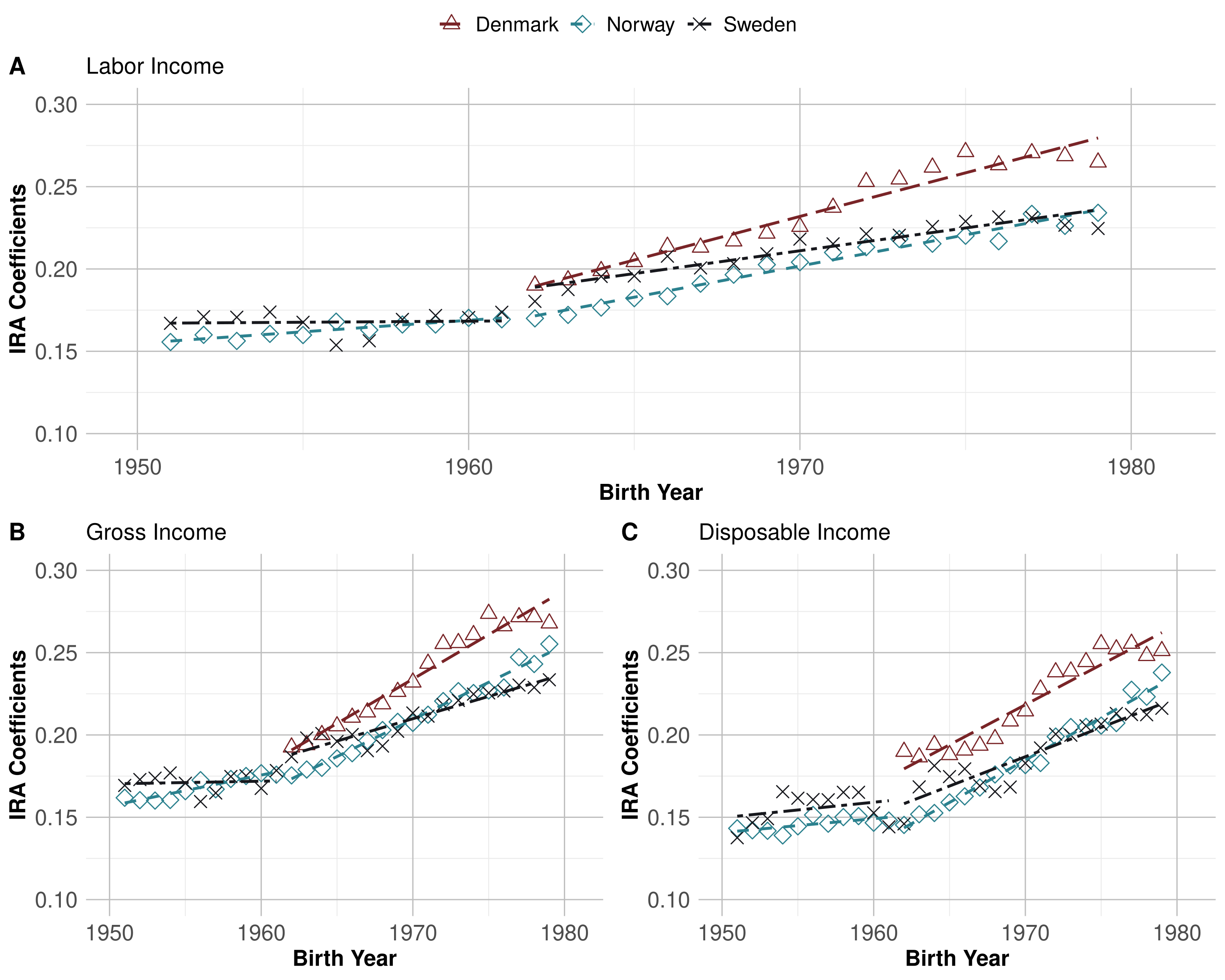}
\end{center}
\caption{Estimates of IRA in Labor, Net-of-tax and Gross Income.}
\label{fig:income_types}
\linespread{1}
\scriptsize{\textit{Notes:} Each panel depicts intergenerational rank associations (eq. \ref{rank_equation}) between parents and children, for each country and by birth year of the child. Panel A shows estimates of the main specification: labor income. In panel B, total factor (gross) income is used, and panel C depicts net-of-tax (disposable) income. Parental income averaged over child ages 17-19, and child income averaged over ages 35-37 in all estimates. ``Birth Year'' refers to the birth year of the \textbf{child} in each parent-child pair.}
\end{figure}

\begin{figure}[H]
\begin{center}
\linespread{1}
\includegraphics[width=1\linewidth]{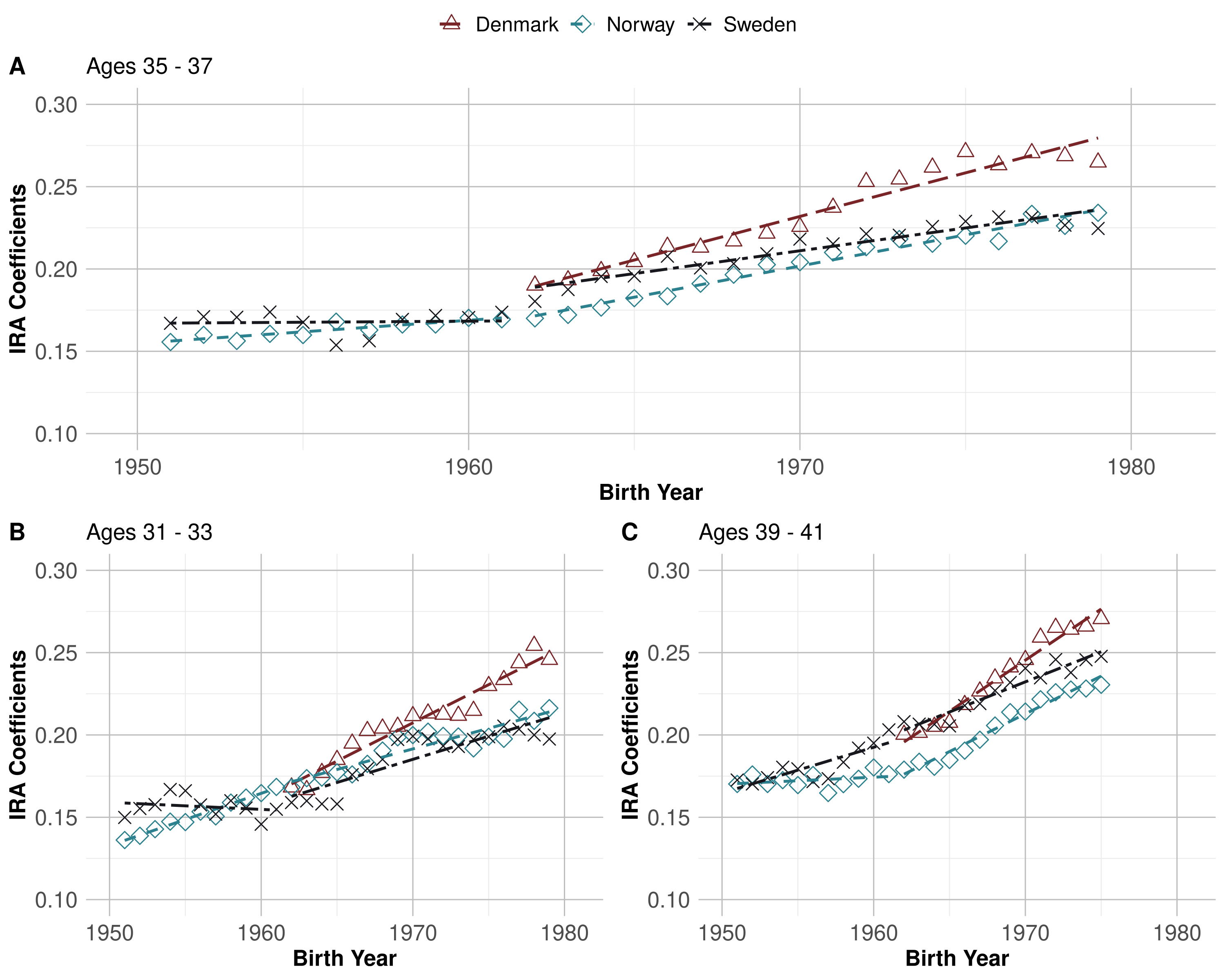}
\end{center}
\caption{Estimates of IRA at Different Ages of the Child (Labor Income).}
\label{fig:ages_income}
\linespread{1}
\scriptsize{\textit{Notes:} Each panel depicts intergenerational rank associations (eq. \ref{rank_equation}) between parents and children, for each country and by birth year of the child. Panel A shows estimates of the main specification: Average income at child ages 35-37. In panel B, child income is measured at ages 31-33, and in panel C, it is measured at ages 39-41. Parental income averaged over child ages 17-19 in all estimations. ``Birth Year'' refers to birth year of the \textbf{child} in each parent-child pair.}
\end{figure}

\begin{figure}[H]
\begin{center}
\linespread{1}
\includegraphics[width=1\linewidth]{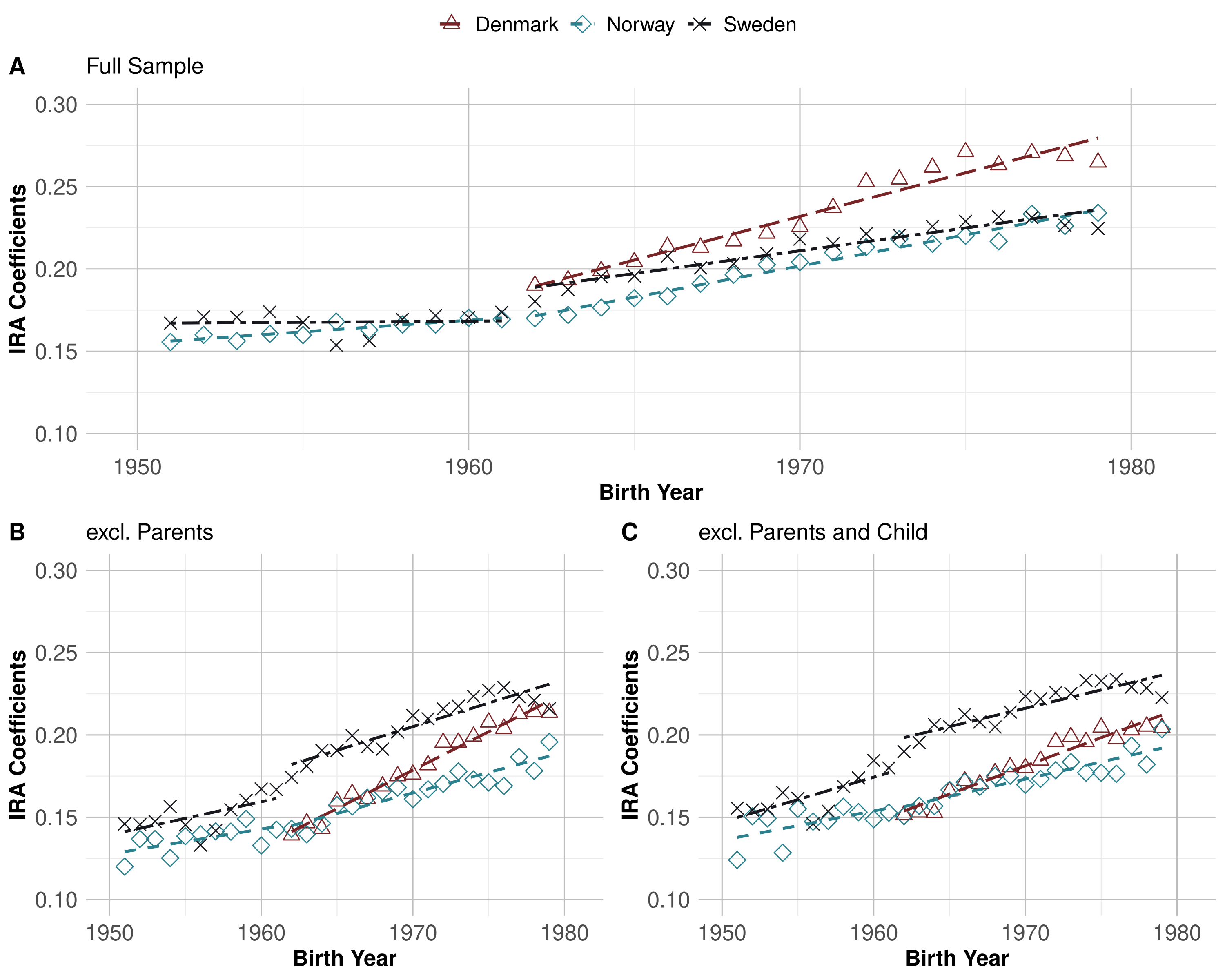}
\end{center}
\caption{Estimates of IRA, Labor Force Participants Only (Labor Income).}
\label{fig:ira_excluding}
\linespread{1}
\scriptsize{\textit{Notes:} This figure shows IRA coefficients by birth year of the child, when removing non-participants from the sample. Panel A shows the baseline estimates on the full sample; Panel B removes non-participants of the child generation; Panel C removes non-participants of the child and parent generations. ``Participation'' is determined by an indicator variable for having annual labor income exceeding 10,000 USD in a year. Parental income averaged over child ages 17-19 in all estimations; child income measured as the average over ages 35-37. ``Birth Year'' refers to the birth year of the \textbf{child} in each parent-child pair.}
\end{figure}

\begin{figure}[H]
\begin{center}
\linespread{1}
\includegraphics[width=1\linewidth]{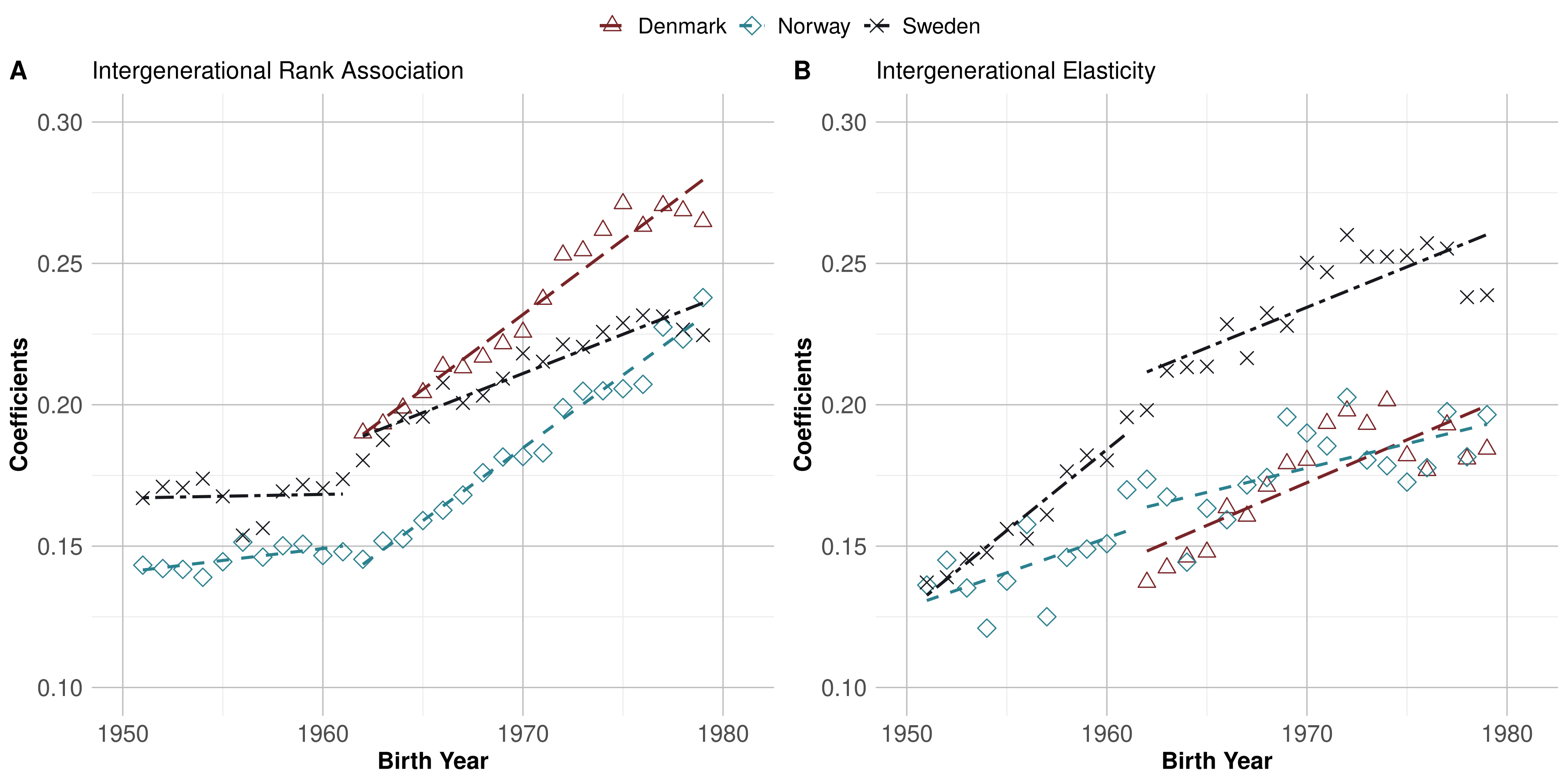}
\vspace{-.3cm}
\end{center}
\caption{Estimates of IRA and IGE in Labor Income.}
\label{fig:ira_ige}
\linespread{1}
\scriptsize{\textit{Notes:} Panel A depicts intergenerational rank associations (eq. \ref{rank_equation}) between parents and children for Sweden, Norway, and Denmark, by birth year of the child. Panel B shows intergenerational income elasticities, i.e., correlations in log income between parent and child pairs (with zero income excluded from analysis). Parental income averaged over child ages 17-19, and child income averaged over ages 35-37 in all estimates. ``Birth Year'' refers to the birth year of the \textbf{child} in each parent-child pair.}
\end{figure}

\begin{figure}[H]
\begin{center}
\linespread{1}
\includegraphics[width=1\linewidth]{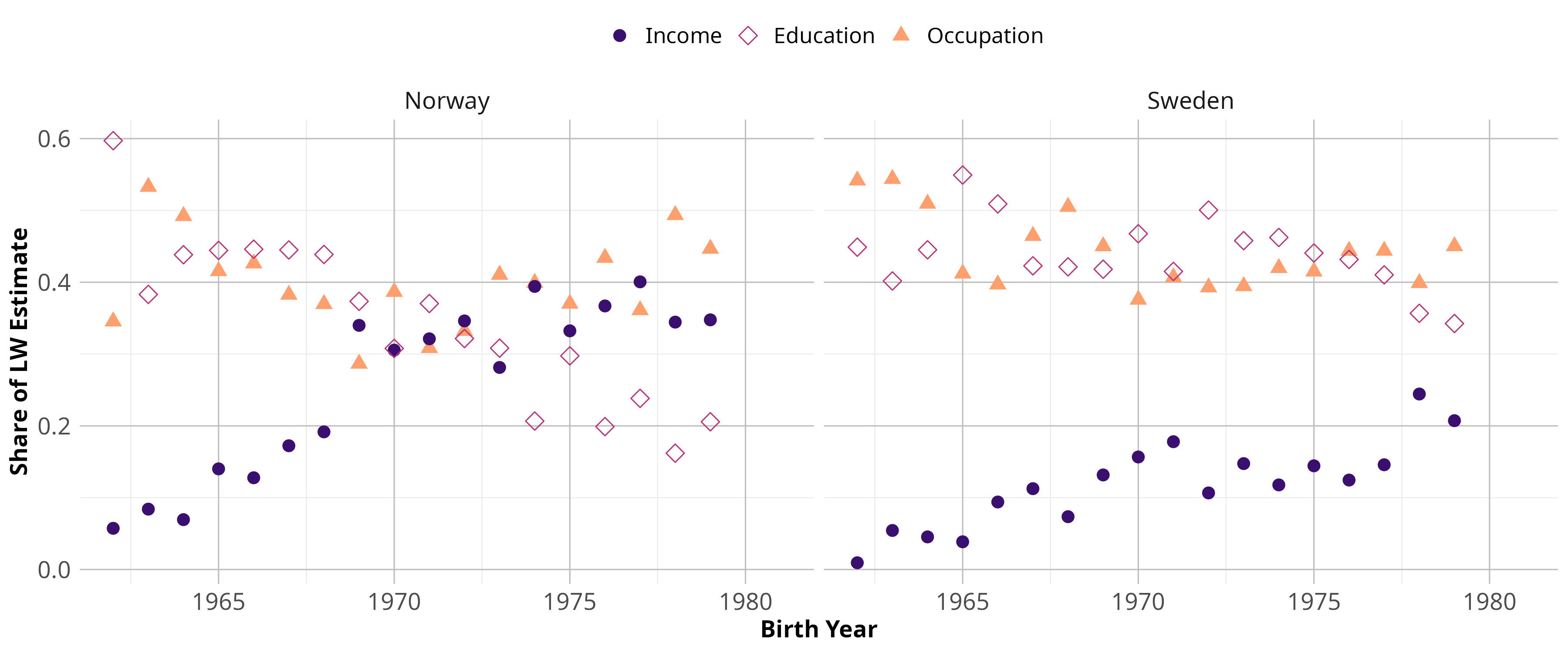}
\end{center}
\caption{The Relative Importance of Maternal Education, Occupation and Realized Income in Mother-Son Correlations in Potential Income.}
\label{fig:LW_contributions}
\linespread{1}
\scriptsize{\textit{Notes:} The figure plots the contribution of the different proxy variables to the overall estimate following \citet{LW2006}. We calculate the contribution of proxy variable $j$ for birth cohort $t$ as $\frac{\rho_{jt} b_{jt}}{\beta^{LW}_t}$, i.e. as the share of the aggregate in year $t$ coming from proxy variable $j$. The proxy variables depicted are education, income, and occupation. The contribution of occupations is calculated by summing over the ten different occupational group dummies. The left panel depicts results for Norway, while the right panel shows results for Sweden (data for Denmark is not available for this exercise).}
\end{figure}

\begin{figure}[H]
\begin{center}
\linespread{1}
\includegraphics[width=1\linewidth]{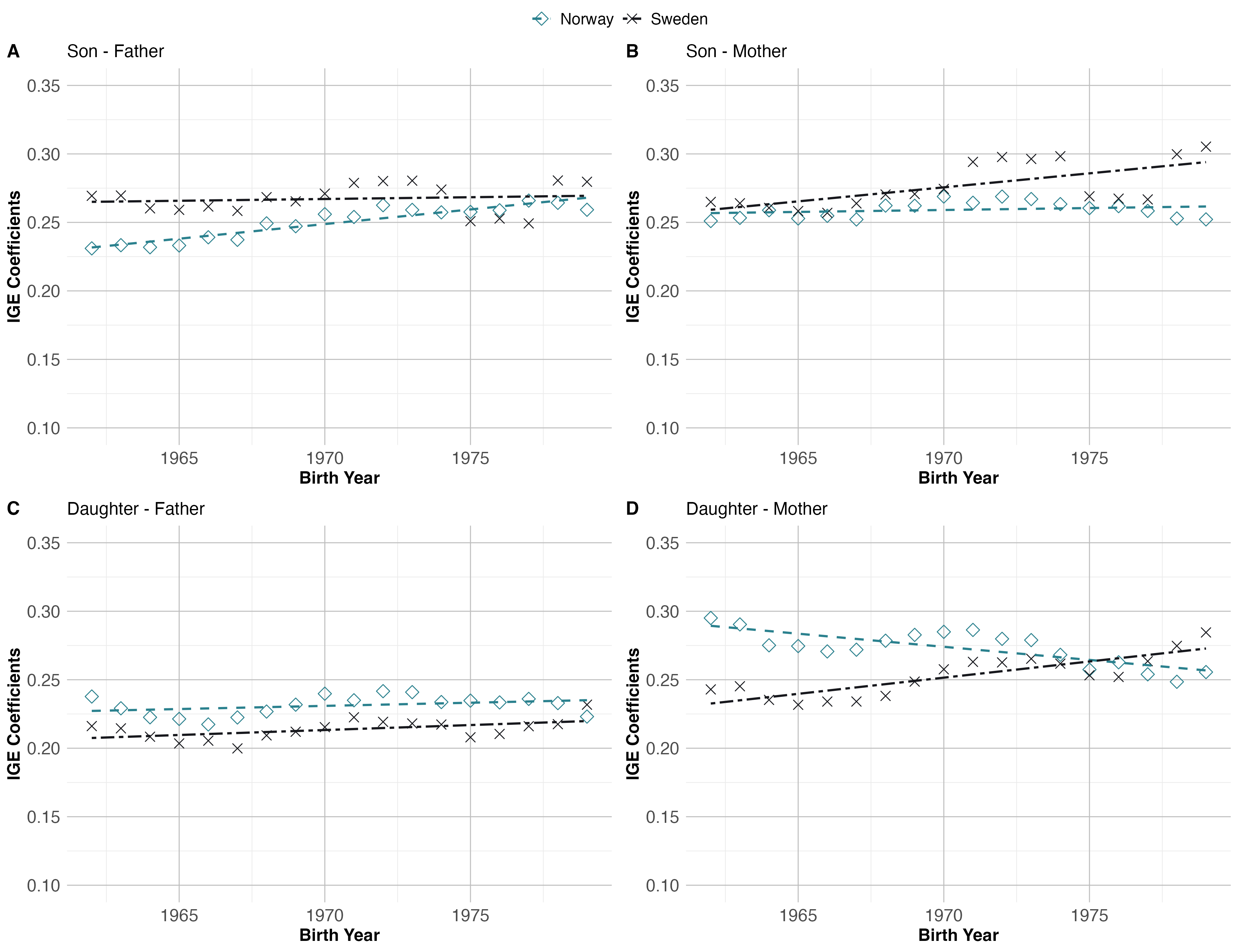}
\end{center}
\caption{Trends in Educational Mobility by Gender of Parent and Child.}
\label{fig:ige_education}
\linespread{1}
\scriptsize{\textit{Notes:} The four panels plot coefficients for the intergenerational elasticity in education for Sweden and Norway, by child year of birth. Each panel provides estimates separately by the gender of the parent and child. Each marker indicates the coefficient of a separate regression, and each line indicates fitted trend lines for the time period 1962 to 1979. “Birth Year” refers to the birth year of the child in each parent-child pair.}
\end{figure}

\begin{figure}[H]
\linespread{1}
\begin{center}
\includegraphics[width=1\linewidth]{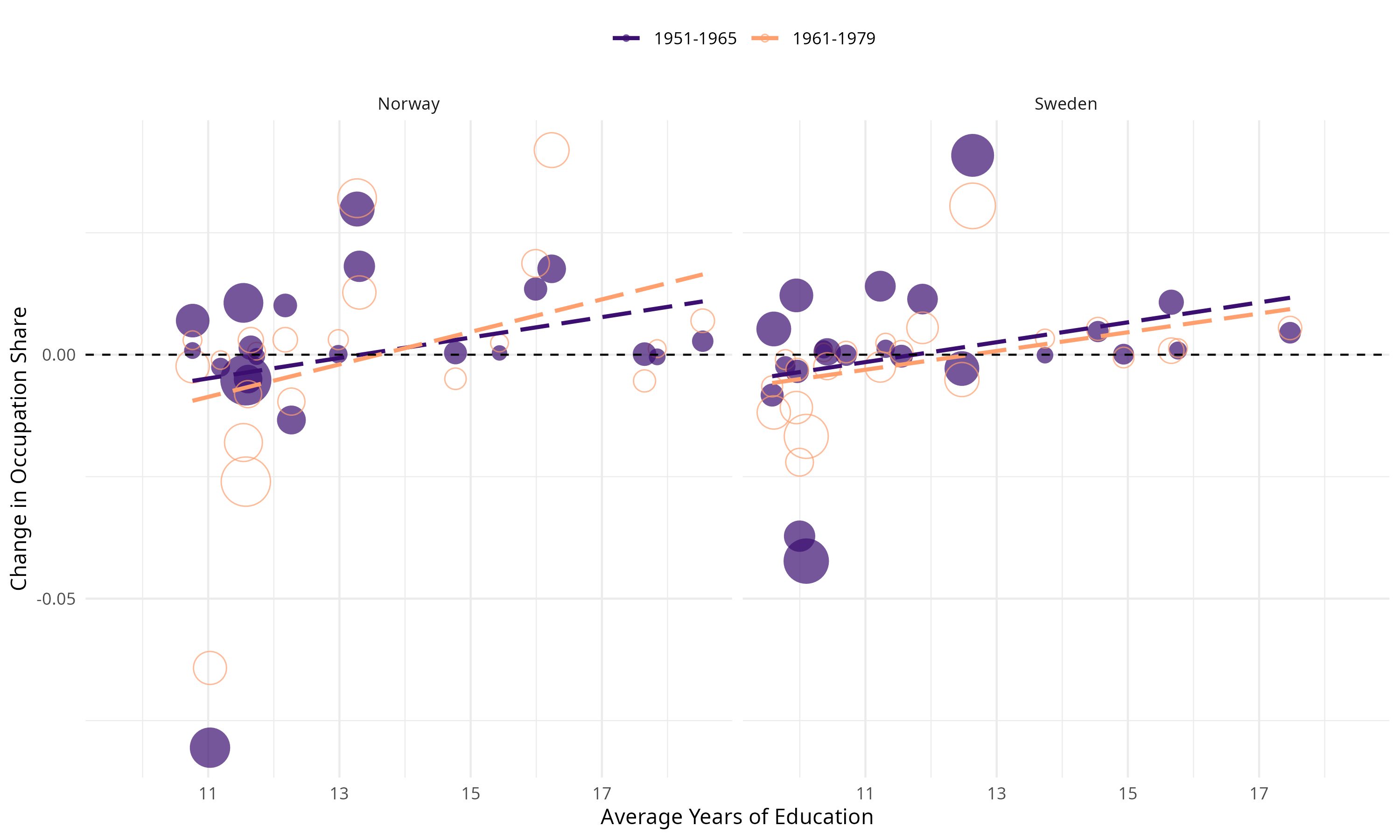}
\end{center}
\caption{Changes in the Occupational Composition of Fathers.}
\label{fig:occ_rank_father}
\linespread{1}
\scriptsize{\textit{Notes:} The figure plots the average years of education within an occupational group (a proxy for the skill level of the occupation) against the change in the share of fathers with a job in that occupational group, from one group of birth cohorts to another. Each dot represents an occupational group, and the size of the dot represents the group size. Purple (filled) circles denote the change from father of the 1951-1955 birth cohorts to father of the 1961-1965 birth cohorts; orange (hollow) circles denote 1961-1965 to 1975-1979. Average years of education calculated in 1975-1979. Dashed lines represent the OLS fitted line between change in occupational share and average years of education, weighted by occupational size; purple for 1951-1965 and orange for 1961-1979. The left panel shows Norway, the right panel shows Sweden. Occupations are grouped following Appendix Table \ref{tab:occ_groups_new}.}
\end{figure}

\begin{figure}[H]
\linespread{1}
\begin{center}
\includegraphics[width=1\linewidth]{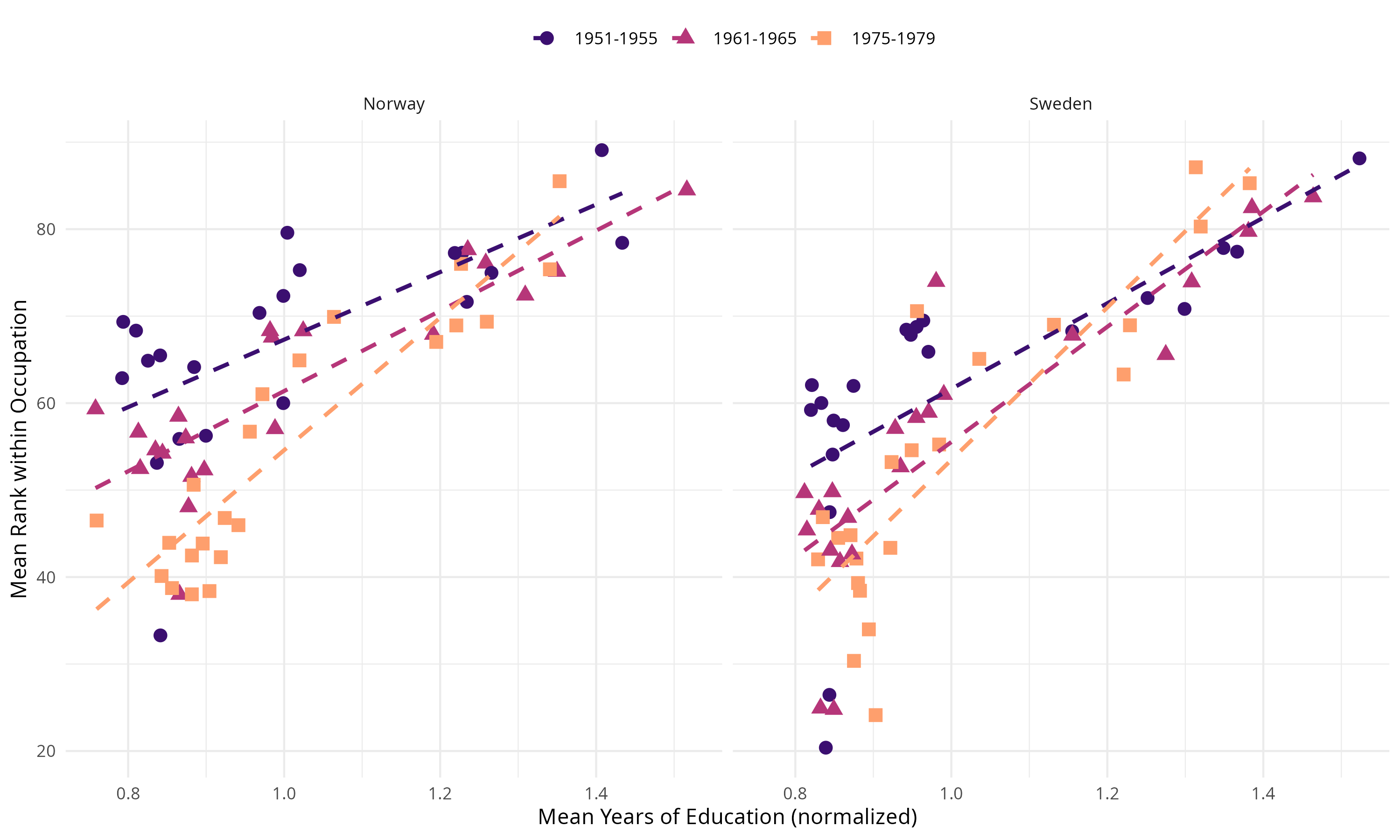}
\end{center}
\caption{Change in Average Income Rank and Education within Occupation Groups, across Birth Cohorts.}
\label{fig:edu_rank_occupations}
\linespread{1}
\scriptsize{\textit{Notes:} The figure shows the years of education relative to the mean in a given time period on the x-axis and the mean income rank of mothers in a respective occupation on the y-axis. The left panel provides the relationship between occupations, educational level, and mean income rank for Norway, while the right panel provides equivalent results for Sweden.}
\end{figure}

\begin{figure}[H]
\begin{center}
\linespread{1}
\includegraphics[width=.9\linewidth]{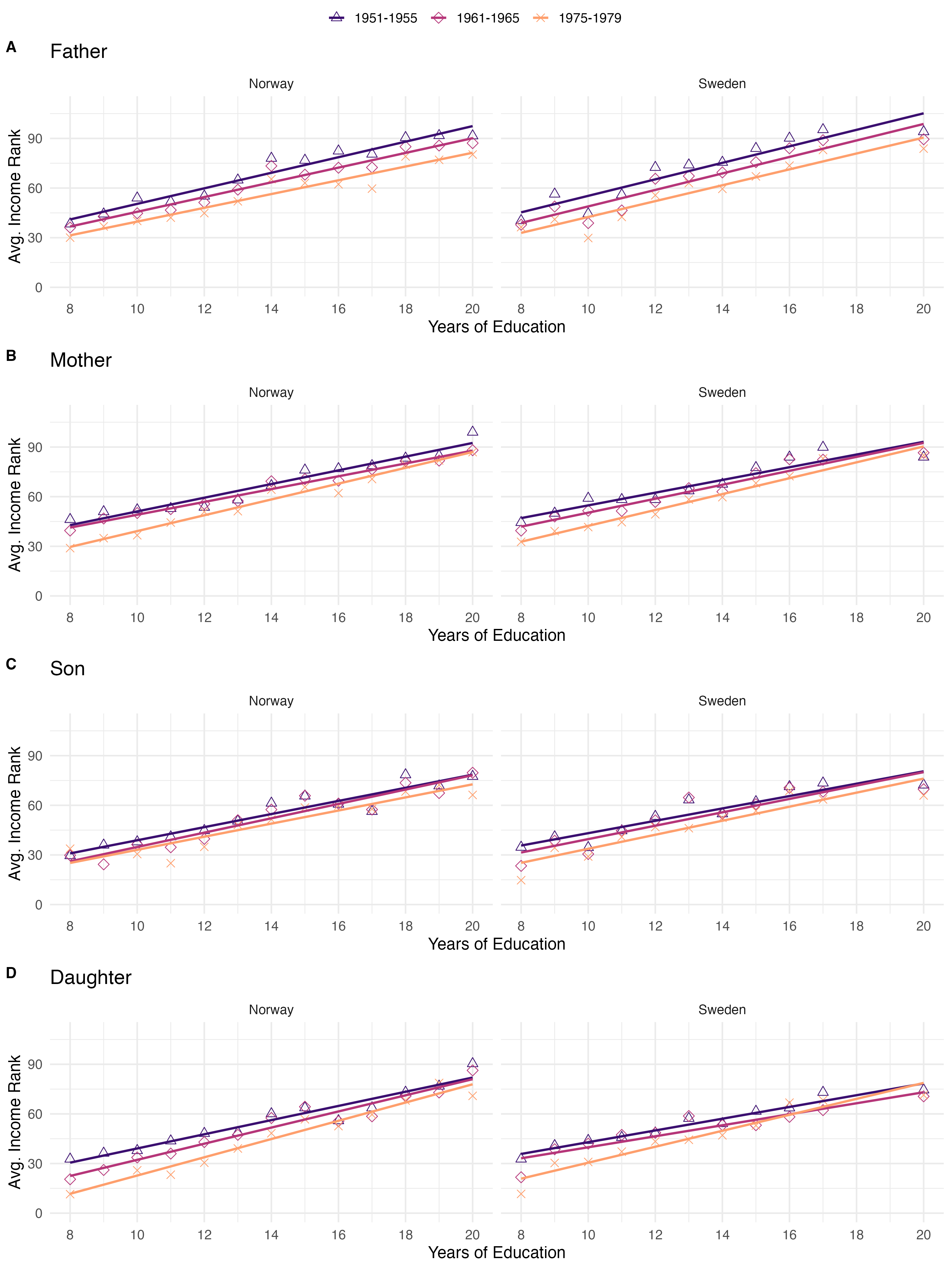}
\end{center}
\caption{Average Income Rank by Years of Education.}
\label{fig:avg_rank_by_eduy}
\linespread{1}
\scriptsize{\textit{Notes:} The figure shows the average income rank (in labor income, ranked separately by gender for sons and daughters) by years of education, for fathers in Panel A, mothers in Panel B, sons in Panel C, and daughters in Panel D. The left panel depicts results for Norway, while the right panel shows results for Sweden (data for Denmark is not available for this exercise).}
\end{figure}

\begin{figure}[H]
\begin{center}
\linespread{1}
\includegraphics[width=1\linewidth]{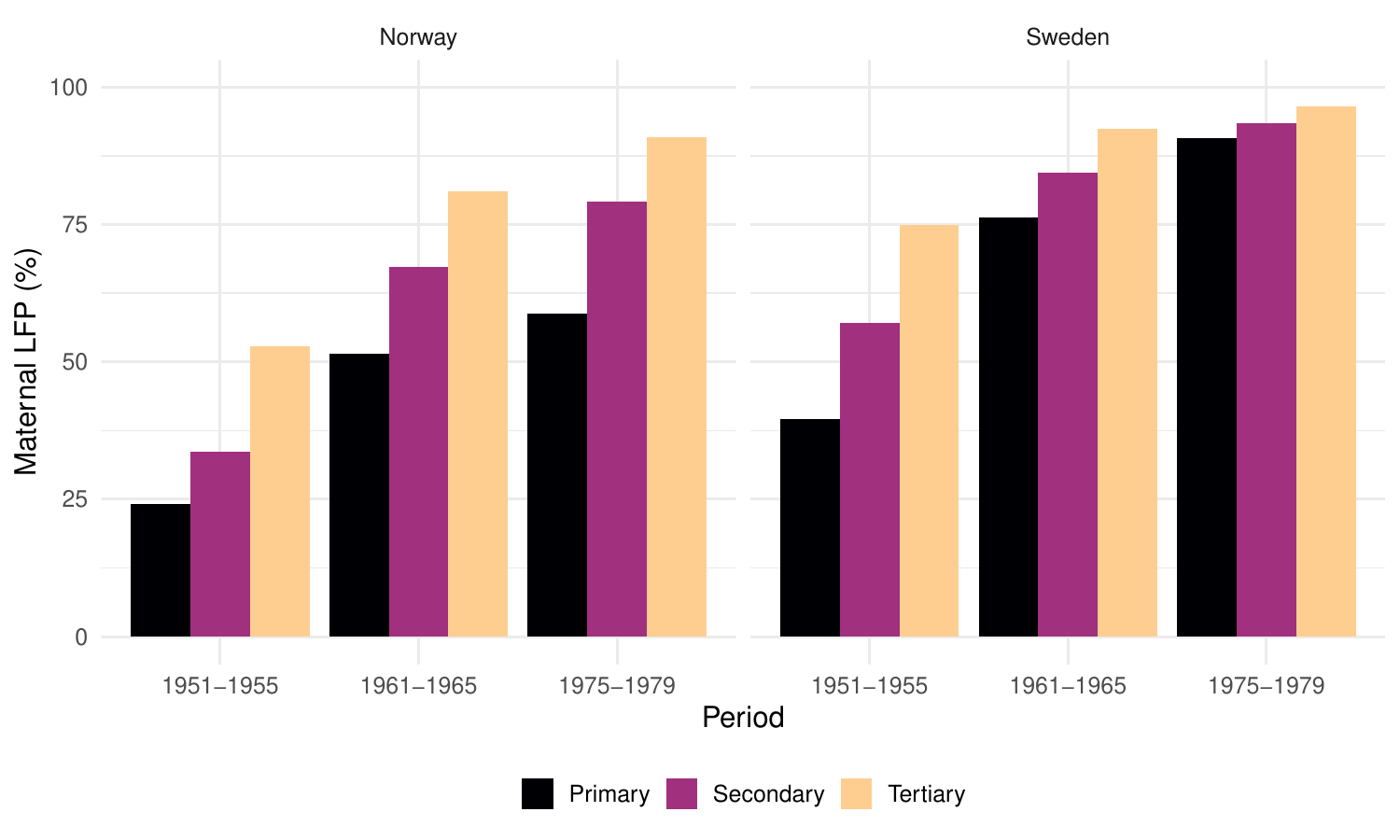}
\end{center}
\caption{Labor Force Participation among Mothers, by Level of Education and Birth Cohort of their Children.}
\label{fig:lfp_mothers_byedu}
\linespread{1}
\scriptsize{\textit{Notes:} The figure shows the labor force participation rate (in \%) among mothers of the 1951-55, 1961-65, and 1975-79 cohorts, respectively, by mothers' level of education. Maternal education is divided into three groups: less than high school (primary), high school degree (secondary), and any post-secondary (tertiary) education. Labor force participation is constructed as an indicator for average annual labor earnings across child ages 17-19 exceeding 10,000 USD (2017).}
\end{figure}

\begin{figure}[H]
\begin{center}
\linespread{1}
\includegraphics[width=1\linewidth]{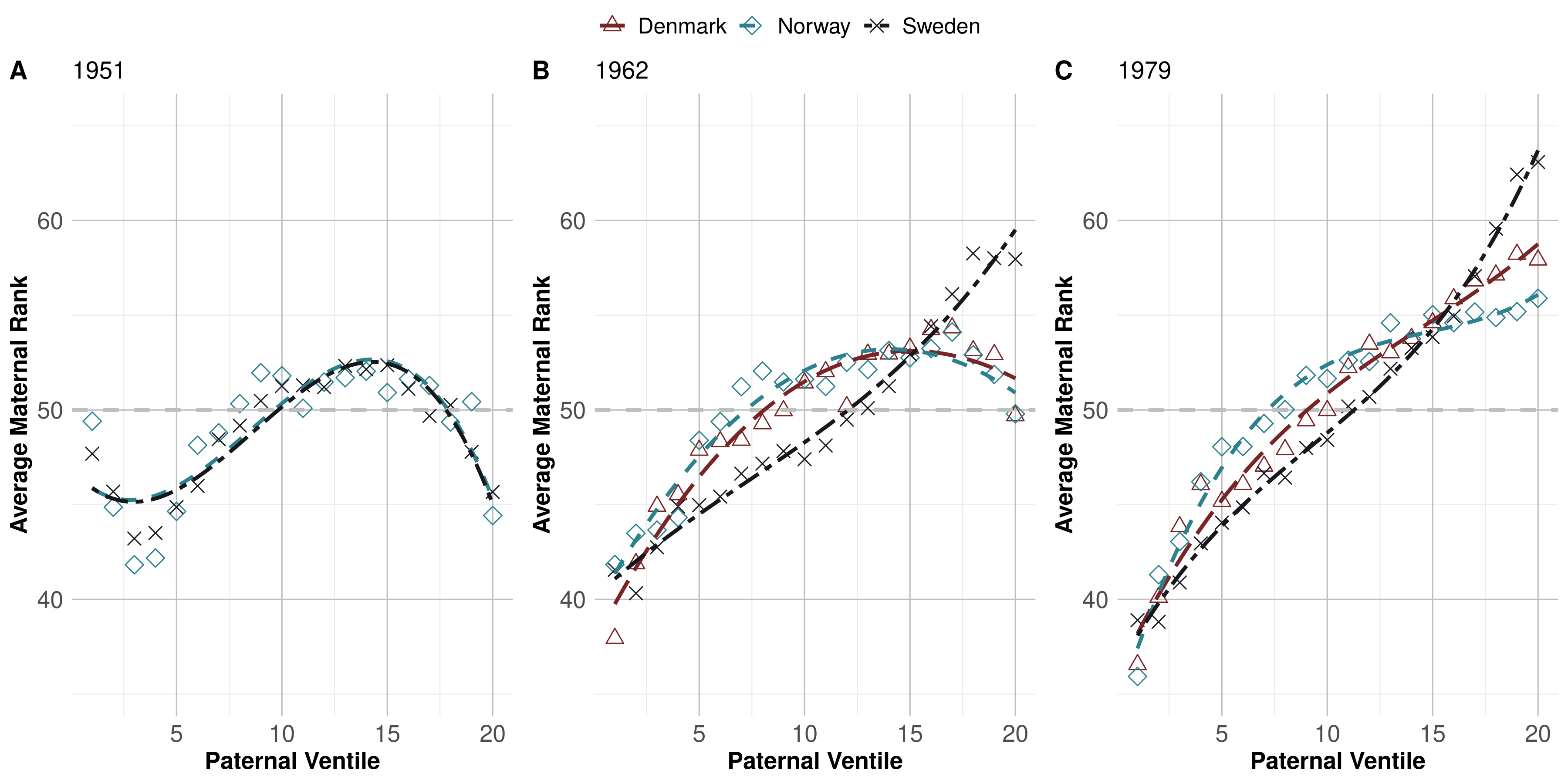}
\end{center}
\caption{Average Maternal Income Percentile Rank by Paternal Income Ventile.}
\label{fig:assort_mating}
\linespread{1}
\scriptsize{\textit{Notes:} The three panels show the average maternal income rank of mothers with children in the same birth cohort, by paternal (within parental pairs) income ventile. Each panel depicts these measures separately by country; for birth cohorts 1951 in Panel A; birth cohort 1962 in Panel B, and birth cohort 1979 in Panel C. The fitted lines are estimated with local polynomial (third order) regressions.}
\end{figure}

\clearpage

\begin{table}[H]
\centering
\caption{Summary Statistics Children 1951 - 1979.}
\label{tab:summary_stats_children}
\begin{threeparttable}
\renewcommand{\arraystretch}{0.5} 
\begin{tabular}{l*{9}{c}}
\toprule
 & \multicolumn{3}{c}{\textbf{Number of Children}} & \multicolumn{3}{c}{\textbf{Son Share Zero}} & \multicolumn{3}{c}{\textbf{Daughter Share Zero}}\\
\cmidrule(lr){2-4} 
\cmidrule(lr){5-7} 
\cmidrule(lr){8-10}
 & \multicolumn{1}{c}{DK} & \multicolumn{1}{c}{NO} & \multicolumn{1}{c}{SE} & \multicolumn{1}{c}{DK} & \multicolumn{1}{c}{NO} & \multicolumn{1}{c}{SE} & \multicolumn{1}{c}{DK} & \multicolumn{1}{c}{NO} & \multicolumn{1}{c}{SE}\\
       & (2)  & (3)    & (4)   & (5)   & (6)    & (7)   & (8)   & (9)   & (10)  \\ 
\midrule
1951 &  & 53,768 & 101,153 &  & 0.02 & 0.01 &  & 0.08 & 0.01\\
1952 &  & 57,474 & 101,048 &  & 0.03 & 0.01 &  & 0.08 & 0.01\\
1953 &  & 59,492 & 100,978 &  & 0.03 & 0.01 &  & 0.08 & 0.01\\
1954 &  & 59,334 & 96,577 &  & 0.03 & 0.01 &  & 0.08 & 0.01\\
1955 &  & 60,253 & 98,415 &  & 0.03 & 0.01 &  & 0.08 & 0.01\\
\addlinespace
1956 &  & 61,008 & 98,904 &  & 0.03 & 0.02 &  & 0.08 & 0.02\\
1957 &  & 60,136 & 97,445 &  & 0.03 & 0.02 &  & 0.08 & 0.02\\
1958 &  & 60,152 & 96,016 &  & 0.03 & 0.02 &  & 0.07 & 0.06\\
1959 &  & 60,297 & 95,185 &  & 0.03 & 0.02 &  & 0.07 & 0.02\\
1960 &  & 59,326 & 92,591 &  & 0.03 & 0.02 &  & 0.06 & 0.02\\
\addlinespace
1961 &  & 59,948 & 94,567 &  & 0.03 & 0.02 &  & 0.06 & 0.02\\
1962 & 69,948 & 59,843 & 96,813 & 0.04 & 0.03 & 0.02 & 0.06 & 0.06 & 0.02\\
1963 & 73,901 & 61,073 & 101,760 & 0.04 & 0.03 & 0.02 & 0.06 & 0.06 & 0.02\\
1964 & 74,955 & 63,298 & 110,549 & 0.04 & 0.03 & 0.02 & 0.06 & 0.05 & 0.02\\
1965 & 76,700 & 63,988 & 110,364 & 0.04 & 0.03 & 0.02 & 0.06 & 0.04 & 0.02\\
\addlinespace
1966 & 79,340 & 64,874 & 110,074 & 0.05 & 0.02 & 0.02 & 0.06 & 0.04 & 0.02\\
1967 & 73,204 & 64,721 & 108,302 & 0.04 & 0.02 & 0.02 & 0.06 & 0.04 & 0.02\\
1968 & 67,242 & 65,963 & 100,853 & 0.04 & 0.02 & 0.02 & 0.06 & 0.04 & 0.02\\
1969 & 64,121 & 66,462 & 95,375 & 0.04 & 0.02 & 0.02 & 0.05 & 0.03 & 0.02\\
1970 & 63,892 & 63,636 & 96,388 & 0.04 & 0.02 & 0.02 & 0.05 & 0.03 & 0.02\\
\addlinespace
1971 & 67,457 & 64,517 & 99,960 & 0.04 & 0.02 & 0.02 & 0.05 & 0.03 & 0.02\\
1972 & 67,514 & 63,452 & 98,766 & 0.04 & 0.02 & 0.02 & 0.06 & 0.02 & 0.02\\
1973 & 64,060 & 60,694 & 97,018 & 0.05 & 0.02 & 0.03 & 0.07 & 0.03 & 0.02\\
1974 & 63,527 & 59,327 & 97,644 & 0.06 & 0.02 & 0.03 & 0.07 & 0.03 & 0.02\\
1975 & 64,604 & 56,014 & 91,500 & 0.06 & 0.02 & 0.03 & 0.07 & 0.03 & 0.02\\
\addlinespace
1976 & 58,824 & 53,354 & 86,260 & 0.06 & 0.03 & 0.03 & 0.07 & 0.03 & 0.02\\
1977 & 55,902 & 51,076 & 84,042 & 0.07 & 0.01 & 0.03 & 0.07 & 0.01 & 0.02\\
1978 & 55,893 & 51,893 & 81,184 & 0.06 & 0.01 & 0.03 & 0.07 & 0.01 & 0.02\\
1979 & 53,578 & 51,791 & 83,740 & 0.06 & 0.01 & 0.03 & 0.07 & 0.01 & 0.02\\
\bottomrule
\end{tabular}
\vspace{0.2cm}
\linespread{1}
\scriptsize{\textit{Notes:} The table provides summary statistics for cohort size and the share of individuals with zero incomes in the child generation. Columns (2)-(4) present the cohort size, columns (5)-(7) indicate the share of sons with zero incomes, and columns (8)-(10) indicate the share of daughters with zero incomes. All statistics are displayed separately by county and birth year of the child.}
\end{threeparttable}
\end{table}

\begin{table}[H]
\centering
\caption{Summary Statistics Parents 1951 - 1979.}
\label{tab:summary_stats_parents}
\setlength{\tabcolsep}{2pt}
\begin{threeparttable}
\footnotesize
\renewcommand{\arraystretch}{0.5} 
\begin{tabular}{l*{15}{c}}
\toprule
 & \multicolumn{3}{c}{\textbf{Number of Children}} & \multicolumn{3}{c}{\textbf{Father ID Missing}} & \multicolumn{3}{c}{\textbf{Father Share Zero}} &\multicolumn{3}{c}{\textbf{Mother ID Missing}} & \multicolumn{3}{c}{\textbf{Mother Share Zero}}\\
\cmidrule(lr){2-4} 
\cmidrule(lr){5-7} 
\cmidrule(lr){8-10}
\cmidrule(lr){11-13}
\cmidrule(lr){11-13}
\cmidrule(lr){14-16}

 & \multicolumn{1}{c}{DK} & \multicolumn{1}{c}{NO} & \multicolumn{1}{c}{SE} & \multicolumn{1}{c}{DK} & \multicolumn{1}{c}{NO} & \multicolumn{1}{c}{SE} & \multicolumn{1}{c}{DK} & \multicolumn{1}{c}{NO} & \multicolumn{1}{c}{SE} & \multicolumn{1}{c}{DK} & \multicolumn{1}{c}{NO} & \multicolumn{1}{c}{SE} & \multicolumn{1}{c}{DK} & \multicolumn{1}{c}{NO} & \multicolumn{1}{c}{SE}\\
       & (2)  & (3)    & (4)   & (5)   & (6)    & (7)   & (8)   & (9)   & (10)  & (11)  & (12)  & (13)  & (14)  & (15)  & (16) \\  
\midrule
\midrule
1951 &  & 53,768 & 101,153 &  & 3,086 & 2,332 &  & 0.03 & 0.10 &  & 1,109 & 487 &  & 0.51 & 0.29\\
1952 &  & 57,474 & 101,048 &  & 2,642 & 1,826 &  & 0.03 & 0.09 &  & 851 & 385 &  & 0.48 & 0.24\\
1953 &  & 59,492 & 100,978 &  & 1,652 & 1,633 &  & 0.03 & 0.08 &  & 488 & 373 &  & 0.44 & 0.21\\
1954 &  & 59,334 & 96,577 &  & 1,449 & 1,480 &  & 0.03 & 0.08 &  & 380 & 254 &  & 0.36 & 0.19\\
1955 &  & 60,253 & 98,415 &  & 1,218 & 1,320 &  & 0.04 & 0.08 &  & 370 & 202 &  & 0.32 & 0.17\\
\addlinespace
\addlinespace
1956 &  & 61,008 & 98,904 &  & 1,154 & 1,259 &  & 0.04 & 0.08 &  & 300 & 174 &  & 0.28 & 0.15\\
1957 &  & 60,136 & 97,445 &  & 1,074 & 1,210 &  & 0.04 & 0.07 &  & 256 & 180 &  & 0.24 & 0.12\\
1958 &  & 60,152 & 96,016 &  & 925 & 1,197 &  & 0.03 & 0.07 &  & 236 & 134 &  & 0.21 & 0.10\\
1959 &  & 60,297 & 95,185 &  & 833 & 1,080 &  & 0.03 & 0.07 &  & 185 & 108 &  & 0.19 & 0.08\\
1960 &  & 59,326 & 92,591 &  & 749 & 1,106 &  & 0.03 & 0.07 &  & 207 & 91 &  & 0.17 & 0.07\\
\addlinespace
1961 &  & 59,948 & 94,567 &  & 754 & 859 &  & 0.03 & 0.07 &  & 180 & 2 &  & 0.16 & 0.06\\
1962 & 69,948 & 59,843 & 96,813 & 1,315 & 826 & 900 & 0.05 & 0.03 & 0.07 & 212 & 175 & 8 & 0.16 & 0.15 & 0.06\\
1963 & 73,901 & 61,073 & 101,760 & 1,255 & 780 & 926 & 0.05 & 0.03 & 0.07 & 177 & 155 & 6 & 0.15 & 0.14 & 0.05\\
1964 & 74,955 & 63,298 & 110,549 & 1,209 & 778 & 972 & 0.05 & 0.03 & 0.07 & 129 & 126 & 4 & 0.14 & 0.13 & 0.05\\
1965 & 76,700 & 63,988 & 110,364 & 1,082 & 857 & 947 & 0.05 & 0.03 & 0.07 & 117 & 99 & 6 & 0.13 & 0.13 & 0.04\\
\addlinespace
1966 & 79,340 & 64,874 & 110,074 & 937 & 852 & 890 & 0.04 & 0.03 & 0.06 & 86 & 82 & 4 & 0.12 & 0.11 & 0.04\\
1967 & 73,204 & 64,721 & 108,302 & 779 & 839 & 861 & 0.04 & 0.03 & 0.06 & 59 & 88 & 7 & 0.11 & 0.10 & 0.03\\
1968 & 67,242 & 65,963 & 100,853 & 341 & 828 & 697 & 0.04 & 0.03 & 0.06 & 12 & 96 & 8 & 0.10 & 0.10 & 0.03\\
1969 & 64,121 & 66,462 & 95,375 & 305 & 813 & 631 & 0.04 & 0.03 & 0.06 & $<$10 & 72 & 6 & 0.10 & 0.09 & 0.03\\
1970 & 63,892 & 63,636 & 96,388 & 306 & 768 & 647 & 0.05 & 0.03 & 0.05 & 11 & 91 & 7 & 0.10 & 0.09 & 0.03\\
\addlinespace
1971 & 67,457 & 64,517 & 99,960 & 288 & 720 & 566 & 0.05 & 0.03 & 0.05 & $<$10 & 90 & 2 & 0.10 & 0.09 & 0.02\\
1972 & 67,514 & 63,452 & 98,766 & 305 & 760 & 567 & 0.05 & 0.04 & 0.05 & $<$10 & 75 & 4 & 0.10 & 0.09 & 0.02\\
1973 & 64,060 & 60,694 & 97,018 & 314 & 793 & 568 & 0.05 & 0.04 & 0.05 & $<$10 & 66 & 5 & 0.10 & 0.09 & 0.02\\
1974 & 63,527 & 59,327 & 97,644 & 294 & 736 & 483 & 0.05 & 0.04 & 0.05 & $<$10 & 88 & 4 & 0.09 & 0.09 & 0.03\\
1975 & 64,604 & 56,014 & 91,500 & 326 & 760 & 464 & 0.06 & 0.04 & 0.05 & $<$10 & 82 & 2 & 0.09 & 0.09 & 0.03\\
\addlinespace
1976 & 58,824 & 53,354 & 86,260 & 310 & 778 & 436 & 0.06 & 0.04 & 0.05 & $<$10 & 64 & 0 & 0.10 & 0.09 & 0.03\\
1977 & 55,902 & 51,076 & 84,042 & 288 & 740 & 400 & 0.06 & 0.04 & 0.05 & $<$10 & 78 & 1 & 0.09 & 0.09 & 0.03\\
1978 & 55,893 & 51,893 & 81,184 & 280 & 767 & 379 & 0.06 & 0.04 & 0.06 & $<$10 & 72 & 0 & 0.09 & 0.08 & 0.03\\
1979 & 53,578 & 51,791 & 83,740 & 290 & 802 & 399 & 0.06 & 0.04 & 0.06 & $<$10 & 96 & 3 & 0.09 & 0.08 & 0.03\\
\bottomrule
\end{tabular}
\vspace{0.2cm}
\linespread{1}
\scriptsize{\textit{Notes:} The table provides summary statistics for cohort size and the share of individuals with zero incomes in the parent generation. Columns (2)-(4) present the cohort size, columns (5)-(7) and (11)-(13) shows the number of observations with missing father ID or mother ID, columns (8)-(10) and (14)-(16) indicates the share of fathers or mothers with zero income. All statistics are displayed separately by country and birth year of the child.}
\end{threeparttable}
\end{table}

\begin{landscape}
\begin{table}[H]
\centering
\caption{IRA Coefficients, Trends and Differences Across Countries and Time.}
\label{tab:ira_coefs}
\begin{threeparttable}
\renewcommand{\arraystretch}{0.5} 
\begin{tabular}{l*{15}{c}}
\toprule
\multicolumn{1}{c}{ } & \multicolumn{2}{c}{1951} & \multicolumn{3}{c}{1962} & \multicolumn{3}{c}{1979} & \multicolumn{3}{c}{Trend 1962-1979} & \multicolumn{3}{c}{$\Delta$ \textit{p}-value} \\
\cmidrule(l{3pt}r{3pt}){2-3} \cmidrule(l{3pt}r{3pt}){4-6} \cmidrule(l{3pt}r{3pt}){7-9} \cmidrule(l{3pt}r{3pt}){10-12} \cmidrule(l{3pt}r{3pt}){13-15}
IRA Spec. & NO & SE & DK & NO & SE & DK & NO & SE & DK & NO & SE & DK-NO & DK-SE & NO-SE\\
\midrule
All & 0.156 & 0.167 & 0.190 & 0.170 & 0.180 & 0.265 & 0.234 & 0.225 & 0.530 & 0.379 & 0.277 & 0.065 & 0.004 & 0.176\\
 &  &  &  &  &  &  &  &  & (0.035) & (0.018) & (0.033) &  &  & \\
Son-Parent & 0.242 & 0.245 & 0.225 & 0.222 & 0.233 & 0.280 & 0.241 & 0.235 & 0.360 & 0.085 & 0.014 & 0.000 & 0.000 & 0.736\\
 &  &  &  &  &  &  &  &  & (0.036) & (0.024) & (0.061) &  &  & \\
Daughter-Parent & 0.146 & 0.158 & 0.197 & 0.173 & 0.169 & 0.276 & 0.262 & 0.240 & 0.592 & 0.552 & 0.428 & 0.363 & 0.067 & 0.020\\
 &  &  &  &  &  &  &  &  & (0.035) & (0.037) & (0.038) &  &  & \\
Son-Father & 0.253 & 0.248 & 0.220 & 0.236 & 0.242 & 0.241 & 0.213 & 0.211 & 0.139 & -0.160 & -0.224 & 0.000 & 0.000 & 0.378\\
 &  &  &  &  &  &  &  &  & (0.035) & (0.022) & (0.060) &  &  & \\
Son-Mother & 0.068 & 0.080 & 0.098 & 0.089 & 0.101 & 0.198 & 0.150 & 0.155 & 0.619 & 0.324 & 0.307 & 0.000 & 0.007 & 0.723\\
 &  &  &  &  &  &  &  &  & (0.026) & (0.026) & (0.041) &  &  & \\
Daughter-Father & 0.137 & 0.139 & 0.175 & 0.144 & 0.152 & 0.216 & 0.195 & 0.192 & 0.342 & 0.325 & 0.239 & 0.778 & 0.570 & 0.421\\
 &  &  &  &  &  &  &  &  & (0.031) & (0.032) & (0.034) &  &  & \\
Daughter-Mother & 0.073 & 0.077 & 0.120 & 0.119 & 0.112 & 0.227 & 0.221 & 0.194 & 0.731 & 0.607 & 0.541 & 0.439 & 0.181 & 0.107\\
 &  &  &  &  &  &  &  &  & (0.036) & (0.037) & (0.034) &  &  & \\
\bottomrule
\end{tabular}
\vspace{0.2cm}
\linespread{1}
\scriptsize{\textit{Notes:} Columns (1)-(7) report the IRA coefficients of separated regressions in the years 1951, 1962 and 1979 separately for Denmark, Norway and Sweden. Columns (8)-(10) report the coefficient of the fitted regression lines of country specific regressions of the IRA coefficient on a linear trend for the years 1962 to 1979. The trend coefficients and corresponding standard errors have been multiplied by 100 in order to avoid too many digits after the separator. Columns (11)-(13) report rounded p-values for the null hypothesis that the slopes for the respective countries (see column header) are equal. Robust standard errors are reported in parentheses.}
\end{threeparttable}
\end{table}
\end{landscape}

\begin{table}[H]
\centering
\caption{IRA Coefficients and Trends - Details (United States).}
\label{tab:psidirafull}
\begin{threeparttable}
\renewcommand{\arraystretch}{0.5} 
\begin{tabular}{l *{5}{p{2.3cm}}}
\toprule
 & Parents   & \multicolumn{2}{c}{Father}   & \multicolumn{2}{c}{Mother}  \\ \cmidrule(l{2pt}r{2pt}){2-2} \cmidrule(l{2pt}r{2pt}){3-4} \cmidrule(l{2pt}r{2pt}){5-6}
              &   \multicolumn{1}{l}{Child}    & \multicolumn{1}{l}{Son} & \multicolumn{1}{l}{Daughter} & \multicolumn{1}{l}{Son} & \multicolumn{1}{l}{Daughter}  \\
\midrule
\multicolumn{1}{l}{\textbf{A: Weights} }\\
\cmidrule{1-1}
Pooled IRA   & 0.317***  & 0.336***  & 0.195***    & 0.097***    & 0.137***  \\
 & (0.017)  & (.022)  & (0.031)   & (0.025)  & (0.029) \\
Trend  $\times$ 100 & 0.603*** & -0.240  & 0.980***  & 0.136     & 1.047***  \\
& (0.149)    & (0.205)   & (0.277)   & (0.253)  & (0.292)         \\
N      & 5,392  & 2,272      & 1,637      & 2,477  & 2,205 \\                       
\midrule
\multicolumn{1}{l}{\textbf{B: No weights} }\\
\cmidrule{1-1}
Pooled IRA   & 0.335***                    & 0.360***                & 0.237***                     & 0.107***                & .152***                      \\
             & (0.013)                     & (0.020)                 & (0.025)                      & (0.021)                 & (0.022)                       \\
Trend  $\times$ 100 & 0.449***                  & -0.263*               & 0.728***                   & 0.268                 & 0.917***                    \\
             & (0.118)                   & (0.178)               & (0.229)                    & (0.202)               & (0.213)                     \\
N      & 5,392  & 2,272      & 1,637      & 2,477  & 2,205 \\                         
\midrule
\multicolumn{1}{l}{\textbf{C: SRC sample} }\\
\cmidrule{1-1}
Pooled IRA   & 0.294***                    & 0.327***                & 0.192***                     & 0.098***                & 0.126***                      \\
              & (0.018)                     & (0.023)                 & (0.0353)                      & (0.026)                 & (0.032)                       \\
Trend  $\times$ 100 & 0.433**                   & -0.393                & 1.156***                   & 0.180                 & 0.727                       \\
              & (0.162)                   & (0.218)               & (0.305)                    & (0.266)               & (0.327)                     \\
N            & 2,927                       & 1,583                   & 904                          & 1,497                   & 1,001                           \\
\bottomrule
\end{tabular}
\vspace{0.2cm}
\linespread{1}
\scriptsize{\textit{Notes:}  The table presents estimates of the IRA and linear trends in the IRA separately for different child-parent combinations. Due to the small sample sizes, trends have been estimated directly on the underlying micro data by regressing cohort-specific child ranks on cohort-specific parent ranks interacted with a linear time trend. The trend coefficients and standard errors have been multiplied by 100 in order to avoid too many digits after the separator. Panel A contains estimates for the full PSID sample using provided sample weights, Panel B uses the full sample without weights and Panel C includes estimates on the nationally representative SRC sample. Standard errors are in parentheses. P-values indicated by * $<$ 0.1, ** $<$ 0.05, *** $<$ 0.01.}
\end{threeparttable}
\end{table}

\begin{table}[H]
\centering
\begin{table}[H]
\centering
\caption{PSID Sample Size with Parent-Child Links by Birth Cohort (United States).}
\label{tab:psidira_count}
\begin{threeparttable}
\renewcommand{\arraystretch}{0.5} 
\begin{tabular}{l *{6}{c}}
\toprule
    & & \multicolumn{1}{c}{\textbf{Parents}} & \multicolumn{2}{c}{\textbf{Father}} & \multicolumn{2}{c}{\textbf{Mother}}  \\
 \midrule
 \multicolumn{1}{c}{Child birth year} & & \multicolumn{1}{c}{Child}   & \multicolumn{1}{c}{Son} & \multicolumn{1}{c}{Daughter} & \multicolumn{1}{c}{Son} & \multicolumn{1}{c}{Daughter}  \\
\midrule
1947 &  & 76  & 27  & 26  & 35  & 39   \\
1948 &  & 107 & 38  & 39  & 48  & 56   \\
1949 &  & 143 & 52  & 44  & 73  & 65   \\
1950 &  & 171 & 57  & 72  & 68  & 99   \\
1951 &  & 218 & 76  & 81  & 101 & 108  \\
1952 &  & 193 & 70  & 78  & 88  & 102  \\
1953 &  & 239 & 87  & 98  & 116 & 117  \\
1954 &  & 235 & 78  & 96  & 101 & 128  \\
1955 &  & 267 & 103 & 98  & 122 & 136  \\
1956 &  & 263 & 86  & 106 & 107 & 147  \\
1957 &  & 247 & 95  & 85  & 126 & 115  \\
1958 &  & 220 & 75  & 95  & 95  & 119  \\
1959 &  & 159 & 79  & 37  & 107 & 46   \\
1960 &  & 177 & 96  & 34  & 121 & 54   \\
1961 &  & 105 & 54  & 27  & 62  & 38   \\
1962 &  & 117 & 50  & 37  & 63  & 53   \\
1963 &  & 125 & 49  & 41  & 64  & 56   \\
1964 &  & 100 & 47  & 30  & 56  & 43   \\
1965 &  & 91  & 42  & 20  & 49  & 40   \\
1966 &  & 88  & 39  & 17  & 52  & 34   \\
1967 &  & 95  & 49  & 21  & 60  & 33   \\
1968 &  & 66  & 32  & 17  & 39  & 26   \\
1969 &  & 99  & 49  & 32  & 55  & 42   \\
1970 &  & 87  & 40  & 20  & 50  & 35   \\
1971 &  & 92  & 40  & 21  & 60  & 32   \\
1972 &  & 111 & 49  & 22  & 63  & 43   \\
1973 &  & 107 & 50  & 20  & 65  & 36   \\
1974 &  & 117 & 51  & 25  & 64  & 48   \\
1975 &  & 128 & 55  & 36  & 70  & 48   \\
1976 &  & 132 & 58  & 37  & 63  & 53   \\
1977 &  & 130 & 66  & 23  & 41  & 21   \\
1978 &  & 138 & 66  & 34  & 27  & 30   \\
1979 &  & 179 & 93  & 43  & 29  & 41   \\
1980 &  & 142 & 65  & 36  & 33  & 26   \\
1981 &  & 148 & 77  & 30  & 40  & 24   \\
1982 &  & 120 & 58  & 27  & 26  & 30   \\
1983 &  & 160 & 74  & 32  & 38  & 42  \\ 
\midrule
Total  & \multicolumn{1}{c}{} & 5,392   & 2,272 & 1,637& 2,477 & 2,205\\
\bottomrule
\end{tabular}
\vspace{0.2cm}
\linespread{1}
\scriptsize{\textit{Notes:} The table presents the number of cohort-specific parent-child links that were used to produce the main results from the PSID survey data.}
\end{threeparttable}
\end{table}
\end{table}

\begin{table}[H]
\centering
\begin{table}[H]
\centering
\caption{IRA Coefficients and Trends - Age 30 (United States).}
\label{tab:psidira30}
\begin{threeparttable}
\renewcommand{\arraystretch}{0.5} 
\begin{tabular}{l *{5}{p{2.3cm}}}
\toprule
 & Parents   & \multicolumn{2}{c}{Father}   & \multicolumn{2}{c}{Mother}  \\ \cmidrule(l{2pt}r{2pt}){2-2} \cmidrule(l{2pt}r{2pt}){3-4} \cmidrule(l{2pt}r{2pt}){5-6}
              &   \multicolumn{1}{l}{Child}    & \multicolumn{1}{l}{Son} & \multicolumn{1}{l}{Daughter} & \multicolumn{1}{l}{Son} & \multicolumn{1}{l}{Daughter}  \\
\midrule
\multicolumn{1}{l}{\textbf{A: Weights} }\\ \cmidrule{1-1}
Pooled IRA   & 0.327***                    & 0.318***                & 0.222***                     & 0.120***                & 0.151***                      \\
              & (0.015)                     & (0.022)                 & (0.026)                      & (0.025)                 & (0.027)                       \\
Trend  $\times$ 100 & 0.643***                  & 0.133                 & 0.661***                   & 0.610**              & 0.571**                     \\
            & (0.129)                   & (0.193)               & (0.220)                    & (0.239)               & (0.262)                     \\
N            & 6,652                      & 2,664                   & 2,109                        & 2,685                   & 2,611                         \\
\midrule
\multicolumn{1}{l}{\textbf{B: No weights} }\\ \cmidrule{1-1}
Pooled IRA   & 0.345***                    & 0.341***                & 0.263***                     & 0.148***                & 0.168***                      \\
     & (0.012)                     & (0.018)                 & (0.021)                      & (0.019)                 & (0.020)                       \\
Trend  $\times$ 100 & 0.457***                  & 0.102                 & 0.510***                   & 0.429**              & 0.567***                    \\
              & (0.101)                   & (0.59)               & (0.183)                    & (0.176)               & (0.181)                     \\
N            & 6,652                      & 2,663                   & 2,109                        & 2,686                   & 2,611                         \\
\midrule
\multicolumn{1}{l}{\textbf{C: SRC sample} }\\ \cmidrule{1-1}
Pooled IRA   & 0.303***                    & 0.310***                & 0.225***                     & 0.097***                & 0.133***                      \\
          & (0.016)                     & (0.023)                 & (0.028)                      & (0.027)                 & (0.030)                       \\
Trend  $\times$ 100& 0.528***                  & 0.020                 & 0.586**                    & 0.661**              & 0.352                       \\
  & (0.146)                   & (0.210)               & (0.245)                    & (0.261)               & (0.307)                     \\
N            & 3,451                      & 1,757                  & 1,161                        & 1,460                   & 1,142                         \\

\bottomrule
\end{tabular}
\vspace{0.2cm}
\linespread{1}
\scriptsize{\textit{Notes:} The table presents estimates of the IRA and linear trends in the IRA separately for different child-parent combinations Children's income is measure at age 30. Due to the small sample sizes, trends have been estimated directly on the underlying micro data by regressing cohort-specific child ranks on cohort-specific parent ranks interacted with a linear time trend. The trend coefficients and standard errors have been multiplied by 100 in order to avoid too many digits after the separator. Panel A contains estimates for the full PSID sample using provided sample weights, Panel B uses the full sample without weights and Panel C includes estimates on the nationally representative SRC sample. Standard errors are in parentheses. P-values indicated by * $<$ 0.1, ** $<$ 0.05, *** $<$ 0.01.}
\end{threeparttable}
\end{table}
\end{table}

\begin{table}[H]
\centering
\begin{table}[H]
\centering
\caption{Comparison of Trends 1962 - 1979.}
\label{tab:trendcomppart}
\begin{threeparttable}
\renewcommand{\arraystretch}{0.5} 
\begin{tabular}{ll *{3}{c}}
\toprule 
\toprule
\addlinespace
 &  & Denmark & Norway & Sweden\\
\midrule
\addlinespace[0.3em]
\multicolumn{2}{l}{\textbf{Panel A: Son - Father}}\\
\cmidrule{1-2}
\hspace{1em}  & IRA & 0.1385 & -0.1598 & -0.2243\\
\hspace{1em} &  & (0.0349) & (0.0222) & (0.0605)\\
\hspace{1em} & LW & 0.1504 & -0.2062 & -0.1898\\
\hspace{1em} &  & (0.0239) & (0.0350) & (0.0613)\\
\hspace{1em} & Difference & -0.0118 & 0.0464 & -0.0346\\
\hspace{1em} &  & (0.0423) & (0.0414) & (0.0861)\\
\addlinespace[0.3em]
\midrule
\multicolumn{2}{l}{\textbf{Panel B: Son - Mother }}\\
\cmidrule{1-2}
\hspace{1em} & IRA & 0.6186 & 0.3244 & 0.3069\\
\hspace{1em} &  & (0.0256) & (0.0262) & (0.0408)\\
\hspace{1em} & LW & 0.2994 & -0.1200 & 0.0175\\
\hspace{1em} &  & (0.0353) & (0.0273) & (0.0495)\\
\hspace{1em} & Difference & 0.3192 & 0.4444 & 0.2894\\
\hspace{1em} &  & (0.0436) & (0.0379) & (0.0642)\\
\midrule
\addlinespace[0.3em]
\multicolumn{2}{l}{\textbf{Panel C: Daughter - Father }}\\
\cmidrule{1-2}
\hspace{1em}  & IRA & 0.3416 & 0.3247 & 0.2388\\
\hspace{1em} &  & (0.0309) & (0.0318) & (0.0339)\\
\hspace{1em} & LW & 0.0658 & -0.0385 & -0.0897\\
\hspace{1em} &  & (0.0324) & (0.0314) & (0.0201)\\
\hspace{1em} & Difference & 0.2758 & 0.3632 & 0.3285\\
\hspace{1em} &  & (0.0447) & (0.0447) & (0.0394)\\
\bottomrule
\end{tabular}
\vspace{0.2cm}
\linespread{1}
\scriptsize{\textit{Notes:} IRA indicates linear trends estimated through all coefficients of the intergenerational rank association. LW specifies linear trends estimated through all coefficients obtained from applying the Lubotsky-Wittenberg method. The trend coefficients and corresponding standard errors have been multiplied by 100 in order to avoid too many digits after the separator. Difference indicates differences between LW and IRA trends and tests the null-hypothesis of equality in trends between the IRA and LW coefficients. Heteroskedasticity robust standard errors are in parentheses.}
\end{threeparttable}
\end{table}
\end{table}

\begin{table}[H]
\centering
\caption{Change in Occupational Structure Among Mothers and Daughters.}
\label{tab:occ_change_list}
\begin{threeparttable}
\renewcommand{\arraystretch}{0.5} 
\begin{tabular}{llclc}
\toprule
 Period     & Occupational group & Change (\%) & Occupational group & Change (\%)     \\ 
\midrule
            &   \multicolumn{2}{c}{Mothers} &   \multicolumn{2}{c}{Daughters}   \\ \cmidrule(lr){2-3} \cmidrule(lr){4-5}
\textbf{Norway}  &    &  &    &         \\ \cmidrule{1-1}
1951-1965   &  Secretary/clerical   & 4.4 (26\%)  & Professionals & 3.1 (23.5\%)  \\ 
1951-1965   &  Care                  & 4.2 (25\%) &  Executive & 2.5 (19.3\%)\\
1951-1965   &  Retail                 & 2.5 (14\%) & Other Administration & 2.5 (19.2\%) \\ 
\addlinespace
1962-1979   &  Care                  &  5.8  (28\%) & Professionals & 5.4 (34.3\%)\\
1962-1979   &  Teachers             &  4.2  (20\%) & Teachers &  5.0 (32\%) \\
1962-1979   &  Professionals           &  2.5 (12\%) & Medicine &  2.1 (13.4\%) \\ 
\addlinespace

\textbf{Sweden}  &    &   &   &        \\ \cmidrule{1-1}
1951-1965   &  Secretary/clerical   & 4.4 (19\%)   & Care  &    5.3 (35\%)  \\
1951-1965   &  Care               & 3.4 (15\%)   &  Lower professionals &  5.0 (33\%)  \\
1951-1965   &  Teachers           & 3.1 (13\%)  & Executives  &  2.9 (19\%)  \\ 
\addlinespace
1962-1979   &  Professionals      &   2.9 (22\%) & Professionals  &  5.1 (48\%)   \\
1962-1979   &  Teachers           &   2.8 (21\%)  & Teachers  &    1.6 (16\%)\\
1962-1979   &  Nurses             &  2.0 (21\%)  & Lower professionals  &  1.6 (15\%) \\ \addlinespace
\bottomrule
\end{tabular}
\vspace{0.2cm}
\linespread{1}
\scriptsize{\textit{Notes:} The occupational groups are described in Table \ref{tab:occ_groups_new}. The birth cohorts again refer to the child's year of birth. ``Change'' refers to the net percentage point growth of the occupational group between 1951-55 and 1962-1965, and 1962-1965 to 1975-1979, respectively. The number in parentheses shows the percentage growth in the occupational group of the gross total increase in all occupational groups among women (i.e., what percentage of the total growth in occupational groups is accounted for by \textit{professionals}).}
\end{threeparttable}
\end{table}

\newpage
\section{A Model of Time-Varying Intergenerational Transmission of Income-Generating Potential}
\label{APP:MODEL}
\setcounter{figure}{0} 
\setcounter{table}{0}  

In our framework, individual income at time $t$, $y^k_{it}$, are determined by two factors; inheritable skills, $x^k_{it}$, and a non-inheritable determinant $\varepsilon^k_{it}$. This generalizes to all fathers, mothers, sons, and daughters, i.e. all $k\in\{F,M,S,D\}$. 

Interpreting these factors in the context of a highly simplified version of the frameworks formulated by \citet{BeckerTomes} and \citet{solon_2004}, we can think of $x^k_{it}$ as representing an aggregate measure of income determinants that can be transmitted across generations such as skills, values, and connections, while $\varepsilon^k_{it}$ represents the value of all other income determinants that are uncorrelated to skills that can be transmitted across generations (it may be instructive --- yet slightly naïve --- to think of this as luck). While $y^k_{it}$ is observable, the split between $x^k_{it}$ and $\varepsilon^k_{it}$ is fundamentally unobservable in the data. A similar approach to dividing income into inheritable and non-inheritable factors is also considered by \citet{collado2022estimating} in a more extensive framework.

We assume that inheritable skills in the parental generation follow a bivariate Gaussian distribution. In particular, we assume that

\begin{equation}
    x_{it}^{F}\sim\mathcal{N}\left(0,1\right), \quad x_{it}^{M}=\left(\psi_{t}x_{it}^{F}+\left(1-\psi_{t}\right)u_{it}^{0}\right)/\Gamma_{t}^{0}
\end{equation}

where $u_{it}^{0}\sim\mathcal{N}\left(0,1\right)$ and superscript '0' effectively denotes that the term refers to the parental generation. Standardizing the variance of maternal skills to one using $\Gamma^0_t$,\footnote{This is a trivial scaling coefficient that ensures that the distribution of maternal skills is standard normal.} $\psi_{t}$ reflects cohort-specific correlations in parental skills, thus measuring assortative mating in the model.

We assume that skills are transmitted passively from the parental generation to the child generation as follows:
\begin{equation}
    x_{it}^{k}=\begin{cases}
\left(\kappa_{t}\left[\alpha_{t}x_{it}^{F}+\left(1-\alpha_{t}\right)x_{it}^{M}\right]+\left(1-\kappa_{t}\right)u^1_{it}\right)/\Gamma^1_{t}, & \text{ for }k=S\\
\left(\kappa_{t}\left[\alpha_{t}x_{it}^{M}+\left(1-\alpha_{t}\right)x_{it}^{F}\right]+\left(1-\kappa_{t}\right)u^1_{it}\right)/\Gamma^1_{t}, & \text{ for }k=D
\end{cases}
\end{equation}

Here, $\kappa_t$ is a measure of intergenerational correlation in inheritable skills --- or the rate at which skills are transmitted --- across generations within a given cohort of children, and $\alpha_t\in[0,1]$ is a coefficient that allows the transmission of inheritable skills within gender to be stronger than the transmission of inheritable skills across gender. Hence, whenever $\alpha_t>1/2$, then sons (daughters) inherit a relatively larger part of their inheritable skills from their fathers (mothers). The opposite holds when $\alpha_t<1/2$. Finally, $u_{it}^1$ is a stochastic element in the determination of skills which allows time-variation in the importance of intergenerational transmission of skills through variation in $\kappa_t$, and $\Gamma^1_t$ is once again a trivial scaling coefficient that ensures that the distribution of skills is standard normal --- with '1' now pointing to the child generation.

Individual income is given by a monotone transformation of a linear index, which is composed of inheritable and non-inheritable determinants:
\begin{equation}
    y_{it}^{k}=F_{t}^{k}\left(\tilde{\phi}_{t}^{k}x_{it}^{k}+\left(1-\tilde{\phi}_{t}^{k}\right)\varepsilon_{it}^{k}\right),\quad\text{for }k\in\{F,M,S,D\},
\end{equation}

where $\tilde{\phi}_{t}^{k}=\nicefrac{\phi_{t}^{k}}{\text{max}\left(\phi_{t}^{F},\phi_{t}^{M}\right)}$ for $k\in\{M,F\}$ in the parental generation and $\tilde{\phi}_{t}^{k}=\nicefrac{\phi_{t}^{k}}{\text{max}\left(\phi_{t}^{S},\phi_{t}^{D}\right)}$ for $k\in\{S,D\}$ in the child generation, respectively. Parameters $\phi_{t}^{F}$ and $\phi_{t}^{M}$ measure the importance of inheritable skills relative to the non-inheritable component in the linear income index for fathers and mothers, respectively. Similarly, $\phi_{t}^{S}$ and $\phi_{t}^{D}$ measure the importance of inheritable skills relative to the non-inheritable component in the linear income index for sons and daughters. Making the simple assumption that the distribution of non-inheritable determinants can be summarized by a standard normal distribution, $\varepsilon_{it}^{k}\sim\mathcal{N}\left(0,1\right)$, the individual income index is also normal.\footnote{Through simulations, it can be verified that composing the individual income index of two sets of Gaussian components, one inheritable and one non-inheritable, replicates the aggregate functional relationship between parental and child income ranks remarkably well.}

When measuring gender-specific intergenerational mobility in individual income ranks, the functional form of the monotone transformation function, $F_t^{k}(\cdot)$, is essentially unimportant; as long as it is monotone in the income index, any rank transformation of the income index will yield the same result as a rank transformation of income. However, in order to find both a joint measure of child income ranks across genders and a measure of joint parental income, the functional form can no longer be disregarded, since this would fail to take into account any gender differences in income distributions. Instead, we obtain the functional forms directly from the data. Exploiting the assumed monotone relationship between the income index and income, we match index ranks to the income distribution observed in the data. This allows us to compute pooled income ranks across genders in the child generation. We also compute a measure of joint parental income, $y_{it}^P$, that takes the true income distribution into account, as follows:
\begin{equation}
y_{it}^{P}=\hat{F}_{t}^{F}\left(\tilde{\phi}_{t}^{F}x_{it}^{F}+\left(1-\tilde{\phi}_{t}^{F}\right)\varepsilon_{it}^{F}\right)+\hat{F}_{t}^{M}\left(\tilde{\phi}_{t}^{M}x_{it}^{M}+\left(1-\tilde{\phi}_{t}^{M}\right)\varepsilon_{it}^{M}\right).
\end{equation}
Here, $\hat{F}_t^{F}(\cdot)$ and $\hat{F}_t^{M}(\cdot)$ are cohort-specific estimates of the functions that map the income index to the income distribution observed in the data.

For each country and cohort, we wish to calibrate a vector of seven decomposition parameters $[\begin{array}{ccccccc}
\psi_{t} & \kappa_{t} & \alpha_{t} & \phi_{t}^{F} & \phi_{t}^{M} & \phi_{t}^{S} & \phi_{t}^{D}\end{array}]^{'}$, from the following five equations:\footnote{The first equation is estimated in ventiles as the equation refers directly to relationship that is depicted in Figure \ref{fig:assort_mating} (higher degrees of rank granularity obscures the visual exposition). The full rank distribution is applied in the last four equations.}
\begin{subequations} \label{eq:subeqn}
    \begin{align}
        \text{Rank}_{it}^{\text{F}} & =\alpha_{1t}+\beta_{1t}\text{Rank}_{it}^{\text{M}}+\varepsilon_{1it} \subeqn{} \label{eq:subeqn1}\\
        \text{Rank}_{it}^{\text{F}} & =\alpha_{2t}+\beta_{2t}\text{Rank}_{it}^{\text{S}}+\varepsilon_{2it} \subeqn{} \label{eq:subeqn2}\\
        \text{Rank}_{it}^{\text{F}} & =\alpha_{3t}+\beta_{3t}\text{Rank}_{it}^{\text{D}}+\varepsilon_{3it} \subeqn{} \label{eq:subeqn3}\\
        \text{Rank}_{it}^{\text{M}} & =\alpha_{4t}+\beta_{4t}\text{Rank}_{it}^{\text{S}}+\varepsilon_{4it}\subeqn{} \label{eq:subeqn4}\\
        \text{Rank}_{it}^{\text{M}} & =\alpha_{5t}+\beta_{5t}\text{Rank}_{it}^{\text{D}}+\varepsilon_{5it} \subeqn{} \label{eq:subeqn5}
    \end{align}
\end{subequations} \vspace{-1.0cm}

In order to avoid underidentification, we make two adjustments. First, we set $\phi_{t}^{F}=\phi_{t}^{S}$. This means that the skill importance in income for mothers and daughters, $\phi_{t}^{M}$ and $\phi_{t}^{D}$, are interpreted relative to that of fathers and sons, respectively. In other words, we assume a generation-specific gender bias in the importance of skills for the determination of income. Secondly, we set $\phi_{t}^{F} = \phi_{t}^{S} = 1$, thereby effectively pinning down the level around which $\kappa_t$ trends over time.\footnote{As skills become better reflected in income, skills need to be transmitted across generations to a lesser extent in order to obtain a given correlation in income over time. Fixing the importance of skills for income among males therefore effectively pins down the skill transmission rate across time for a given intergenerational correlation in income.} The vector of decomposition parameters that are now left for us to calibrate across countries and cohorts is then given by $\left[\begin{array}{ccccccc}
\psi_{t} & \kappa_{t} & \alpha_{t} & 1 & \phi_{t}^{M} & 1 & \phi_{t}^{D}\end{array}\right]^{'}$.

\subsection{Calibrating the Parameters of the Model}  \label{APP:cali}

Calibrating the model allows us to understand how country-specific changes in intergenerational mobility can be decomposed into changes in the rate at which inheritable skills manifest themselves in labor income among mothers and daughters relative to fathers and sons, and changes in assortative mating on skills among parents. 

The calibration exercise can be summarized in four steps. First, we simulate data using the model that was outlined above and random values for the five model parameters. These data represent the initial period (1951 in Sweden and Norway, and 1962 in Denmark). Second, we estimate the slope coefficients in the five equations from the simulated data and compare these to those from our ``real'' data. Third, we utilize a form of gradient descent algorithm in order to improve the fit of the simulated data, by adjusting the model parameters. The model fit is evaluated by the difference between the slope coefficients in the ``real'' and the simulated data. Fourth, we re-simulate the data based on these adjusted model parameters. Steps two to four are repeated iteratively until a stopping criterion is met. This stopping criterion is a function of the quality of the model fit as well as the rate of convergence. When the final set of calibrated model parameters for a given cohort is found, we move on to the next cohort of children (say, 1952 in Sweden and Norway, and 1963 in Denmark) and utilize the final set of calibrated model parameters from the previous period as the starting point for the new period.

Specifically, the calibration process was carried out as follows. Each set of country-cohort model parameters for trend decomposition are --- loosely described --- calibrated in the following steps:
\begin{enumerate} \vspace{-0.15cm}
    \item If the cohort is the first cohort of observation for a given country (say, 1951 in Sweden and Norway and 1962 in Denmark), draw a random set of values for the parameters $\psi_{t},\kappa_{t},\alpha_{t},\phi_{t}^{M}$ and $\phi_{t}^{D}$. If the cohort is not the first cohort of observation, initialize the algorithm with the optimal set of parameters from the last cohort associated with the same country. These become the 'search parameters' until they are replaced by parameters that are likely superior in terms of fitting the true data as outlined in the next steps.
    
    \item Simulate 100,000 parent-child pairs\footnote{The same 100,000 'families' are drawn in order to ensure that 'stochastic' differences in simulations neither affect the convergence process nor the evolution of calibrated parameters over the cohorts} and repeat the following procedure until there is a sufficiently close match between empirical rank associations as outlined in equations \eqref{eq:subeqn1}-\eqref{eq:subeqn5} and the simulated counterparts:\footnote{Or stop the algorithm early if it stops converging, i.e. it seems that a much better match cannot be achieved.} \begin{enumerate}
        \item Compute skills and income for all individuals (father, mother, son and daughter) using the set of 'search parameters' along with randomly drawn values for $x^k_{it}$ and $\varepsilon^k_{it}$.
        \item Compute rank associations as listed in equations as outlined in equations equations \eqref{eq:subeqn1}-\eqref{eq:subeqn5} from the simulated data and match them to the associated rank associations as estimated in the real data.
        \item If the convergence criterion is not met, adjust the parameters using a customized variation of gradient descent. These now become the 'search parameters'.
    \end{enumerate}
\end{enumerate}

\begin{figure}
\linespread{1}
\begin{center}
\includegraphics[width=1\linewidth]{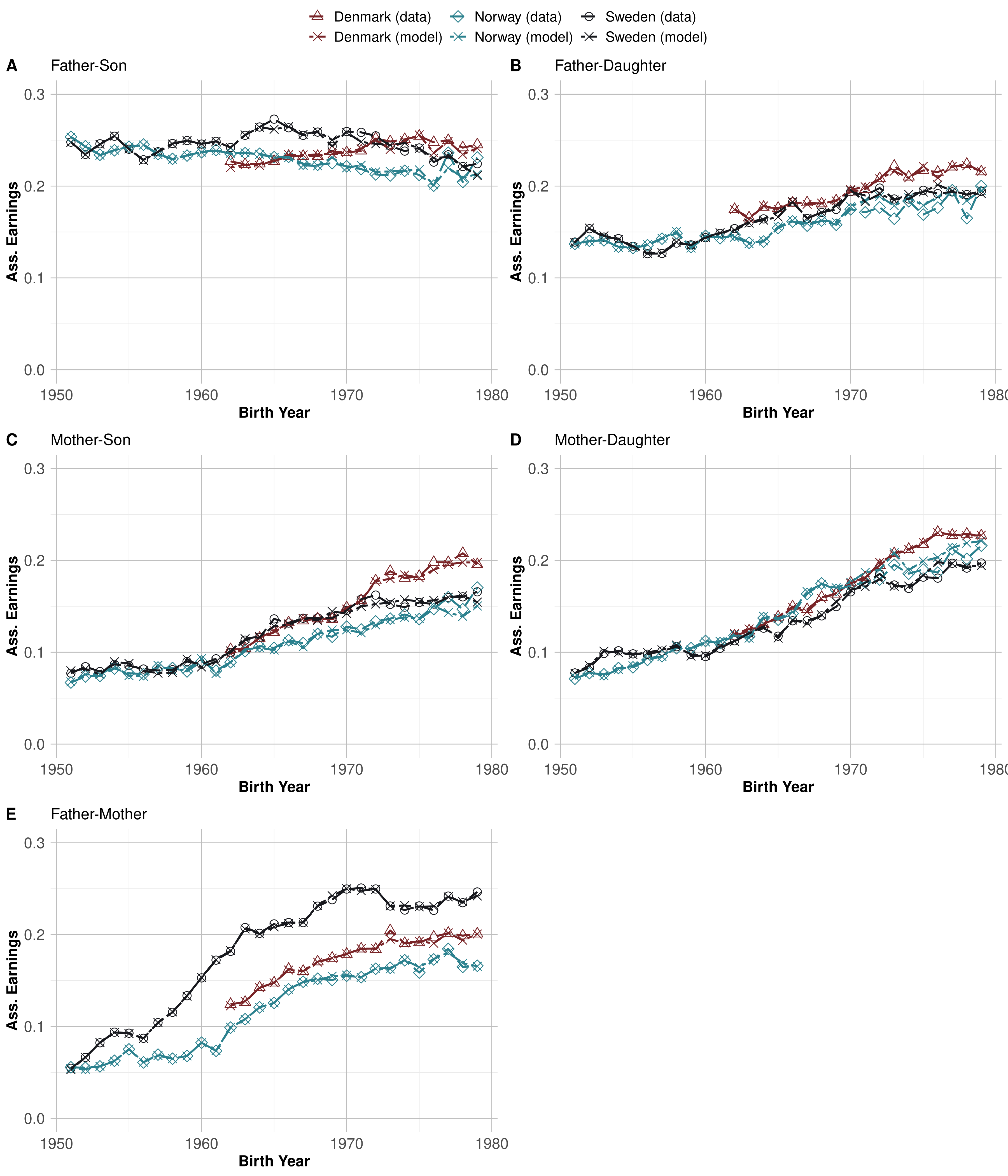}
\end{center}
\caption{Validation of Calibration Exercise.}
\label{fig:matching}
\scriptsize{\textit{Notes:} Each panel displays the empirical association between two income as observed in the data as well as the implied empirical association between the same two types of income in the simulated data as calibrated in the decomposition model. ``Birth Year'' refers to birth year of the \textbf{child} in each parent-child pair.}
\end{figure}

In Figure \ref{fig:matching}, we illustrate how the implied empirical association between the two types of income in the (calibrated) simulated data compares to the empirical association between the same two income as observed in the data. These figures validate the quality of the calibration exercise.

\subsection{Decomposition method} \label{app:decomp}

In order to decompose changes in rank associations into effects associated with changes in the modeling parameters, we compute ``counterfactual'' income associations through the following process. First, we re-simulate the model using our calibrated time-specific parameters while holding \textit{one} parameter fixed at the calibrated value from some baseline period (say, cohort 1962).\footnote{We also allow the aggregate gender-specific income distributions that we obtained from the data to vary over time.} Second, we re-compute the intergenerational rank associations with this ``counterfactual'' set of model parameters. Differences in estimated intergenerational rank associations are attributed to the parameter that was held fixed.

Specifically, we first define $\tilde{\beta}_{t}$ as the rank association between joint parental income and child income obtained from the simulated data (using the calibrated set of parameters). Hence we may define $\tilde{\beta}_{t}\equiv\beta\left(\psi_{t},\kappa_{t},\alpha_{t},\phi_{t}^{M},\phi_{t}^{D}\right)$. Then, we define $\tilde{\beta}_{t,\underline{t}}^{b}$ in a similar fashion, but we fix one parameter $b_{t}\in\left(\psi_{t},\kappa_{t},\alpha_{t},\phi_{t}^{M},\phi_{t}^{D}\right)$ to the calibrated value in period $\underline{t}$. For instance, $\tilde{\beta}^{\psi_{\underline{t}}}_t\equiv\beta\left(\psi_{\underline{t}},\kappa_{t},\alpha_{t},\phi_{t}^{M},\phi_{t}^{D}\right)$. Finally, the part of the trend in $\tilde{\beta}_t$ that can be attributed to parameter $b$ is simply the difference in trend between $\tilde{\beta}_t$ and $\tilde{\beta}^{b}{t,\underline{t}}$, while the part of the actual trend in $\beta_t$ that can jointly be attributed other factors than decomposition parameters and changes in the aggregate gender-specific income distributions is the difference in trend between $\beta_t$ and $\tilde{\beta}_t$.

\end{document}